\documentclass[preprint,pra,amsfonts]{revtex4-1}
\usepackage{graphicx}
\usepackage{times}
\usepackage{color}
\usepackage[T1]{fontenc}
\usepackage{amssymb}
\usepackage{amsmath}

\begin{document}
\title{An introduction to quantum measurements with a historical motivation}
\author{Leonardo Andreta de Castro$^{1}$}
\affiliation{Department of Electrical and Computer Engineering, Duke University, Durham, NC, 27708.}
\author{Ol\' impio Pereira de S\' a Neto$^{2}$}
\affiliation{Coordena\c{c}\~ao de Ci\^{e}ncia da Computa\c{c}\~ao, Universidade Estadual do Piau\'i, CEP 64202-220, Parna\'iba, PI, Brazil.}
\author{Carlos Alexandre Brasil$^{3,4}$}
\email{carlosbrasil.physics@gmail.com}
\affiliation{Department of Physics and Materials Science (FCM), S\~ao Carlos Institute of Physics (IFSC), University of S\~ao Paulo (USP), PO Box 369, CEP 13560-970, S\~ao Carlos, SP, Brazil}
\affiliation{Departamento Acad\^{e}mico de Ci\^{e}ncias da Natureza (DACIN), Campus Corn\'elio Proc\'opio (CP),
Universidade Tecnol\' ogica Federal do Paran\' a (UTFPR), Avenida Alberto Carazzai, 1640, CEP 86300-000, Corn\' elio Proc\' opio, PR, Brazil}

\begin{abstract}
We provide an introduction to the theory of quantum measurements that is centered on the pivotal role played by John von Neumann's model.
This introduction is accessible to students and researchers from outside the field of foundations of quantum mechanics and presented within a historical context.
We first explain the origins and the meaning of the measurement problem in quantum theory, and why it is not present in classical physics.
We perform a chronological review of the quantization of action and explain how this led to successive restrictions on what could be measured in atomic phenomena, until the consolidation of the orthodox interpretation of quantum mechanics.
The clear separation between quantum system and classical apparatus that causes these restrictions is subverted in von Neumann's paradigmatic model of quantum measurements,
a subject whose concepts we explain, while also providing the mathematical tools necessary to apply it to new problems.
We show how this model was important in discussing the interpretations of quantum mechanics and how it is still relevant in modern applications.
In particular, we explain in detail how it can be used to describe weak measurements and the surprising results they entail.
We also discuss the limitations of von Neumann's model of measurements, and explain how they can be overcome with POVMs and Kraus operators.
We provide the mathematical tools necessary to work with these generalized measurements and to derive master equations from them.
Finally, we demonstrate how these can be applied in research problems by calculating the Quantum Zeno Effect.
\end{abstract}

\maketitle

\section{Introduction}

Quantum mechanics caused a profound change to our understanding of nature, giving it an inherently probabilistic character.
Previously, classical physics favored the view of Pierre-Simon Laplace (1749-1827) that the future of the universe could be exactly predicted from its current conditions \cite{Laplace}.
The only obstacles would be our lack of complete knowledge of every particle that exists, and the difficulty in solving all the necessary differential equations.
The conflict between the new quantum physics and the previous paradigm motivated the famous dictum attributed to Albert Einstein (1879-1955) that \textit{''God does not throw dice''} \cite{Pais}.

Today, the non-deterministic character of quantum theory is not a mere curiosity, but a fact that has impact in society.
For example, the developing technology of quantum computation uses the superposition states of quantum mechanics as an edge over classical computers \cite{Shor}.
Techniques such as  measurement-based quantum computation \cite{Briegel,Briegel1} make it even clearer that being acquainted with the theory of measurement is now important even for applied research.

The significance of this subject contrasts with how its discussion was for a long time discouraged among physicists and relegated to the domain of philosophy, something that is reflected in some textbooks still in use~\cite{Eisberg1,Messiah,Schiff,Eisberg2,Merzbacher}.
Only more recent textbooks began to feature these foundational questions prominently, such as Ballentine \cite{Ballentine3}, Peres \cite{Peres}, and Griffiths \cite{Griffiths2}.

In this article, we provide an introduction to the quantum theory of measurements that is accessible to students and researchers from outside the field of foundations quantum mechanics, while also giving a historical context.
Our main innovation is to center our exposition on the model of measurements proposed by John von Neumann (1903-1957), whose long-lasting influence is not often emphasized enough.
Indeed, his work is still important for current topics such as weak measurements, quantum algorithms, and even the many-worlds interpretation of quantum mechanics.

We begin the article with a qualitative overview of the problem of measurement in quantum mechanics.
In Sec. 2, we present a broad discussion of measurements in classical physics, which will begin to be subverted by the initial developments of quantum mechanics, presented in Sec. 3.
In Sec. 4, we explain the introduction of the wave function, which would lead to a statistical interpretation of quantum mechanics, which is the subject of Sec. 5.
The current complete orthodox formulation of the measurement problem is given in Sec. 6 and Sec. 7, which deal with the principle of complementarity and uncertainty relations, respectively.

After this qualitative introduction, we provide all the conceptual and mathematical tools to apply his description of measurements.
In Sec. 8 we explain the concepts involved in von Neumann's model and the required mathematical calculations.
Sec. 9 deals with an important concept that is implicit in von Neumann's model---entanglement---and the apparent paradoxes it implies.
Further discussions of these foundational questions using von Neumann's model led to the many-worlds interpretation and the much-researched decoherence program, as shown in Sec. 10.
As an example of a more recent development, in Sec. 11 we talk about the inclusion of the time variable in the measurement process and the surprising phenomena related to weak measurements.

Despite its wide range of applications, von Neumann's model sometimes is not sufficient to describe a quantum measurement.
After this qualitative introduction, we present a typology of measurements in Sec. 12 and explain why a complete measurement described by von Neumann's model is called projective.
We generalize these into POVMs in Sec. 13, and present Kraus operators as a means of describing their dynamics.
These are used as tools to derive master equations for finite-time measurements in Sec. 14, which are subsequently used in Sec. 15 to describe an example of application, the Quantum Zeno Effect.

\section{Measurements before quantum physics}

Measurements that are part of our everyday life are assumed to be simple and intuitive processes, where the measurement apparatus---the timer, the ruler, the scale---remains distinct from the measured object.
Sometimes, however, the interaction between the apparatus and the object has to be taken into account.
For example, when we use a spring scale, we have to make sure that it will not deform excessively and violate Hooke's law.

There are a few much more fundamental cases in classical physics where this interaction is important:
\begin{enumerate}
	\item If we wish to determine experimentally the electric field of a charge distribution by means of a probe charge \cite{Purcell}, the latter must be as small as possible so that its interaction with the distribution will not cause a displacement of its constituent charges and significantly alter the system. To avoid these problems, the definition of the electric field assumes that the charges of the distribution are fixed.

	\item When we measure the temperature of a body with a thermometer, we implicitly use the zeroth law of thermodynamics \cite{Sears}: we put the thermometer in contact with the system whose temperature we wish to determine and wait until they both reach thermal equilibrium.
	However, the thermometer must be chosen so that the temperature of the system does not suffer great variation before the equilibrium is reached.
\end{enumerate}
Moreover, in the theory of relativity \cite{Eisberg2,French1}, even measurements of space and time have to take into account the relative movement of the system with respect to the measurer, in stark contrast with Newtonian physics.

However, in all the cases mentioned above, the effects of the measurement apparatus over the measured system can be theoretically calculated and minimized with the improvement of the equipments and the techniques.
This is not the case in quantum physics, where the measurement process is often considered intrinsically probabilistic.
Measurements no longer find a result that can be predicted with absolute certainty, except for very specific cases.
We will explain from the beginning how this situation came to be.

\section{The Old Quantum Theory}

The first step towards overcoming the classical view of measurements happened with the emergence of the \textit{Old Quantum Theory}\cite{Bucher,Hermann,Jammer,Feldens,Tomonaga} a heuristic approach to atomic phenomena that made explicit use of elements from
classical physics together with \emph{ad hoc} postulates.
Despite this apparently precarious character, it provided an explanation for the black body spectrum, the specific heat of solids, the photoelectric effect, and the atomic spectrum \cite{Castro,Abro,Duncan1,Duncan2,Parente,Studart,Jammer}.
However, each of these developments also played a role in restricting what could be measured in an atomic system.

\subsection{Quantization of energy}

The Old Quantum Theory began with the hypothesis of quantization proposed by Max Planck (1858-1947) in 1900 as a solution to the problem of \emph{black body} radiation \cite{Planck1orig,Planck2orig,Haar}.
This was a term coined by Gustav Kirchhoff (1824-1887) to designate a body that absorbs all the radiation incident upon it.
Many previous attempts had been made to explain the spectrum of such a body based on classical physics, to no avail \cite{Jammer,Kuhn}.
Planck's success was characterized by himself as ``\emph{an act of desperation [done because] a theoretical explanation [to the black--body--radiation spectrum] had to be supplied at all cost, whatever the price}''  \cite{Jammer}.

What Planck was willing to sacrifice to obtain his solution was nothing less than the continuity of the values of the exchanged energy inside the black body.
Planck's theory restricted the radiation in the interior of the cavity to discrete amounts (quanta) of energy proportional to its frequency $\nu$ multiplied by a constant $h$ (now known as Planck's constant):
\[
	E = h \nu.
\]

Within the next decade, this hypothesis would lead to the recognition of discontinuities in atomic physical processes in general \cite{Kuhn}.
A particularly important contributor to this development was Albert Einstein, who would apply the quantization to other phenomena, such as the specific heat of solids~\cite{Beck} and the photoelectric effect \cite{Arons,Beck}.
About the latter, what Einstein proposed was that:
\begin{quote}
	``[T]he energy of a light ray spreading out from a point source is not continuously distributed over an increasing space but consists of a finite number of energy quanta which are localized at points in space, which move without dividing, and which can only be produced and absorbed as complete units.'' \cite{Arons}
\end{quote}
While for Planck the quantization was restricted to cavities, Einstein extrapolated it to free electromagnetic radiation \cite{Kuhn}, thus introducing the quantum of light, or \emph{photon}.
In this way, quantities that could be found in a continuous range of values in classical physics became restricted to discrete amounts in quantum mechanics.

\subsection{The quantized atom}

In the beginning of the 20th century, matter was explained by different atomic models based on classical physics \cite{Jammer,Abro}.
A well-known example is the atom proposed by Joseph John Thomson (1856-1940) \cite{Thomson1,Thomson2}, which in his own words was ``\emph{built up of large numbers of negatively electrified corpuscles revolving  around the centre of a sphere filled with uniform positive electrification}'' \cite{Thomson1}.
Ernest Rutherford (1871-1937) was responsible for the next famous development:
\begin{quote}
``I supposed that the atom consisted of a positively charged nucleus of small dimensions in which practically all the mass of the atom was concentrated. The nucleus was supposed to be surrounded by a distribution of electrons to make the atom electrically neutral,
and extending to distances from the nucleus comparable with the ordinary accepted radius of the atom.'' \cite{Rutherford}
\end{quote}
Rutherford's model was preceded by the similar \textit{Saturnian model} by Hantaro Nagaoka (1865-1950) \cite{Nagaoka,Nagaoka1,Nagaoka2}.
There was also the model by Arthur Erich Haas (1884-1941), who made the first attempt to apply Planck's quantum of action to the constitution of the atom \cite{Hermann}, and with this managed to calculate the \textit{Rydberg constant} that characterizes many atomic spectra---albeit with a wrong numerical factor \cite{Hermann,Jammer,Kuhn}.

A more successful attempt at quantizing the atom was the one made in 1913 by Niels Bohr (1885-1962).
An assistant to Rutherford since 1912, Bohr grew concerned that there were no means to conciliate the atomic model proposed by his mentor with Newton's mechanics and Maxwell's electrodynamics.
He then introduced what Werner Heisenberg (1901-1976) would call ``\textit{the most direct expression of discontinuity in all atomic processes}'' \cite{Heisenberg2} by restricting the electron orbits to quantized \emph{stationary stable states}.
Bohr stated that it made no sense to ask where the electron was to be found during the transition between these orbits, the so-called \textit{quantum jumps} \cite{Laloe}.

An important tool used by Bohr was the \emph{correspondence principle}, formulated in distinct forms by him, Planck and Heisenberg \cite{Abro,Espagnat,Jammer,Kuhn,Liboff,Makowski}.
This principle states that when we take the limits of some basic parameters of quantum mechanics we must find the classical results again \cite{Jammer,Liboff,Makowski,Hassoun,Bhattacharyya}.
Bohr used this to find the dependence between the electronic frequencies of emission/absorption and translation around the nucleus, obtaining the correct value for Rydberg's constant \cite{Bohr1}.

An improvement over Bohr's model would be attained through phase-space analysis, something that is not surprising, since Planck's constant $h$ \cite{Ishiwara3,Castro,Abro,Bucher,Eckert,Haar} has dimensions of action.
The works of Arnold Sommerfeld (1868-1951) \cite{Sommerfeld1,Sommerfeld2,Sommerfeld3,Sommerfeld4}, William Wilson (1875-1965) \cite{Wilson,Wilson2}, Jun Ishiwara (1881-1947) \cite{Ishiwara,Ishiwara2}, and Bohr \cite{Waerden} led to the restriction of the stationary states of atomic systems to those that satisfied~\cite{Ishiwara3,Castro}:
\begin{equation}
	\oint p \; \mathrm{d}q = n h, \label{WISB}
\end{equation}
where $q$ is a position coordinate, $p$ its conjugated momentum, and $n$ is an integer.
The imposition of Eq. (\ref{WISB}) to the hydrogen atom allowed the explanation of the fine structure of its spectrum \cite{Sommerfeld4}.
According to Planck, this was ``\emph{an achievement fully comparable with that of the famous discovery of the planet Neptune whose existence and orbit was calculated by Leverrier before the human eye had seen it}'' \cite{Planck5}.
The theory was being successful, but the idea of measuring the trajectory of the particle between its discrete allowed states had to be abandoned.

\subsection{Wave-particle duality}

The old controversy about whether light is a particle or a wave was rekindled when the x-rays were discovered in 1895 by Wilhelm Conrad R\"{o}ntgen (1845-1923).
Some facts suggested x-rays had a particle-like character: they were emitted with the acknowledgedly corpuscular $\alpha$ and $\beta$ radiations, and energetic considerations led some to believe that they were particles \cite{Jammer}.
However, diffraction experiments suggesting also a wave-like character prompted the experimentalist William Henry Bragg (1862-1942) to say:
\begin{quote}
	``The problem becomes, as it seems to me, not to decide	between two theories of x-rays, but to find, as I have said elsewhere, one theory which possesses the capacity of both.'' \cite{Jammer}
\end{quote}

Among the experimentalists who worked with x-rays, Maurice de Broglie (1875-1960) was responsible for one of the first rigorous determinations of the charge of the electron \cite{Abro}.
Maurice had already considered an analogy between electrons and x-rays, implying a possible fundamental unity of radiation and matter \cite{Martins} which influenced his younger brother Louis de Broglie (1892-1987).
Louis had worked with telegraphy and communication during World War I, becoming familiarized with wave mechanics \cite{Martins,deBroglie2}.
After the war, he came to work together with his brother on the analysis of the properties of the x-rays and read the works of Einstein, Bohr, and Sommerfeld \cite{Martins} about the quanta of action.
This led him to propose wave properties for matter in the Ph.D. thesis he defended in 1924 \cite{deBroglieOrig,deBroglieTranslated}, which some consider ``\textit{the most influential and successful doctoral dissertation in the history of physics}'' \cite{MacKinnon}.

De Broglie postulated that material particles should be associated to a plane and infinitely extended wave, whose energy and momentum were connected with the frequency and wave length by the same relations that Einstein had used.
He associated the momentum $p$ of a particle to its wave length $\lambda$,
\[
	p = \frac{h}{\lambda},
\]
which would be justified by the electron diffraction experiments and allowed an explanation of Bohr's theory of electronic orbits \cite{Jammer,Martins}.
In his theory, when the orbit length is equal to an integer number of electron wavelengths, the orbit is stable.
To de Broglie, the wave guided the electron's trajectory, \emph{without probabilistic considerations}---wave and particles were two aspects of the system coexisting at the same level, independent of the measurement apparatus \cite{Jammer2,Martins}.

Einstein had contact with de Broglie's ideas via Paul Langevin (1872-1946), who participated in de Broglie's thesis examination before meeting Einstein at the Fourth Solvay Conference in 1924 \cite{Jammer}.
He would put this hypothesis to practice right in the next year, stating in an article about an ideal gas:
\begin{quote}
	``The way we can associate a (scalar) undulatory field to a material particle or to a system of material particles was shown to us in a remarkable work by Mr. L. de Broglie.''  \cite{Einstein1orig}
\end{quote}
Thus Einstein associated a wave character to gas particles, following the reverse path of his 1905 article that attributed corpuscular features to light.

Alongside the works of de Broglie and Einstein, the last attempt to explain electronic transitions and solve the ``paradox'' of the wave-particle duality would be made by Bohr, together with Hendrik Anthony Kramers (1894-1952) and John Clark Slater (1900-1976), in what would be known as the BKS theory \cite{BKS}.
The authors associated the electromagnetic wave inside the atom with the probability of electronic transitions, and hypothesized a \emph{virtual radiation field} that would be responsible for communication between far away atoms and inducing spontaneous transitions.
This unprecedented concept of a \emph{probability wave}\cite{Heisenberg} would become a central feature in quantum mechanics.
However, the same would not be true for the BKS theory.
When applied to the Compton scattering, BKS predicted that the recoil direction of the electron would exhibit a statistical distribution, something that was soon refuted experimentally \cite{Jammer}.

To conclude, Old Quantum Mechanics was a heuristic theory strongly based on classical physics that explained a few recently-observed phenomena using hypotheses such as the quantization of action, the wave-particle duality, and the suggestion of the existence of a field of probabilistic character.
These facts imposed a few constraints on which measurable quantities had meaning in quantum theory: some observables only had discrete values, and there was no traceable transition between them.
It was the prelude to the problem of measurement that would emerge with the complete quantum theory.

\section{The rise of the wave function}

De Broglie's waves had a lasting influence in modern quantum mechanics after they were transformed into the modern concept of wave function by Erwin Schr\"odinger (1887-1961).
Schr\"odinger used the analogy proposed by William Rowan Hamilton (1805-1865) between the motion of a point mass under the influence of a potential $V\left(x,y,z\right)$ and the propagation of light rays in a medium with refractive index $n\left(x,y,z\right)$ \cite{Jammer,Joas,Koberle,Wessels}.
Hamilton showed \cite{Abro,Fetter} that the optical path and the mechanical \emph{action} $S$ obey the same variational principle.
In terms of a Hamiltonian function $H(q_k,p_k,t$), which depends on the coordinates $q_k$ and their associated momenta $p_k$, the principle can be expressed as the Hamilton-Jacobi equation:
\begin{equation}
	H\left(q_{k},p_k, t\right)=- \frac{\partial S}{\partial t},\label{HJ}
\end{equation}
where the momenta are given by:
\begin{equation}
	p_{k}=\frac{\partial S}{\partial q_{k}}.
	\label{momenta}
\end{equation}

Hamilton's formalism was treated solely as a formal analogy, lacking deeper physical meaning \cite{Jammer}.
Nevertheless, Schr\"odinger was convinced that it represented not only a robust mathematical theory, but something real in nature.
Reading Einstein's 1925 article led him to find de Broglie's thesis, which he proceeded to study systematically \cite{Martins}.
Schr\"{o}dinger then used the optical-mechanical analogy\cite{Schroedinger2} as a heuristic argument to derive a wave equation for matter, introducing the \emph{wave function} $\psi$, related to the action via \cite{Schroedinger1}:
\begin{equation}
	S = K \ln\psi,
	\label{psi}
\end{equation}
where $K$ is a constant to be determined.
If one replaces Eq. (\ref{psi}) in Eq. (\ref{HJ}), one obtains the Schr\"odinger equation, to this day one of the pinnacles of quantum mechanics:
\begin{equation}
	H(q_k,p_k,t) \psi = - K \frac{\partial \psi}{\partial t},
	\label{Schroedinger}
\end{equation}
where the momenta are found replacing Eq. (\ref{psi}) in Eq. (\ref{momenta}):
\begin{equation}
	p_{k} \psi=K \frac{\partial \psi}{\partial q_{k}}.
\end{equation}

Currently, we choose $K=-i \hbar$, where $\hbar=h/2\pi$,
but Schr\"odinger initially chose $K=\hbar$ because he at first considered $\psi$ to be real, before having to review this idea.
In the beginning, he did not ascribe any physical interpretation to $\psi$, but he did remark that:
\begin{quote}
	``It is, of course, strongly suggested that we should try to connect the function $\psi$ with some \emph{vibration process} in the atom, which would more nearly approach reality than the electronic orbits, the real existence of which is being very much questioned today.'' \cite{Schrodinger}
\end{quote}

The problem of interpreting the function $\psi$ persisted despite the success of the Schr\"odinger equation in calculating the values of energies in experimental situations.
To Schr\"{o}dinger, the wave function was as much a real entity as the electromagnetic waves described by the Maxwell's equations \cite{Jammer2,Laloe}.
However, this realistic interpretation faced some obstacles, such as the fact that $\psi$ is a complex entity that depends on the set of observables chosen for its representation (the most evident example, at that time, was the case of position and momentum representations).

In 1926, Schr\"{o}dinger argued for the necessity of attributing electromagnetic meaning to $\psi$, because the electron, in transitions, emits electromagnetic waves whose energy is equal to the difference between two eigenvalues.
Schr\"{o}dinger then established that $\left|\psi\right|^{2}$ obeys a continuity equation (known today as \emph{probability current continuity} \cite{CohenTannoudji}) and proposed that it multiplied by the electric charge provides the \emph{density of charge}.
In this view, charge would be distributed in space, unlike de Broglie's view that it should be located at a given point \cite{Martins}.
This new diffuse character of the charge of a particle would open space for a probabilistic interpretation where the particle itself would not be properly localized.

\section{The probabilistic interpretation}

The introduction of the wave function gave a diffuse character to the quantum particles, associating with them also wave--like features.
This was the first step towards making their measurable quantities no longer as well defined as in classical physics, but rather probabilistic phenomena.
This change was due to the contribution of two influential physicists: Max Born (1882-1970) and Paul Dirac (1902-1984).

\subsection{Born's rule}

Max Born took a position at the University of G\"{o}ttingen in 1921, where he established a research program of elaborating quantum theories for those simpler atomic problems that had a great amount of empirical data available \cite{Im}.
Analyzing the problem of an electron beam colliding with an atom, Born provided a probabilistic interpretation of $\psi$ in a work of just a few pages and a mainly qualitative argument, consisting of the asymptotic analysis of the elastic collision of an electron with an atomic core\cite{BornOrig,WheelerZurek}.

Born considers a beam with $F$ particles per unit time colliding with an atom.
The number of particles $\mathrm{d}n$ that are deflected in the infinitesimal solid angle $\mathrm{d}\Omega$ is obtained via the \emph{differential scattering cross section}, $\sigma(\theta,\phi)$, which depends on the potential of the atom:
\begin{equation}
	dn=F \sigma\left(\theta,\phi\right) d\Omega.
	\label{dn}
\end{equation}

Initially, the beam is not affected by the interaction with the atom and its wave function is assumed to be the plane wave $e^{ikz}$, where $z$ is the direction of propagation.
After the scattering, and when the beam is distant from the target, it will be composed of a part proportional to the initial wave function and a part containing a scattering contribution:
\[
	\psi_{\text{scatter}} = f_{k} \left(\theta,\phi\right) \frac{e^{ikr}}{r}.
\]
The scattering term has the form of a spherical wave, and is inversely proportional to the distance in order to conserve the flux of $|\psi|^2$, a condition Schr\"odinger had shown to be necessary.
Likewise, the part of this flux that is deviated from the main beam to a certain direction is given by the cross section $\sigma(\theta,\phi)$ (calculations can be seen in detail in Ref. \cite{CohenTannoudji}):
\[
	\sigma\left(\theta,\phi\right)=\left|f_{k}\left(\theta,\phi\right)\right|^{2},
\]
so that Eq. (\ref{dn}) becomes:
\[
	\mathrm{d}n = F \left| f_{k} \left(\theta,\phi\right) \right|^{2} \mathrm{d}\Omega.
\]

Therefore, the number of particles detected in the solid angle by unit of time will be proportional to the squared modulus of the scattered wave function:
\begin{equation}
	\mathrm{d} n \propto \left| \psi_\text{scatter} (r)\right|^2.
	\label{Born}
\end{equation}

For a single particle, Born interpreted the squared modulus of the wave function as the probability of measuring it in that direction after the scattering.
He summarized how this interpretation inserted uncertainty into the nature of measurement as follows:
\begin{quote}
	``One gets no answer to the question, `what is the state after the collision', but only to the question `how probable is a specified outcome of the collision'.'' \cite{WheelerZurek}
\end{quote}

Born published a considerably longer paper in  the same year\cite{BornLong} where he stated the current paradigm of the evolution of the wave function in quantum mechanics:
\begin{quote}
	''[T]he motion of the particles follows laws of probability, but the probability itself propagates in harmony with the causal law.'' \cite{Ludwig}
\end{quote}
In this paper he also took a step further and attributed the probabilistic interpretation not to the transition, but to the stationary states.
Born started with the time-independent Schr\"odinger equation, considering that $\psi$ in Eq. (\ref{Schroedinger}) is an eigenfunction of the time derivative:
\begin{equation}
	\hat{H} \psi_{n} (\mathbf{x}) =E_{n}\psi_{n} (\mathbf{x}).\label{eqautoval}
\end{equation}
Here, the system is considered non-degenerate and discrete, and $\psi_n$ are orthonormal eigenfunctions of the Hamiltonian:
\begin{equation}
	\int_V \mathrm{d}^3 x \; \psi_{n} \left(\mathbf{x}\right) \psi_{m}^{*} \left(\mathbf{x}\right) = \delta_{n,m}, \label{normalizacao}
\end{equation}

Born assumes the system is submitted to a potential $U\left(\mathbf{x}\right)$, so that Eq. (\ref{eqautoval})
takes the following expanded form (adapting the notation):
\begin{equation}
	\frac{\hbar^{2}}{2m}\nabla^{2}\psi_n (\mathbf{x}) +\left[ E_n - U(\mathbf{x}) \right] \psi_n(\mathbf{x})=0,
\end{equation}
where $m$ is the particle mass.
Now, multiplying by $\psi_{m}^{*}$ and integrating over the volume $V$, we obtain:
\begin{equation}
	\int_V \mathrm{d}^3 x \left\{ \frac{\hbar^{2}}{2m} \psi_{m}^{*}(\mathbf{x}) \nabla^{2} \psi_{n} (\mathbf{x}) + \left[ E_n - U(\mathbf{x}) \right] \psi_{m}^{*} (\mathbf{x}) \psi_{n} (\mathbf{x}) \right\} = 0. \label{a1}
\end{equation}
In the first term on the left-hand side, we use one of \emph{Green's identities} \cite{Butkov,Morse}:
\begin{equation}
	\int_{V} \mathrm{d}^3 x\; \psi_{m}^{*}(\mathbf{x}) \nabla^{2} \psi_{n} (\mathbf{x}) =
	\int_{S} \mathrm{d}a\;  \psi_{m}^{*}(\mathbf{x}) \mathbf{\hat n} \cdot \boldsymbol\nabla \psi_{n} (\mathbf{x})
	- \int_{V} \mathrm{d}^3 x\; \boldsymbol\nabla \psi_{m}^{*}(\mathbf{x}) \cdot \boldsymbol\nabla \psi_{n} (\mathbf{x}),
\label{a2}
\end{equation}
where $S$ is a surface that involves the volume $V$.
We will consider that the wave functions vanish at this surface---which could be as distant as necessary---so that when we replace Eq. (\ref{a2}) in Eq. (\ref{a1}), we obtain the expression:
\begin{equation}
	\int_V \mathrm{d}^3 x \left\{ - \frac{\hbar^{2}}{2m} \boldsymbol\nabla \psi_{m}^{*}(\mathbf{x}) \cdot \boldsymbol\nabla \psi_{n} (\mathbf{x}) + \left[ E_{n} - U(\mathbf{x}) \right] \psi_{m}^{*} (\mathbf{x}) \psi_{n} (\mathbf{x}) \right\} = 0.
\end{equation}
Using the normalization from Eq. (\ref{normalizacao}), we find:
\begin{align}
	\int_V \mathrm{d}^3 x \left\{ \frac{\hbar^{2}}{2m} \boldsymbol\nabla \psi_{m}^{*}(\mathbf{x}) \cdot \boldsymbol\nabla \psi_{n} (\mathbf{x}) + U(\mathbf{x}) \psi_{m}^{*} (\mathbf{x}) \psi_{n} (\mathbf{x}) \right\} = &
	E_n \int_V \mathrm{d}^3 x\; \psi_{m}^{*} (\mathbf{x}) \psi_{n} (\mathbf{x}) \nonumber \\
	= & E_n \delta_{n,m}.
	\label{a3}
\end{align}
The terms on the left-hand side can be interpreted, respectively, as the volumetric integrals of the kinetic and potential energies, so that ``\emph{each energy level {[}$E_{n}${]} can therefore be regarded as a space integral of the energy density of the eigenvibrations}'' \cite{Ludwig}.

Following a similar reasoning for any normalized wave function $\psi$, the energy $E$ of the particle associated to it should be the sum of the integrals of the kinetic and potential energy densities:
\begin{eqnarray}
	\int_{\mathbb{R}^3} \mathrm{d}^3 x \left\{ \frac{\hbar^{2}}{2m} \left|\boldsymbol\nabla\psi\right|^{2} + U(\mathbf{x}) \left|\psi(\mathbf{x}) \right|^{2} \right\} & = & E.
	\label{b3}
\end{eqnarray}
This wave function can always be decomposed into a superposition of eigenfunctions $\psi_n(\mathbf{x})$:
\begin{equation}
	\psi (\mathbf{x}) = \sum_n c_n \psi_n (\mathbf{x}),
	\label{dirac1}
\end{equation}
which can be substituted into Eq. (\ref{b3}) and compared to Eq. (\ref{a3}), yielding:
\begin{equation}
	E = \sum_r \left| c_n \right|^{2} E_n. \label{a4}
\end{equation}

According to Eq. (\ref{a4}), the total energy is obtained by weighing the energy of each eigenfunction by $\left|c_{n}\right|^{2}$.
Hence, the $\left|c_{n}\right|^{2}$ are probabilities associated with finding the system in the eigenstate $\psi_n$.

The interpretation of the square modulus of the wave function as the probability of finding the particle in a region of space, as in Eq. (\ref{Born}), or with the probability of finding it in a certain eigenstate, as in Eq. (\ref{a4}), is called \textit{Born's rule}.
This is still an important part of quantum theory, although it seems to contradict the classical idea of measurement as the process of acquiring a value that already exists in nature.
Despite that, Born still saw the particle as a point mass, possessing at each instant a definite position and a definite momentum, like a classical entity \cite{Jammer2}.
The wave function $\psi$ would represent our knowledge about the physical system, and not the system itself.

\subsection{Dirac's contribution}

While theoretical quantum mechanics was being intensely developed in G\"ottingen by Born, Heisenberg, Kramers, and Pauli, who were in continuous communication with Bohr in Copenhagen \cite{Gottfried}, England had not contributed much.
This started to change in July 1925, after Heisenberg gave a seminar in Cambridge that was attended by Ralph Howard Fowler (1889-1944), who asked for more details and forwarded the received material to a young researcher under his supervision, Paul Dirac, with the question: ``\emph{What do you think of this?}''.
Dirac was on holiday in Bristol, his homeland, and would declare later:
\begin{quote}
	``[I]t needed about ten days or so before I was really able to master it. And I suddenly became convinced that this would provide the key to understanding the atom.''
\end{quote}
When his first article on the subject reached Germany, there was such a great outburst that Born would declare afterwards it had been \textit{``one of the great surprises of my scientific life''} \cite{Gottfried}.

Dirac's long-lasting influence is evident in the fact that he introduced many of the notations and conventions still in use in modern quantum mechanics---for example, bras and kets.
His articles, extremely clear even today, contrast with the difficult texts of the G\"ottingen group, published in German with rare English translations.

For our discussion of the measurement problem, Dirac made the significant contribution of
modifying Eq. (\ref{dirac1}) by adding an extra time-varying term to the potential for $t>0$, so that the coefficients $c_n$ would vary in time for $t>0$.
He called these new time-dependent coefficients $a_n (t)$:
\begin{equation}
	\psi = \sum_r a_{n}\left(t\right) \psi_{n}. \label{dirac2}
\end{equation}

Replacing Eq. (\ref{dirac2}) in Eq. (\ref{Schroedinger}), Dirac found the differential equations for each of the $a_{n}\left(t\right)$.
In his own words,
\begin{quote}
	``We shall consider the general solution [Eq. (\ref{dirac1})] to represent an assembly of the undisturbed atoms in which $\left|c_{n}\right|^{2}$ is the number of atoms in the \emph{n}th state, and shall assume that [Eq. (\ref{dirac2})] represents in the same way an assembly of the disturbed atoms, $\left|a_{n}\left(t\right)\right|^{2}$ being the number in the $n$th state at any time $t$. We take $\left|a_{n}\right|^{2}$ instead of any other function of $a_{n}$ because, as will be shown later, this makes the total number of atoms remain constant.''\cite{Dirac3}
\end{quote}
Indeed, Dirac found that the number of atoms in the $n$th state is $N_{n}=a_{n}a_{n}^{*}$ and demonstrated that $\dot{N}=\sum_{n}\dot{N}_{n}=0$.
Therefore, using an approach independent from the one used by Born, Dirac presented a similar interpretation of the coefficients of the expansion of the wave function in its eigenstates, although he did not use the term \emph{probability}.

\section{The complementarity principle}

As the probabilistic interpretation was adequate for experimental analyses, it became the hegemonic orthodox interpretation.
In this interpretation, the act of measurement influences the observed system \cite{Abro,Bunge,Hartle,Laloe}, disturbing it in a way that cannot be completely predicted by quantum theory \cite{Abro,Ballentine,Bunge,Hanson}.
Therefore, the wave functions are mere symbolic entities used to make probabilistic predictions about what is observed in the lab, under specific experimental conditions \cite{Ballentine}.
The complete character of the theory would be based on Bohr's pragmatic criterion: the description of the phenomena offered by quantum theory would encompass everything that is \emph{possible} to be experimentally measured \cite{Abro,Bunge,Laloe,Stapp},
as encapsulated in von Neumann's statement:
\begin{quote}
	``Although we believe that after having specified [the wave function] we know the state completely, nevertheless, only statistical statements can be made on the values of the physical quantities involved.'' \cite{Neumann}
\end{quote}
Yet, von Neumann considered the wave function just a part of a theoretical scheme and not an entity that represents a real aspect of nature \cite{Neumann}.

The orthodox interpretation is sometimes called \textit{Copenhagen interpretation},  although it was neither created by Niels Bohr nor developed in Copenhagen.
The term originated \cite{Howard} in the 1955 text ``\emph{The development of the interpretation of the quantum theory}'' \cite{Pauli} by Heisenberg, who would in a later book state that this interpretation actually emerged at the 1927 Solvay conference in Brussels \cite{Heisenberg}.
One of Bohr's closest collaborators, Leon Rosenfeld (1904-1974), even said that ``\emph{...we in Copenhagen do not like at all [the phrase `Copenhagen interpretation']}''\cite{Stapp}.

Despite this misnomer, Bohr's influence is sufficiently great for his view of measurement to be considered a paradigm in the orthodox interpretation.
To Bohr, we cannot treat the measurement process without necessarily taking into account the interaction between the system and the measurement apparatus\cite{Bohr4}---this is the concept of \emph{indivisibility} \cite{Charria}.
As the measurement apparatus is a macroscopic system that is classical by nature \cite{Espagnat,Jammer,Jammer2}, it must be necessarily described by classical physics \cite{Jammer,Omnes1}.
From this need to describe the experimental apparatus in classical language, Bohr derived his concept of complementarity \cite{Bohr7}.

Complementarity was so vital to Bohr's thinking that he would even suggest the expansion of it to other areas such as biology and anthropology \cite{Bohr7}.
In quantum mechanics, Bohr derived the concept of complementarity from the necessity to use the mutually-exclusive ideas of ``wave'' and ``particle''\cite{Jammer}, which he saw as mere abstractions when applied to quantum systems \cite{Abro,Bunge}.
A system can manifest its corpuscular or wave-like character \emph{depending on the experimental set-up used to observe it}.
This does not mean that it is a wave in certain occasions and a particle in others; the system is an entity of ``undefined'' nature that can reveal wave-like features when interacting with a given measurement apparatus, and particle-like features for another
measurement apparatus---the results would be mutually complementary, not exclusive \cite{Abro,Jammer}.

\begin{figure}[htb]
\includegraphics[width=0.495\textwidth]{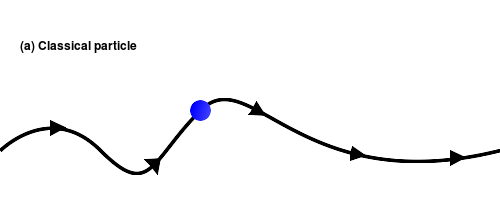}
\includegraphics[width=0.495\textwidth]{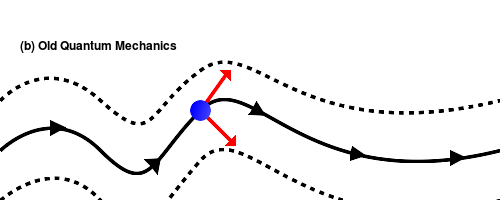}
\includegraphics[width=0.495\textwidth]{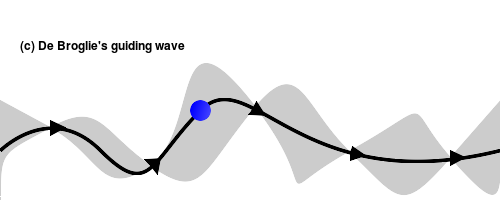}
\includegraphics[width=0.495\textwidth]{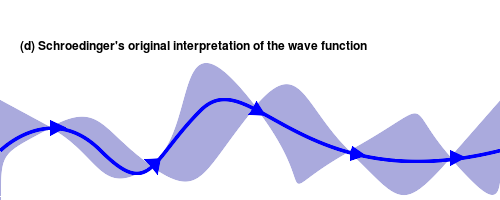}
\includegraphics[width=0.495\textwidth]{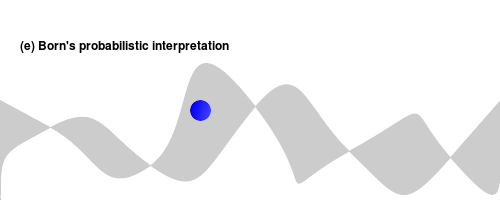}
\includegraphics[width=0.495\textwidth]{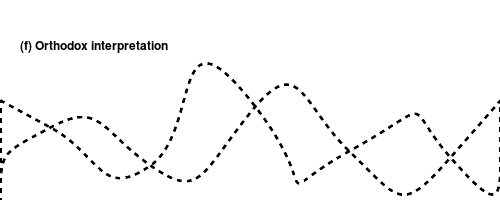}
\caption{Summary of the concepts of subatomic particles:
(a) Classically, they could move in a continuous trajectory in a potential.
(b) In the Old Quantum Theory, only certain stable trajectories were allowed, with quantum jumps occurring between them.
(c) De Broglie's waves would guide the particle in its trajectory.
(d) Schr\"odinger's original interpretation of the wave function would have the particle's charge distributed throughout it.
(e) To Born, the wave function would encapsulate the probabilities associated with the particle, which would have precise position and momentum at every instant.
(f) In the established orthodox interpretation, the wave function is an abstraction representing our maximum knowledge: position and momentum of the particle are only defined if associated with an experimental apparatus.}
\label{figure}
\end{figure}

As such, the complementarity principle does not apply to the isolated system, but only to the complete set including the measurement apparatus \cite{Ballentine,Bunge,Espagnat,Jammer,Laloe,Omnes1,Omnes2}.
As opposed to some later interpretations, Bohr saw the measurement as a consequence of the experimental set-up, having nothing to do with the consciousness of the observer.
Moreover, Bohr saw the wave packet collapse (the reduction of the wave function to the eigenstate corresponding to the measured eigenvalue) as a mere ``\emph{artifact of the formalism}'' \cite{Henderson,Howard}.
Despite this, Bohr believed that quantum mechanics could be applied to individual particles, rather than being just an ensemble theory:
\begin{quote}
	``To my mind, there is no other alternative than to admit that, in this field of experience, we are dealing with individual phenomena and that our possibilities of handling the measuring instruments allow us only to make a choice between the different complementary types of phenomena we want to study.'' \cite{Bohr7}
\end{quote}

Many authors \cite{Jammer2,Stapp} emphasize that Bohr's writings were not always clear enough.
Einstein, for example, said Bohr's clear thinking became obscure when written \cite{Jammer2}.
However, his position was summarized in his own words during an enlightening debate between him and Heisenberg:
\begin{quote}
	``That is the whole paradox of quantum theory.
	On the one hand, we establish laws that differ from those of classical physics; on the other, we apply the concepts of classical physics quite unreservedly whenever we make observations, or take measurements or photographs.
	And we have to do just that because, when all is said and done, we are forced to use language if we are to communicate our results to other people.'' \cite{Heisenberg1}
\end{quote}

\section{The uncertainty principle}

Any discussion of the measurement problem would be incomplete without mentioning Heisenberg's uncertainty principle.
Heisenberg was a doctoral student under the supervision of Arnold Sommerfeld, who had a singular view of physics and who sought to explain spectroscopic phenomena  based solely on experimentally-measured physical quantities \cite{Seth1,Seth2}.
This point of view was inherited by Heisenberg, who built his theory taking into account the way the parameters are determined experimentally---this assumption is called \emph{rule of restriction} \cite{Reichenbach}.
For example, he considered that the concept of electron orbit was inadequate due to the spatial dimensions involved, small enough that any observation would require a wavelength with such an energy that it would remove the electron from its orbit \cite{Heisenberg2}.
In his words:
\begin{quote}
	``[O]ne ought to ignore the problem of electron orbits inside the atom, and treat the frequencies and amplitudes associated with the line intensities as perfectly good substitutes.
	In any case, these magnitudes could be observed directly, and ... physicists must consider none but observable magnitudes when trying to solve the atomic puzzle.'' \cite{Heisenberg1}
\end{quote}

Based on this, Heisenberg concluded that the quantum theory demanded a revision of kinematic and mechanical concepts \cite{HeisenbergOrig,WheelerZurek}.
His analysis was based on the canonical commutation relation:
\begin{equation}
	\left[ \hat{p} , \hat{q} \right]= -i\hbar.
	\label{canonical}
\end{equation}
To illustrate his point, Heisenberg used an example that became famous \cite{Abro,Heisenberg2,Jammer,Neumann}:
an electron interacting with radiation of a sufficiently short wavelength $\lambda$ to be observed under an optical microscope.
This wavelength will define the uncertainty of the position observed by the microscope, $\Delta x\sim\lambda$.
But the electron will also suffer a Compton recoil \cite{Eisberg1,Eisberg2}, that implies an uncertainty in the momentum of magnitude $\Delta p\sim \hbar/\lambda$.
Thus, Heisenberg obtained the famous formula:
\begin{equation}
	\Delta x \Delta p \sim \hbar,
	\label{incDxDp}
\end{equation}
which constitutes a formulation of the \emph{uncertainty principle}.
What Eq. (\ref{incDxDp}) is expressing is that the measurement apparatus causes uncontrollable perturbations on the system being measured, making it impossible to obtain simultaneously the values of two conjugated variables with arbitrary precision \cite{Heisenberg2,Heisenberg,Reichenbach}.
Despite this formulation of the uncertainty principle not being very rigorous---for example, Heisenberg omits the angular aperture of the microscope lens---the general idea was confirmed by new uncertainty relations obtained shortly later by Ditchburn \cite{Ditchburn}, Robertson \cite{Robertson,Robertson1,Robertson2}, and Heisenberg himself \cite{Heisenberg2}.

More evidently than Bohr's complementarity principle, the uncertainty relations represent an essential distinction between quantum theory and classical physics \cite{Abro}.
Until then, we assumed that, despite the experimental apparatus being able to affect the system and the result of the measurement, we could improve the experimental set-up and take precautions to make the distortion of the result arbitrarily small.
But, after this point, a minimum perturbation became impossible to eliminate when two conjugate variables are measured.
Even if the imperfections were removed by a highly trained and competent observer with a perfectly-tuned equipment, a fundamental limitation would still exist \cite{Abro}.

\begin{table}[htb]
\begin{tabular}{|c|c|c|c|}
\hline
	\bf Year &
	\bf Theoretical Development &
	\bf Restrictions to Measurements &
	\bf Refs. \\
\hline \hline
	\bf 1900 &
	\begin{tabular}{c}
		Planck quantizes the energy \\
		exchanges in a black body
	\end{tabular} &
	\begin{tabular}{c}
		Energy inside a cavity \\
		only could be exchanged \\
		in discrete amounts
	\end{tabular} &
	(\onlinecite{Planck1orig}), (\onlinecite{Planck2orig}), {[}\onlinecite{Haar}{]} \\
\hline
	\bf 1905 &
	\begin{tabular}{c}
	Einstein quantizes free \\
	electromagnetic radiation
	\end{tabular} &
	\begin{tabular}{c}
		Light only could be \\
		detected in discrete amounts
	\end{tabular} &
	(\onlinecite{photonOrig}), {[}\onlinecite{Arons}{]} \\
\hline
	\bf 1913 &
	Bohr's atomic model  &
	\begin{tabular}{c}
		Trajectory of the electron \\
		between orbits cannot \\
		be measured
	\end{tabular} &
	(\onlinecite{Bohr1}), (\onlinecite{Bohr2}), (\onlinecite{Bohr3}) \\
\hline
	\bf 1924 &
	\begin{tabular}{c}
		de Broglie attributes \\ a wave-like \\
		character to matter
	\end{tabular}  &
	--- &
	(\onlinecite{deBroglieOrig}), [\onlinecite{deBroglieTranslated}] \\
\hline
	\bf 1926 &
	\begin{tabular}{c}
	Schr\"odinger introduces \\
	wave functions
	\end{tabular} &
	\begin{tabular}{c}
		The charge of a particle \\
		may not be localized
	\end{tabular} &
	(\onlinecite{Schroedinger1}), (\onlinecite{Schroedinger2}), {[}\onlinecite{Schrodinger}{]} \\
\hline
	\bf 1926 &
	Born's rule &
	\begin{tabular}{c}
		Measurements become \\
		probabilistic phenomena
	\end{tabular}  &
	(\onlinecite{BornOrig}), (\onlinecite{BornLong}), {[}\onlinecite{WheelerZurek}{]} \\
\hline
	\bf 1927 &
	\begin{tabular}{c}
		Heisenberg introduces the  \\
		uncertainty principle
	\end{tabular} &
	\begin{tabular}{c}
		Measurements of \\
		non-commuting	observables \\
		have limited precision
	\end{tabular} &
	(\onlinecite{HeisenbergOrig}), {[}\onlinecite{WheelerZurek}{]} \\
\hline
	\bf 1928 &
	\begin{tabular}{c}
		Bohr introduces the \\
		complementarity principle
	\end{tabular} &
	\begin{tabular}{c}
		Measurements results \\
		depend on the experimental \\
		set-up
	\end{tabular} &
	(\onlinecite{Bohr4}) \\
\hline
\end{tabular}
\caption{Summary of the early developments of quantum mechanics and how each step affected how we see measurements. We also list in the last column which of our references correspond to the original papers (in parenthesis) and to their English translations [in square brackets].}
\label{summary}
\end{table}

However, it is important to know what the uncertainty principles still allow us to do.
We will remark on two important subtle aspects \cite{Heisenberg2}:
\begin{enumerate}
	\item The relations do not rule out the arbitrarily precise knowledge of one observable alone.
	For example, we can know completely, at a specific instant, the position of a particle, because we can have $\Delta x\rightarrow0$ while $\Delta p\rightarrow\infty$.

	\item If we know the value of the position at an instant $t_{1}$, and we measure the value of the momentum at an instant $t_{2}$, with $t_{2}>t_{1}$, then we can calculate both values for $t_{1}<t<t_{2}$ with greater precision than the allowed by the uncertainty relations.
	However, we cannot use that information to predict future measurements, as the second measurement of the momentum affected the position of the particle.
	In Heisenberg's words, \textit{``[i]t is a matter of personal belief whether such a calculation concerning the past history of the electron can be ascribed any physical reality or not''} \cite{Heisenberg2}.
\end{enumerate}

To Bohr, the uncertainty relations were a confirmation of his view: ``\emph{in quantum mechanics, we are not dealing with an arbitrary renunciation of a more detailed analysis of atomic phenomena, but with a recognition that such an analysis is in principle excluded}'' \cite{Bohr7}.
Complementing, Heisenberg stated:
\begin{quote}
	``The demand to `describe what happens' in the quantum-theoretical process between two successive observations is a contradiction \emph{in adjecto}, since the word `describe' refers to the use of the classical concepts, while these concepts cannot be applied in the space between the observations; they can only be applied at the points of observation.''\cite{Heisenberg}
\end{quote}

Like Bohr, Heisenberg believed the problem began with language itself, given that we have only the language of classical physics to describe both quantum and non-quantum phenomena.
The wave function would represent our language to deal with the experimental conditions,
and for this reason would change discontinuously during the measurement process, when our knowledge about the systems changes.
But, contrary to Bohr, Heisenberg believed the quantum theory was an ensemble theory
\cite{Jammer,Jammer2}:
\begin{quote}
	``The probability function does---unlike the common procedure in Newtonian mechanics---not describe a certain event but, at least during the process of observation, a whole ensemble of possible events.'' \cite{Heisenberg}
\end{quote}

In short, this new prevailing view of the measurement process divided the world in a quantum part, totally inaccessible to the experimentalist, and a classical part, which we can access via the experimental record.
The interaction between the system and the classical apparatus exerts a fundamental role in obtaining the results, preventing measurements more precise than a certain fundamental limit.

\section{von Neumann's model of measurements}

So far, we have described quantum measurements in a mostly qualitative fashion,
as the orthodox interpretation would see them as an artifact of language,
their precise description lying outside the scope of scientific inquiry.
Despite that, a model by John von Neumann,
where quantum measurements are derived from the explicit interaction between
the quantum system and the macroscopic apparatus, has had a lasting influence.

With a PhD in mathematics and a degree in chemical engineering, John von Neumann began his work with David Hilbert (1862-1943) at the University of G\"ottingen in 1926.
He used his mentor's discussions about the mathematical foundations of quantum mechanics as a starting point for a series of papers that would become the basis of his 1932 book \cite{Jammer}.
There, von Neumann distinguished the two ways in which a wave function can evolve:
\begin{itemize}
\item \textbf{Process 1:} Reduction, when a measurement is performed.

\item \textbf{Process 2:} Unitary evolution according to the Schr\"odinger equation.
\end{itemize}
An important feature of Process 1 is its violation of causality.
Planck had already stated that causality was a ``\emph{a heuristic principle}'' because ``\emph{it is never possible to predict a physical occurrence with unlimited precision}'' \cite{Planck3}.
In the specific domain of quantum mechanics, Heisenberg had concluded that his uncertainty principle would only allow causal laws to be defined for isolated systems, which are not disturbed by an external classical observer \cite{Heisenberg2}.
Therefore, the presence of an observation would be the source of randomness in Process 1.

To avoid this problem, von Neumann's model uses as much as possible of Process 2's determinism to describe the interaction between the quantum system and the measurement apparatus.
To see how this works, suppose that the quantum system initially interacts with a single atom of the apparatus.
This interaction will be quantum in nature, as well as the interaction between other microscopic parts of the apparatus with this first atom.
In principle, then, we can model the beginning of the measurement, before Process 1 takes place, as a unitary evolution.
But in order for it to be a true measurement, this evolution will have to satisfy a few conditions, as we will see bellow.

\subsection{The dynamics of the measurement}

The measurement model introduced by von Neumann requires the following four assumptions, stated clearly by David Bohm (1917--1992) \cite{Bohm}:
\begin{enumerate}
	\item ``\emph{One obtains information by studying the interaction of the system of interest, which we denote hereafter by [$S$], with the observing apparatus, which we denote by [$M$]}''.

	\item ``\emph{Before the experiment begins, the observing apparatus [$M$] and the system [$S$] under observation are, in general, not coupled}''.

	\item ``\emph{After the interaction has taken place, the state of the apparatus [$M$] must be correlated to the state of the system [$S$] in a reproducible and reliable way}''.

	\item ``\emph{It is, in principle, always possible to design an apparatus that measures any given variable without changing that variable during the course of measurement}''.
\end{enumerate}

Assumption 1 simply says that we are interested in describing the evolution of a measurement apparatus $M$ observing a quantum system $S$, although we are not explaining why the measurement of an observable $\hat{S}^{\left(S\right)}$ yields a specific result.

According to Assumption 2, the quantum system $S$ can initially be described by its own state vector $\left|\phi^{\left(S\right)} (0)\right\rangle$, which can be expanded in terms of the eigenstates $\left|s_{n}\right\rangle$ of $\hat S^{(S)}$:
\begin{equation}
	\left|\phi^{\left(S\right)} (0)\right\rangle
	=\sum_{n}c_{n}\left|s_{n}\right\rangle,
	\;
	\sum_{n}\left|c_{n}\right|^{2} = 1,
	\label{expvN}
\end{equation}
where the $c_n$ are constant complex scalars, and where we are assuming discrete eigenstates.

We are also tacitly assuming that the measurement apparatus $M$ can be initially associated to a wave function $\left| m_0 \right\rangle$ of its own.
This would not be permissible under orthodox quantum mechanics, where the measurement apparatus necessarily had to be described by classical physics \cite{Jammer2}, but
Bohm argues that von Neumann's assumption is valid because quantum theory ``\emph{should be able to describe the process of observation itself in terms of the wave functions of the observing apparatus and those of the system under observation}'',
otherwise ``\emph{quantum theory could not be regarded as a complete logical system}'' \cite{Bohm}.
In this case, the joint initial state is \emph{von Neumann's pre-measurement state}:
\begin{equation}
	\left|\phi^{\left(S+M\right)}\left(0\right)\right\rangle =
	\sum_{n} c_{n} \left|s_{n}\right\rangle \left|m_{0}\right\rangle.
	\label{premedidavN}
\end{equation}

Next, to satisfy Assumption 3, we associate a state vector $\left| m_{n} \right\rangle$ to the state of the measurement apparatus when the eigenvalue $s_n$ has been observed.
If we assume that the states $\left\{ \left| m_n \right\rangle \right\}$ are orthonormal, they will not to be confused with each other, and we can propose the following unitary evolution operator via Process 2:
\[
	\hat \Delta = \sum_{n,p} \left| s_n \right\rangle \left| m_{n+p} \right\rangle
	\left\langle s_n \right| \left\langle m_p \right|,
\]
where the sum in the subscript is calculated \emph{modulo} the total number of states of $S$.
This interaction means that:
\begin{equation}
	\hat{\Delta} \left|s_n \right\rangle \left|m_p\right\rangle
	=
	\left|s_{n}\right\rangle \left|m_{n+p}\right\rangle,
\end{equation}
so that when we apply $\hat{\Delta}$ to Eq. (\ref{premedidavN}), we find:
\begin{equation}
	\hat{\Delta} \left|\phi^{\left(S+M\right)}\left(0\right)\right\rangle
	=
	\sum_{n} c_{n} \left|s_{n}\right\rangle \left|m_{n}\right\rangle.
	\label{medidavN}
\end{equation}
We call Eq. (\ref{medidavN}) \emph{von Neumann's measurement states}.
They tell us that each state $\left|m_{n}\right\rangle$ of the measurer is correlated to a state $\left|s_{n}\right\rangle$ of the system, and has a probability $\left|c_{n}\right|^{2}$ of being found---according to the Born--Dirac rule.

This joint evolution of the system and the measurement apparatus already contains all the necessary elements for a measurement, up to the moment when Process 1 takes place.
However, the exact dynamics before and after this point depend on the Hamiltonian.

\subsection{A specific interaction Hamiltonian}
The unitary evolution $\hat \Delta$ representing Process 2 is governed by a Hamiltonian $\hat H$:
\[
	\hat{H}
	= \hat{H}^{\left(S\right)} + \hat{H}^{\left(M\right)} + \hat{H}^{\left(S+M\right)},
\]
where $\hat{H}^{\left(S\right)}$ contains the terms that act only on the system,  $\hat{H}^{\left(M\right)}$ has the terms that act only on the measurement apparatus, and the \emph{interaction Hamiltonian} $\hat{H}^{\left(S+M\right)}$ encompasses the terms that act on both.
von Neumann assumes that this interaction term predominates.

\begin{figure}[bht]
\begin{center}
	\includegraphics[width=0.49\textwidth]{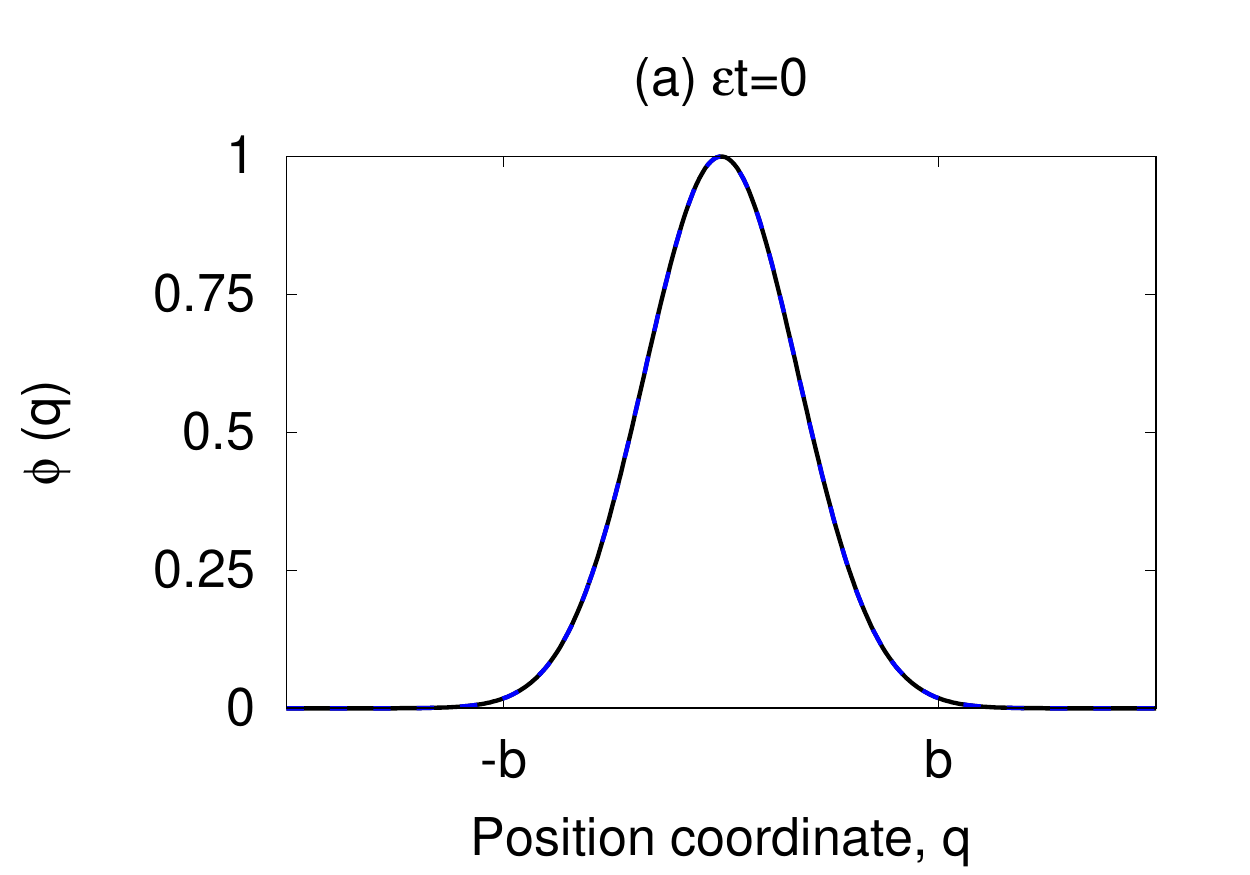}
	\includegraphics[width=0.49\textwidth]{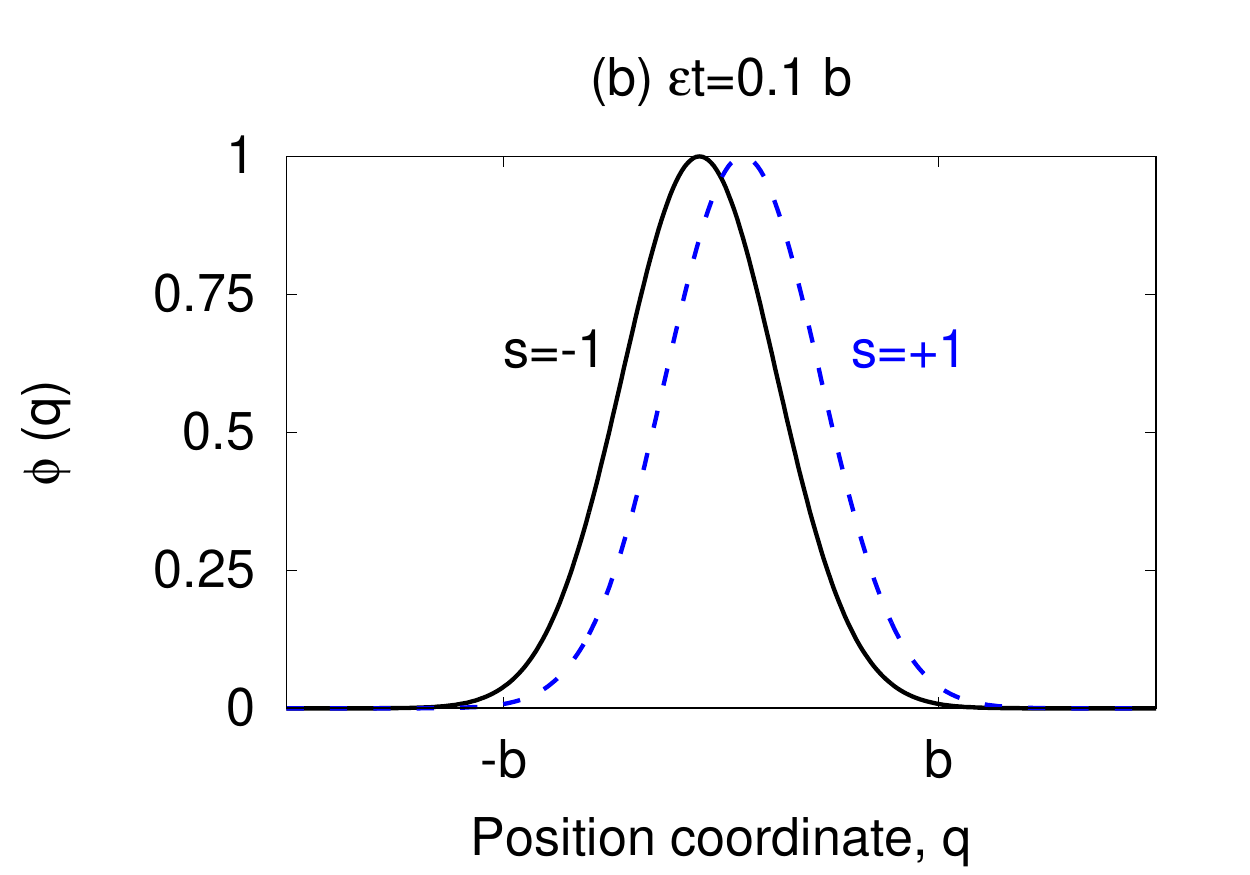}
	\includegraphics[width=0.49\textwidth]{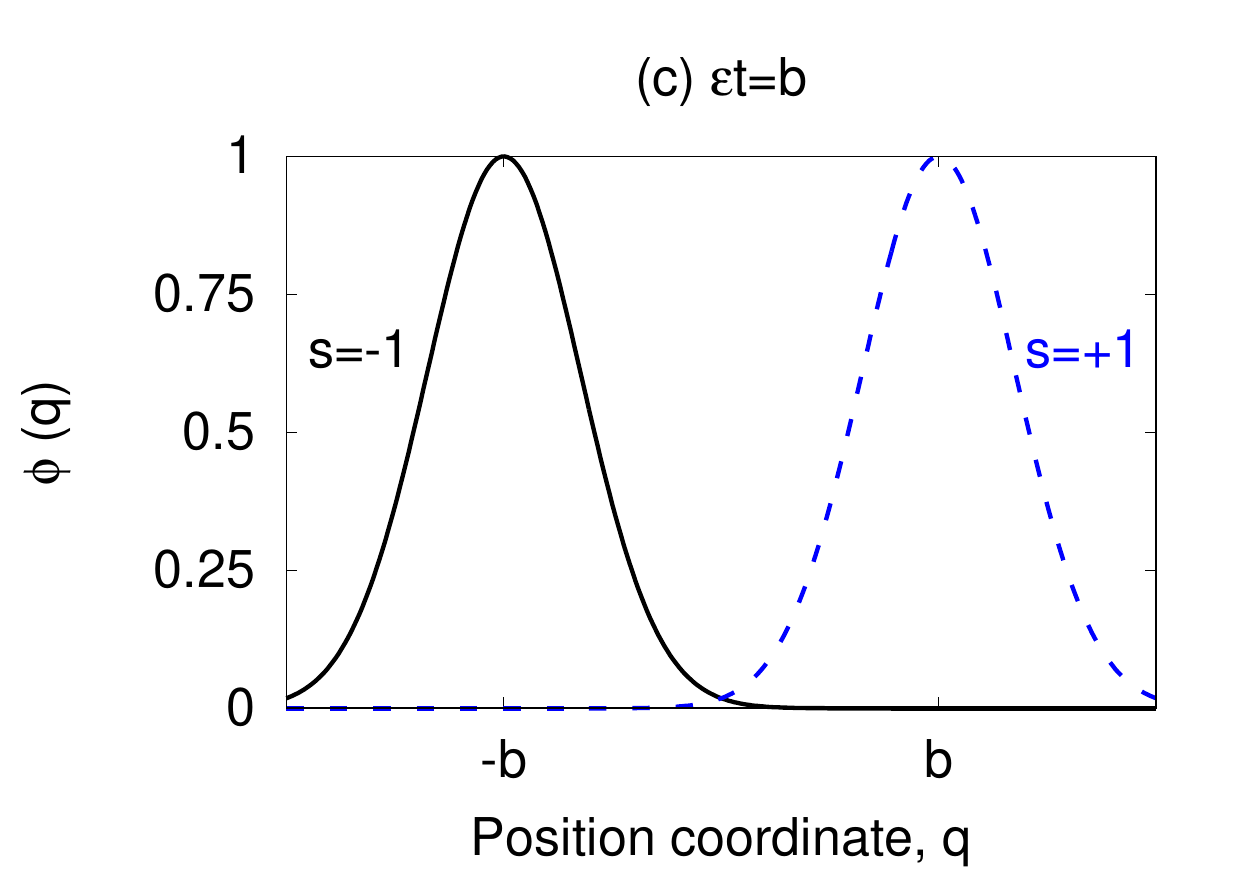}
\end{center}
	\caption{Evolution of the wave function of the measurement apparatus, originally with a compact support of approximate extent $[-b,b]$. (a) At $t=0$ there is only one wave function. (b) A small period of time after the beginning of the measurement, the wave functions corresponding to different measurements of $s$ become different, but still overlap. (c) After sufficient time, the wave functions cease to overlap and the measured states become orthogonal. \label{vnPointer}}
\end{figure}

According to Assumption 4, this measurement can be repeated any number of times in sequence and still yield the same result.
As we demonstrate in Appendix \ref{subsec:Non-demolition}, this is true if we impose the condition:
\begin{equation}
	\left[\hat{H}^{\left(S+M\right)},\hat{S}^{\left(S\right)}\right]=\hat{0},
	\label{QND}
\end{equation}
as long as the term $\hat H^{(S)}$ is irrelevant.
To satisfy Eq. (\ref{QND}), we choose:
\begin{equation}
	\hat{H}^{\left(S+M\right)}
	= \varepsilon \hat{S}^{\left(S\right)} \hat{M}^{\left(M\right)},
	\label{intvN}
\end{equation}
\noindent where $\varepsilon$ is a real constant, and $\hat M^{(M)}$ is a Hermitian operator that acts only on the measurement apparatus.
For $\hat M^{(M)}$, von Neumann chose a momentum operator $\hat P^{(M)}$ conjugated to a certain position operator $\hat Q^{(M)}$, which could represent where is the pointer that indicates the measurement result.
In this case, the time evolution operator for a measurement that lasts for a period $\tau$ will be:
\[
	\hat \Delta =
	\exp \left\{
		- i \frac{\tau}{\hbar} \hat H
	\right\}
	=
	\exp \left\{
		- i \frac{\varepsilon}{\hbar} \tau \hat S^{(S)} \hat P^{(M)}
	\right\}.
\]
Writing the initial state of the apparatus in terms of the eigenbasis of $\hat Q^{(M)}$, which we represent by $\left\{ \left| q \right\rangle \right\}$:
\[
	\left| m_0 \right\rangle =
	\int_{-\infty}^\infty \mathrm{d} q \;
	\phi_0 (q) \left| q \right\rangle,
\]
we can evolve the initial state of the system from Eq. (\ref{premedidavN}) into:
\begin{equation}
	\hat \Delta \left| \phi^{(S+M)} (0) \right\rangle =
	\sum_n c_n \int_{-\infty}^\infty \mathrm{d} q \;
	\exp \left\{
		- i \frac{\varepsilon}{\hbar} \tau s_n \frac{\hbar}{i} \frac{\partial}{\partial q}
	\right\}
	\phi_0 (q) \left| s_n \right\rangle \left| q \right\rangle.
	\label{finalm0}
\end{equation}
If we expand the exponential into a power series, we see a Taylor expansion of $\phi_0(m)$:
\begin{equation}
	\exp \left\{
		- i \frac{\varepsilon}{\hbar} \tau s_n \frac{\hbar}{i} \frac{\partial}{\partial q}
	\right\}
	\phi_0 (q)
	= \sum_{k=0}^{\infty} \frac{1}{k!} \left( -\varepsilon \tau s_n \right)^k
	\frac{\partial^k}{\partial q^k} \phi_0(q)
	= \phi_0 \left( q - \varepsilon \tau s_n \right).
	\label{taylor}
\end{equation}
Hence, the measurement state given in Eq. (\ref{medidavN}) can be written in the form:
\[
	\hat \Delta \left| \phi^{(S+M)} (0) \right\rangle
	= \sum_n c_n
	\int_{-\infty}^{\infty} \mathrm{d} q \;
	\phi_0 \left( q- \varepsilon \tau s_n \right)
	\left| s_n \right\rangle \left| q \right\rangle.
\]
Therefore, the final state of the measurement apparatus corresponding to the measurement result $s_n$ will be the original wave function displaced by an amount $\varepsilon \tau s_n$:
\[
	\left| m_n \right\rangle =
	\int_{-\infty}^{\infty} \mathrm{d} q \;
	\phi_0 \left( q - \varepsilon \tau s_n \right) \left| q \right\rangle.
\]

As long as $\phi_0(q)$ is a function with a compact support smaller than $\varepsilon \tau \Delta s$, where $\Delta s$ is the minimum distance between the eigenvalues, the states $\left| m_n \right\rangle$ will be orthogonal, as demanded by Assumption 3.
In this way, the apparatus has distinguishable states for each outcome, a fact illustrated in Fig. \ref{vnPointer}.

\subsection{The general interaction Hamiltonian}
Despite the lasting influence of von Neumann's model of measurements, his description is not as clear as its reformulation published in Bohm's 1951 textbook\cite{Bohm}.
This book was written as part of Bohm's efforts to understand quantum theory and to explain it to beginners while he delivered a course on the subject at Princeton \cite{Peat}.
In fact, a major portion of it is dedicated to explaining experiments and arguments using mathematics that would be accessible even for a first-year physics student.
The book was soon adopted by many universities and even received praise from Einstein, who would consider it the clearest possible exposition of the orthodox interpretation of quantum mechanics \cite{Peat}.

Like von Neumann, Bohm considered that only the interaction term $\hat{H}^{\left(S+M\right)}$ is relevant, but went further, justifying this fact with the assumption that $\tau$ is short enough for this measurement to be \emph{impulsive}.
Bohm also generalized the form of the Hamiltonian from Eq. (\ref{intvN}).
He only assumed that, being a Hermitian operator, $\hat M^{(M)}$ must be diagonalizable:
\[
	\hat{M}^{\left(M\right)}
	=\int_{-\infty}^{\infty} \mathrm{d}m\; f\left(m\right) \left|m\right\rangle \left\langle m\right|.
\]
Here, $\{ \left| m \right\rangle \}$ are orthonormal eigenvalues and $f(m)$ is a real function that yields the continuous set of eigenvalues.
If the eigenvalues are discrete, then $f(m)$ has the form:
\[
	f(m) = \sum_n f(m_n) \delta (m-m_n).
\]

\begin{figure}[htb]
	\includegraphics[width=0.495\textwidth]{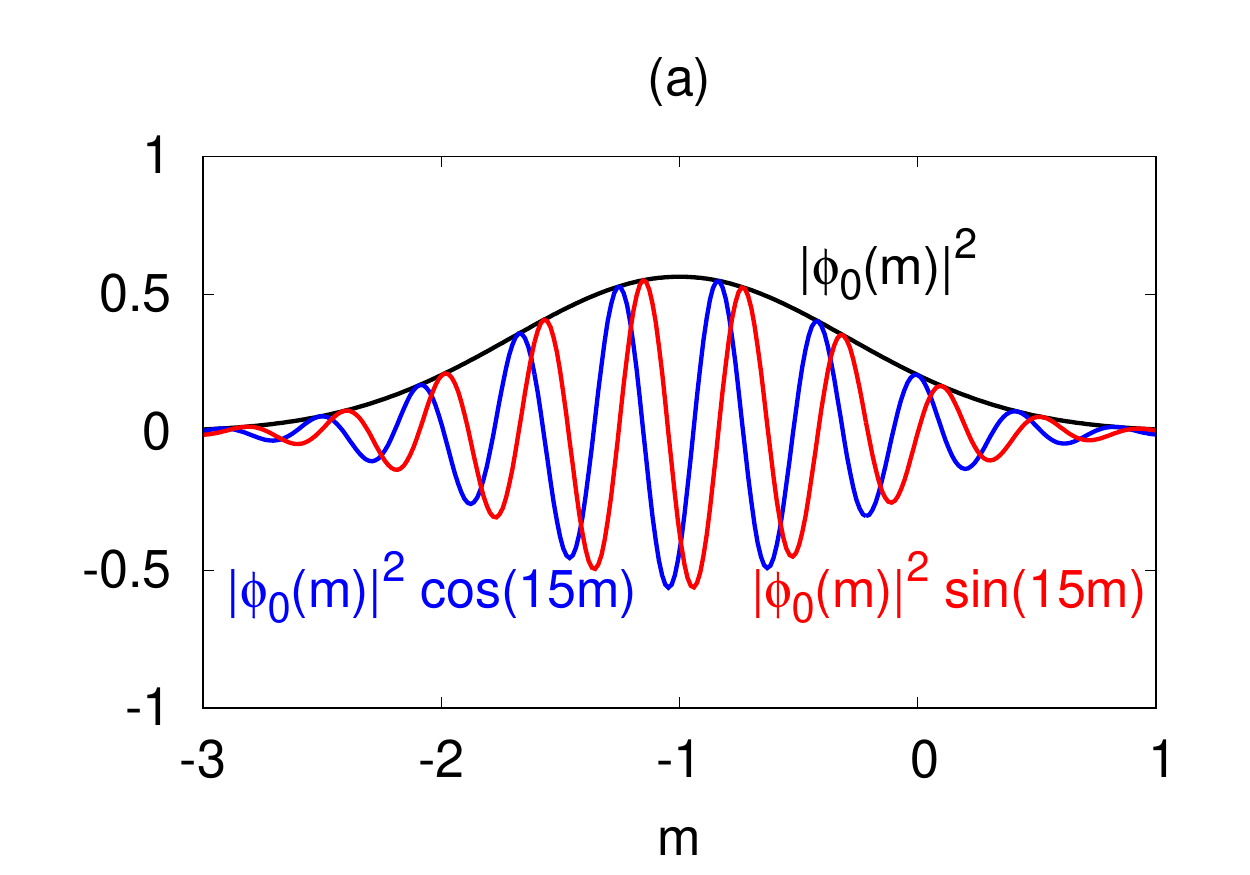}
	\includegraphics[width=0.495\textwidth]{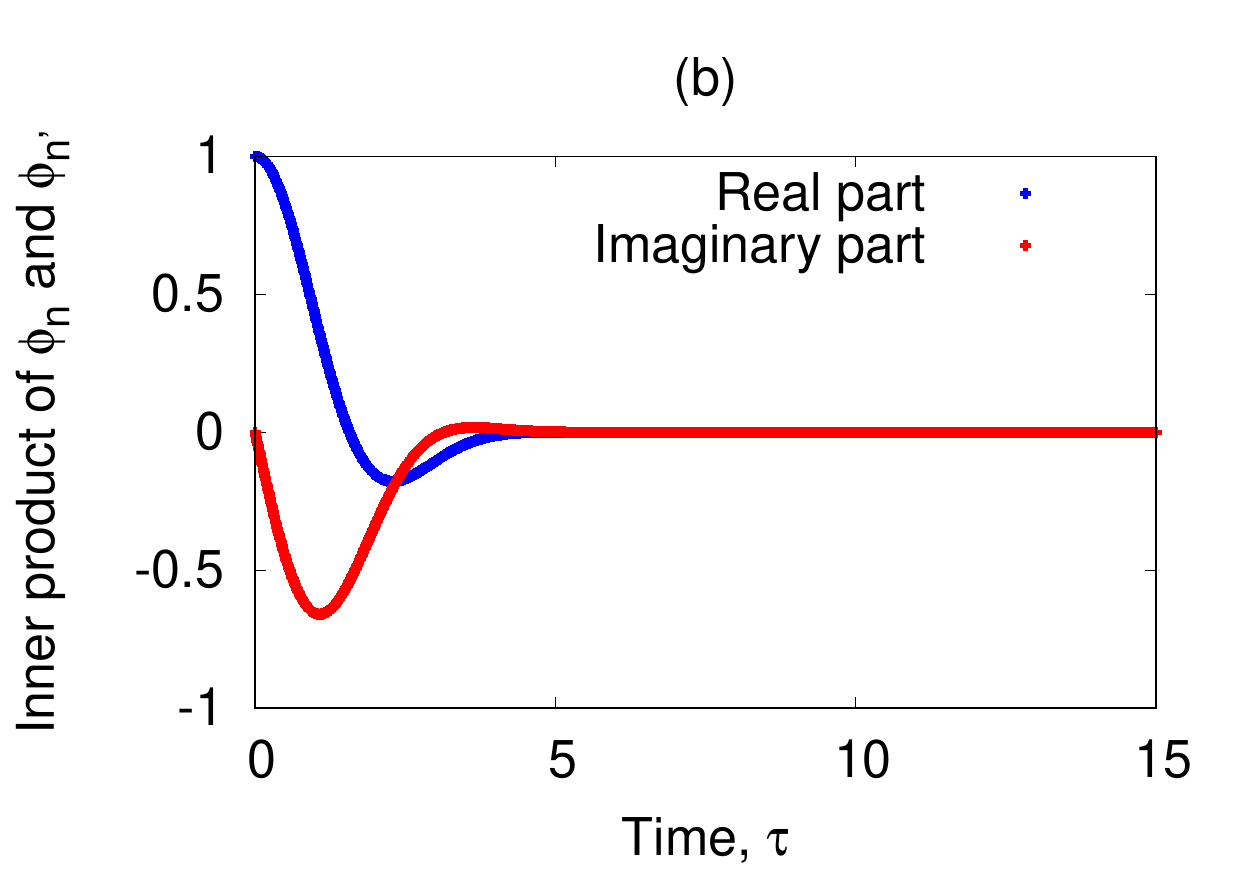}
	\caption{How a probability distribution $\left| \phi_0 (m) \right|^2$, chosen as a Gaussian significantly different from zero only in a region $\Delta m=4$, vanishes when multiplied by rapidly oscillating trigonometric functions.
	Here, we chose $f(m)=m$ and $\varepsilon (s_n-s_{n^\prime})/\hbar =1$.
	Fig. 2(a) shows how the multiplication of $\left| \phi_0 (m) \right|^2$ by $\cos(15m)$ and $\sin(15m)$ creates an approximately equal number of positive and negative regions.
	Fig. 2(b) shows the numerical calculation of the real and imaginary parts of the integrals on the right-hand side of Eq. (\ref{eq:InnerProduct}), which vanish  when the trigonometric functions oscillate faster.
	Hence, the functions become practically orthogonal when $\tau>\Delta m = 4$.
	\label{BohmOrthogonal} }
\end{figure}

We can re-write the pre-measurement state from Eq. (\ref{premedidavN}) in terms of this basis $\left| m \right\rangle$:
\begin{equation}
	\left|\phi^{\left(S+M\right)} (0)\right\rangle
	=
	\sum_n c_n \left|s_n\right\rangle
	\int_{-\infty}^{\infty} \mathrm{d}m\; \phi_0\left(m\right) \left|m\right\rangle,
	\label{eq:InitialState}
\end{equation}
where $\phi_0(m)$ is the initial wave function of the apparatus.
The time evolution due to the Hamiltonian from Eq. (\ref{intvN}) after the short time interval $\tau$ is given by:
\[
	\hat{\Delta}
	=
	\exp\left\{
		-i \frac{\varepsilon}{\hbar} \tau \hat{S}^{\left(S\right)} \hat{M}^{\left(M\right)}
	\right\}.
\]
Applied to Eq. (\ref{eq:InitialState}), this $\hat \Delta$ results in the measurement state from Eq. (\ref{medidavN}):
\[
	\hat \Delta
	\left|\phi^{\left(S+M\right)} (0)\right\rangle
	= \sum_n c_n
	\int_{-\infty}^{\infty} \mathrm{d}m\;
	\phi_n \left(m\right) \left|s_n\right\rangle \left|m\right\rangle,
\]
where $\phi_n \left(m\right)$ are the wave functions associated with each outcome of the measurement:
\[
	\phi_n \left(m\right)
	\equiv e^{-i\varepsilon\tau s_n f\left(m\right)/\hbar} \phi_0\left(m\right).
\]
The inner product between the wave functions $\phi_n \left(m\right)$ corresponding to different measurement outcomes is:
\begin{multline}
	\int_{-\infty}^{\infty}\mathrm{d}m
	\;\phi_{n}\left(m\right) \phi_{n^{\prime}}^{*}\left(m\right)
	= \int_{-\infty}^{\infty} \mathrm{d}m\;
	\left| \phi_0\left(m\right) \right|^{2}
	\cos\left[ \frac{\varepsilon}{\hbar} \tau \left(s_n-s_{n^\prime}\right) f\left(m\right) \right] \\
	-i \int_{-\infty}^{\infty} \mathrm{d}m\;
	\left| \phi_0\left(m\right) \right|^{2}
	\sin\left[ \frac{\varepsilon}{\hbar} \tau \left(s_n-s_{n^\prime}\right) f\left(m\right) \right].
	\label{eq:InnerProduct}
\end{multline}
Now, suppose the probability distribution $\left| \phi_0\left(m\right) \right|^{2}$ has a compact support of length $\Delta m$.
In this case, the strength $\varepsilon \tau/\hbar$ of the impulsive interaction will make the sines and cosines oscillate so fast whenever $s_n \ne s_{n^\prime}$ that, when multiplied by the comparatively slowly-varying positive function $\left|\phi_0 \left(m\right)\right|^{2}$, the resulting integrals in Eq. (\ref{eq:InnerProduct}) will vanish, as can be seen in Fig. \ref{BohmOrthogonal}.
Hence, Bohm argues that the states of the apparatus corresponding to different eigenstates of the observable ``\emph{are very nearly orthogonal}'' \cite{Bohm}:
\[
	\left| m_n \right\rangle =
	\int_{-\infty}^\infty \mathrm{d}m\; \phi_n (m) \left| m \right\rangle,
	\;
	\left\langle m_n \right| \left. m_{n^\prime} \right\rangle
	\approx \delta_{n,n^\prime}.
\]

In both Bohm's and von Neumann's versions of this measurement, while the complete state of the system can be represented by a single state vector, the state of the individual system $S$ can be any of the set $\left\{ \left| s_n \right\rangle \right\}$.
In order to represent this situation, we will need to go beyond wave functions and use the density matrix formalism.

\subsection{Density matrices}

Density matrices were introduced during the development of quantum statistical mechanics, which required the description of systems whose state might not be completely known.
They were independently \cite{Espagnat} introduced in 1928 by John von Neumann \cite{Neumann} and Lev Davidovich Landau (1908--1968) \cite{Landau}.
Here, we will show Landau's reasoning to derive this mathematical object---von Neumann's derivation is more elaborate and can be seen in modern notation in Appendix \ref{vnMatrix}.

Suppose we have a general state that involves both $S$ and $M$:
\[
	\left| \Psi^{(S+M)} \right\rangle =
	\sum_{n,p}
	c_{n,p} \left| s_n \right\rangle \left| m_p \right\rangle,
	\;
	\sum_{n,p} \left| c_{n,p} \right|^2 = 1,
\]
 but we are only interested in observables of the system $S$.
Landau notes that any such observable $\hat{S}^{\left(S\right)}$ will have the following expectation value, according to Born's rule:
\[
	\left\langle \hat{S}^{\left(S\right)}\right\rangle
	=\left\langle \Psi^{(S+M)} \right|\hat{S}^{\left(S\right)} \left| \Psi^{(S+M)} \right\rangle
	= \sum_{n,p} \sum_{n^{\prime},p^{\prime}} c_{n^{\prime},p^{\prime}}^{*} c_{n,p}
	\left\langle s_{n^\prime} \right| \left\langle m_{p^{\prime}} \right|
	\hat{S}^{\left(S\right)} \left|s_n \right\rangle \left|m_p \right\rangle.
\]
As $\hat{S}^{\left(S\right)}$ only acts on the $S$ system and the $\left\{ \left|m_p\right\rangle \right\} $ are orthonormal, we have:
\begin{equation}
	\left\langle \hat{S}^{\left(S\right)}\right\rangle
	= \sum_{n,p}\sum_{n^\prime} c_{n,p}c_{n^{\prime},p}^{*}
	\left\langle s_{n^{\prime}}\right|\hat{S}^{\left(S\right)}\left|s_{n}\right\rangle.
	\label{eq:ExpectationValue}
\end{equation}
We then define the reduced density matrix of the system $S$ as:
\begin{equation}
	\hat{\rho}^{\left(S\right)}
	\equiv
	\sum_{n,n^\prime}
	\rho_{n,n^{\prime}}\left|s_{n}\right\rangle \left\langle s_{n^{\prime}}\right|,
	\;
	\rho_{n,n^{\prime}}
	\equiv
	\sum_{p} c_{n,p}c_{n^{\prime},p}^{*},
	\label{rho}
\end{equation}
so that Eq. (\ref{eq:ExpectationValue}) can be re-written as a trace (sum of diagonal elements of a matrix):
\begin{equation}
	\left\langle \hat{S}^{\left(S\right)}\right\rangle
	=
	\sum_{n,n^{\prime}} \left\langle s_{n^{\prime}}\right|\rho_{n,n^{\prime}}\hat{S}^{\left(S\right)}\left|s_{n}\right\rangle
	=
	\sum_{n} \left\langle s_n \right| \hat \rho^{(S)} \hat{S}^{\left(S\right)} \left|s_{n}\right\rangle
	\equiv \mathrm{Tr}\left\{ \hat{\rho}^{\left(S\right)}\hat{S}^{\left(S\right)}\right\}.
	\label{expectation}
\end{equation}
This matrix can be used to find the expectation values of all observables of $S$ and thus contains all the information we can know about this system, without including the unnecessary information from $M$.

States that can be represented by either density matrices or state vectors are called \emph{pure states}.
Simply choose $\rho_{n,n^\prime} = c_n c_{n^\prime}^*$ in Eq. (\ref{rho}), so that:
\begin{equation}
	\hat{\rho}^{\left(S\right)}
	=\left|\Psi\right\rangle \left\langle \Psi\right|,
	\;
	\left| \Psi \right\rangle =
	\sum_n c_n \left| s_n \right\rangle.
	\label{puro}
\end{equation}
However, the density matrix can also be a convex combination of pure states, in which case we have a \emph{mixture}:
\begin{equation}
	\hat{\rho}^{\left(S\right)}
	=\sum_{i}p_{i}\left|\Psi_{i}\right\rangle \left\langle \Psi_{i}\right|,
	\;
	\left| \Psi_i \right\rangle \equiv
	\frac{1}{\sqrt{p_i}}
	\sum_n c_{n,i} \left| s_n \right\rangle,
	\;
	p_i \equiv
	\sum_n \left| c_{n,i} \right|^2.
	\label{nao-puro}
\end{equation}
Eq. (\ref{nao-puro}) represents an ensemble where there is a probability $p_i$ of finding each state $\left|\Psi_{i}\right\rangle$.
We can see this from the fact that the expectation value of the mixture is the weighed average of the expectation values of the individual states:
\begin{equation}
	\left\langle \hat{S}^{(S)} \right\rangle
	= \mathrm{Tr} \left\{ \hat \rho^{(S)} \hat S^{(S)} \right\}
	= \sum_i p_i \left\langle \Psi_i \right| \hat S^{(S)} \left| \Psi_i \right\rangle.
	\label{eq:ExpectationS}
\end{equation}
The different states in the mixture do not interfere in a quantum manner with each other, making this an \emph{``incoherent superposition''}\cite{Fano}.

In the case of the measurement state from Eq. (\ref{medidavN}), we have $c_{n,p}=\delta_{n,p}c_{n}$, which means that the density matrix only has diagonal elements, called \emph{populations}.
All the off-diagonal terms, called \emph{coherences}, vanished after the measurement:
\begin{equation}
	\left(\begin{array}{cccc}
		\rho_{1,1} & \rho_{1,2} & \dots & \rho_{1,n} \\
		\rho_{2,1} & \rho_{2,2} & \dots & \rho_{2,n} \\
		\vdots & \vdots & \ddots & \vdots \\
		\rho_{n,1} & \rho_{n,2} & \dots & \rho_{n,n}
	\end{array}\right)
	\longrightarrow
	\left(\begin{array}{cccc}
		\rho_{1,1} & 0 & \dots & 0\\
		0 & \rho_{2,2} & \dots & 0\\
		\vdots & \vdots & \ddots & \vdots\\
		0 & 0 & \dots & \rho_{n,n}
	\end{array}\right).\label{decoerencia}
\end{equation}
By the end of the measurement, the initial state becomes an incoherent mixture, a process known as \emph{decoherence}.
We will explore this further in Sec. 3.

\section{Entanglement and paradoxes}

Implicit in the discussion of von Neumann's model of measurements is the idea that the measurement apparatus $M$ and the observed quantum system $S$ end up correlated in a particularly quantum manner, called entanglement.
This concept was developed in the 1930s and is still one of the most surprising features of quantum mechanics, responsible for three apparent paradoxes that we will discuss in this section: Wigner's friend, the EPR problem, and Schr\"odinger's cat.

\subsection{Wigner's friend}

Previously, we considered two systems $S$ and $M$ interacting.
Suppose we now add a new system $E$, whose basis $\left\{ \left|e_{n}\right\rangle \right\}$ becomes correlated to the states of $M$ in the same way $M$ became correlated to $S$:
\begin{equation}
	\left|s_{n}\right\rangle \left|m_{0}\right\rangle \left|e_{0}\right\rangle
	\longrightarrow
	\left|s_{n}\right\rangle \left|m_{n}\right\rangle \left|e_{0}\right\rangle
	\longrightarrow
	\left|s_{n}\right\rangle \left|m_{n}\right\rangle \left|e_{n}\right\rangle.
	\label{chain}
\end{equation}
Should we consider $E$ to be the measurer of the joint system $S+M$?
Or is this process invalid because $M$ was already what we had defined as the measurer?
And what if we append another system after $E$?
Is there no end to number of candidate systems that can claim to be ``real'' measurer?

This the question of \emph{von Neumann's chain} \cite{Espagnat}.
Heisenberg had considered this problem before, but concluded it was irrelevant to the predictions made by quantum theory:
\begin{quote}
	``It has been said that we always start with a division of the world into an object, which we are going to study, and the rest of the world, and that this division is to some extent arbitrary.
	It should indeed not make any difference in the final result if we, e. g., add some part of the measuring device or the whole device to the object and apply the laws of quantum theory to this more complicated object. It can be shown that such an alteration of the theoretical treatment would not alter the predictions concerning a given experiment'' \cite{Heisenberg}.
\end{quote}
In practical terms, it makes no difference which system is the ``true'' responsible for the measurement.

While Heisenberg's solution is mathematically correct, this problem was put into more dramatic terms independently by Hugh Everett\cite{DeWitt2} (1930--1982) and Eugene Wigner \cite{WheelerZurek} (1902--1995).
They asked: what if the system $E$ is the scientist performing the observation?
Can a human being be in a superposition?
As Everett's complete work remained unpublished for more than a decade, this problem became known as \emph{Wigner's friend}.

Wigner presents a situation where a physicist (``Wigner's  friend'') is performing an experiment that consists of taking note of whether a flash is emitted from a quantum system.
In case the flash is emitted, Wigner's friend will record the result corresponding to the eigenstate $\left|s_1\right\rangle $;
otherwise, his friend records a result corresponding to the eigenstate $\left|s_2\right\rangle$.
We assume that the system is initially in the superposition state:
\begin{equation}
	\frac{1}{2}\left|s_1\right\rangle
	+ \frac{\sqrt{3}}{2} \left|s_2\right\rangle.
	\label{sistema}
\end{equation}
Suppose that Wigner's friend is in an isolated lab and that, after he has performed the measurement, Wigner approaches him to ask what the result was.
The question is whether both the measured quantum system and Wigner's friend will be in a superposition state before Wigner enters the room and asks about the result.
If we suppose that $\left|e_1\right\rangle $ is the state of Wigner's friend corresponding to recording the result $\left|s_1\right\rangle$ and $\left|e_2\right\rangle $ is the state corresponding to recording $\left|s_2\right\rangle$, then, following Eq. (\ref{chain}):
\begin{equation}
	\left(\frac{1}{2}\left|s_1\right\rangle
	+\frac{\sqrt{3}}{2}\left|s_2\right\rangle \right)\left|e_0\right\rangle
	\longrightarrow
	\frac{1}{2}\left|s_1\right\rangle \left|e_1\right\rangle
	+\frac{\sqrt{3}}{2}\left|s_2\right\rangle \left|e_2\right\rangle,
	\label{amigo}
\end{equation}
\noindent where $\left| e_0\right\rangle$ is the state of the friend before any measurement.

Wigner's conclusion was that consciousness privileged the human observer, so that there is no longer a superposition from the moment that his friend reads the result.
Wigner does not know what the state is when he enters the room, but it has already been determined \cite{Hartle,WheelerZurek}---the probabilities of each possible result do not indicate a quantum superposition, but merely Wigner's ignorance.

Of course, formulating the problem in this manner would be unacceptable in orthodox quantum mechanics, where a macroscopic system such as a human being should not be considered in quantum terms.
But, should this be possible, Bohm would argue that the superposition disappears much before that:
\begin{quote}
	``the interaction between the observer and his apparatus is such that statistical fluctuations arising from the quantum nature of the interaction are negligible in comparison with experimental error.
	It is therefore correct for us to approximate the relation between the investigator and his observing apparatus in terms of the simplified notion that these are two separate and distinct systems interacting only according to the laws of classical physics.'' \cite{Bohm}
\end{quote}
In this way, Bohm proposes a natural manner of breaking von Neumann's chain and separating the observer from the measurement apparatus.

However, this solution was not unanimous and the discussion of whether the measurement apparatus or even the scientist could be described according to quantum mechanics would be important for foundational questions in the future.

\subsection{The EPR problem}
\label{EPRsection}

Albert Einstein (1879--1955), Boris Podolsky (1896--1966), and Nathan Rosen (1909--1995) published in 1935 an article that proposed what would be known as the EPR problem, an attempt to decide whether the current formulation of quantum mechanics could be considered both \emph{correct} and \emph{complete}.
The authors defined completeness as the condition that every element of physical reality must have a counterpart in the physical theory.
According to them, an \emph{element of physical reality} corresponds to the situation where it is possible to predict the value of some physical quantity with certainty without perturbing the system in any way.

We will analyze the EPR problem from Bohm's simplified re-formulation \cite{Bohm}, which provided a clearer analysis in terms of the spin \cite{Jammer2,Jauch,Omnes1}.
Assume that we have a single spin-$1/2$ particle, and let $\left\{ \left|+\right\rangle ,\left|-\right\rangle \right\}$ be the eigenbasis of the operator $\hat{\sigma}_{z}$:
\begin{equation}
	\hat{\sigma}_{z}\left|\pm\right\rangle =\pm\left|\pm\right\rangle
	\label{basez}
\end{equation}
while $\left\{ \left|+_{x}\right\rangle ,\left|-_{x}\right\rangle \right\}$ is the eigenbasis of $\hat{\sigma}_{x}$:
\begin{equation}
	\left|\pm\right\rangle =\frac{\left|+_{x}\right\rangle \pm\left|-_{x}\right\rangle }{\sqrt{2}}.
	\label{bases}
\end{equation}

If a particle is in one of the eigenstates $\left| \pm \right\rangle$, the inherent uncertainty in the measurement of $\hat\sigma_z$ will be zero and this will be an element of reality -- but the value of $\hat\sigma_x$ will be undefined.
In the terminology employed by EPR, when the value of the measurement of $\hat\sigma_z$ is unknown, then $\hat\sigma_x$ has no physical reality.

Now, suppose that two particles are initially in the \emph{singlet} state:
\begin{equation}
	\frac{\left|+-\right\rangle -\left|-+\right\rangle }{\sqrt{2}}
	=
	\frac{\left|+_{x}-_{x}\right\rangle -\left|-_{x}+_{x}\right\rangle }{\sqrt{2}}.
	\label{singlet}
\end{equation}
This state is invariant under rotations, which means that, no matter what measurement we perform on the first particle, the second particle will be found on the \emph{opposite state} of the same observable.
If the first particle is measured in $\hat \sigma_z$, the second particle will be in the other eigenstate of $\hat \sigma_z$.
If the first particle is measured in $\hat \sigma_x$, the second particle will be in the other eigenstate of $\hat \sigma_x$.
Therefore, the state of the second particle is determined by a measurement of the first particle that does not act on it directly.
Particles displaying this property are known as an \emph{EPR pair}\cite{Nielsen}.

Due to this fact, the authors conclude that, despite being \emph{correct}, the quantum description of reality given by wave functions  is not \emph{complete} \cite{Jammer2}.
In their words,
\begin{quote}
	``This makes the reality of {[}$\hat \sigma_z${]} and {[}$\hat \sigma_x${]} depend upon the process of measurement carried out on the first system, which does not disturb the second system in any way. No reasonable definition of reality could be expected to permit this.'' \cite{EPR}
\end{quote}

Bohr soon replied \cite{Bohr6}, emphasizing the complementary character of quantum mechanics.
As he had stated multiple times, complementarity resides exactly in the impossibility of considering a system individually, because quantum theory deals with the joint set of the system and the measurement apparatus \cite{Jammer2,Mehra1}.
For this reason, Bohr concluded that quantum theory is not only correct, but also complete, not in the sense sought by EPR, but in the sense that it allows us to know everything that can be know \cite{Abro}.

Nevertheless, the \emph{``spooky actions at a distance''}\cite{BornEinstein} between these two particles would remain intriguing scientists, and lead to the formulation of the important concept of \emph{entanglement} by Schr\"odinger in the following year.

\subsection{Schr\"odinger's cat}

Schr\"odinger formalized the concept of entanglement in two works\cite{Schrodinger1,Schrodinger2} published soon after the EPR paper \cite{EPR}.
This is a term he coined for the situation where the state of one quantum system depends on the state of another quantum system.
In his words:
\begin{quote}
	``When two systems, of which we know the states by their respective representatives, enter into temporary physical interaction due to known forces between them, and when after a time of mutual influence the systems separate again, then they can no longer be described in the same way as before, viz. by endowing each of them with a representative of its own.
	I would not call that \emph{one} but rather \emph{the} characteristic trait of quantum mechanics, the one that enforces its entire departure from classical lines of thought.
	By the interaction the two representatives (or $\psi$-functions) have become entangled.'' \cite{Schrodinger1}
\end{quote}
This phenomenon was implied when we discussed von Neumann's measurements and density matrices.
Indeed, in von Neumann's model, the measured quantum system $S$ and the measurement apparatus $M$ become \emph{entangled}.

Schr\"odinger further proposes that ``\emph{the [EPR] paradox could be avoided}'' if the ``\emph{the knowledge of the phase relations}'' between the different states had ``\emph{been entirely lost in consequence of the process of separation}'' of the two systems, because ``\emph{this would mean that not only the parts, but the whole system, would be in the situation of a mixture, not of a pure state}'' \cite{Schrodinger2}.
Schr\"{o}dinger associates the truly quantum features of a system to the interference terms between its superposed states, which disappear  after measurement, as we saw in Eq. (\ref{decoerencia}).

To illustrate his point in an eloquent manner, he describes the problem that would be known as \emph{Schr\"{o}dinger's cat}:
\begin{quote}
	``A cat is penned up in a steel chamber, along with the following diabolical device (which must be secured against direct interference by the cat): in a Geiger counter there is a tiny bit of radioactive substance, \emph{so} small, that \emph{perhaps} in the course of one hour one of the atoms decays, but also, with equal probability, perhaps none; if it happens, the counter tube discharges and through a relay releases a hammer which shatters a small flask of hydrocyanic acid.
	If one has left this entire system to itself for an hour, one would say that the cat still lives \emph{if }meanwhile no atom has decayed.
	The first atomic decay would have poisoned it.
	The $\psi$-function of the entire system would express this by having in it the living and the dead cat (pardon the expression) mixed or smeared out in equal parts.'' \cite{catpaper,WheelerZurek}
\end{quote}
The controversial point here is to assume that the superposition of alternatives in a microscopic system (the atoms of radioactive substance) would also imply a superposition in the macroscopic system (the cat) \cite{Jauch}.
Discussions of this kind would spark interesting new interpretations of quantum mechanics, and eventually experimental proposals to create macroscopic ``Schr\"odinger cat states''.

\section{Many worlds and decoherence}
\label{subsec:From-relative-states}

As mentioned in the previous section, during his doctoral work, Everett became interested in the problem that would be known as Wigner's friend.
But, unlike Wigner, who invoked the observer's consciousness as the end point where the system could no longer be in a superposition, Everett assumed that the system, the measurement apparatus, and the observers would all be entangled and embedded in a universal wave function \cite{Everett,Wheeler,Freitas}.
In Everett's words, we are asked to
\begin{quote}
	``assume the universal validity of the quantum description, by the complete abandonment of Process 1.
	The general validity of pure wave mechanics, without any statistical assertions, is assumed for all physical systems, including observers and measuring apparata.
	Observation processes are to be described completely by the state function of the composite system which includes the observer and his object-system, and which at all times obeys the wave equation (Process 2).'' \cite{DeWitt2}
\end{quote}

The proposal that the observers themselves could be in a superposition state was met with skepticism.
Bryce DeWitt (1923--2004) made the following objection after reading the manuscript Everett submitted for publication in 1957:
\begin{quote}
	``As Everett quite explicitly says: `With each succeeding observation... the observer state $\ll$branches$\gg$ into a number of different states.'
	The trajectory of the memory configuration of a real physical observer, on the other hand, does not branch.
	I can testify to this from personal introspection, as can you. I simply do not branch.'' \cite{DeWitt3}
\end{quote}
In response, Everett added a footnote to his paper clarifying that the subjective experience of each observer would correspond to simply seeing a measurement outcome, because each branch would be unable to sense the others:
\begin{quote}
	``[S]eparate elements of a superposition individually obey the wave equation with complete indifference to the presence or absence (`actuality' or not) of any other elements.
	This total lack of effect of one branch on another also implies that no observer will ever be aware of any `splitting' process.'' \cite{Everett}
\end{quote}
Therefore, in a measurement state such as Eq. (\ref{medidavN}), each possible state of the observer $\left|m_{n}\right\rangle$ will only be aware of a single state of the system, its \emph{relative state} $\left|s_{n}\right\rangle$.
It would be as if each of these branches inhabited a different ``world'' inside our universe.
For this reason, this interpretation would be known alternatively as the \emph{relative-state interpretation} and as the \emph{many-worlds interpretation}.

However, Everett's interpretation suffers from the problem of basis ambiguity \cite{Brasil1,Lombardi}.
We can illustrate it by defining a new orthonormal basis for the measurement apparatus:
\[
	\left|m_{n}\right\rangle _{\pm}
	=
	\frac{\left|m_{2n}\right\rangle \pm\left|m_{2n+1}\right\rangle }{\sqrt{2}},
\]
in terms of which the measurement state from Eq. (\ref{medidavN}) can be written as:
\begin{multline*}
	\hat\Delta \left|\phi^{\left(S+M\right)} (0) \right\rangle
	=\sum_{n}\left(\frac{c_{2n}\left|s_{2n}\right\rangle
	+c_{2n+1}\left|s_{2n+1}\right\rangle }{\sqrt{2}}\left|m_{n}\right\rangle_{+} \right. \\
	\left.
	+\frac{c_{2n}\left|s_{2n}\right\rangle-c_{2n+1}\left|s_{2n+1}\right\rangle }{\sqrt{2}}\left|m_{n}\right\rangle _{-} \right).
\end{multline*}
If we can represent the system in this different basis, why is it that, given a fixed experimental set-up, the measured observable is always the one corresponding to the eigenstates $\left|s_{n}\right\rangle $ of the observable $\hat{S}$?
Why do we not observe it in the eigenstates $\left(c_{2n}\left|s_{2n}\right\rangle \pm c_{2n+1}\left|s_{2n+1}\right\rangle \right)/\sqrt{2}$ of a different observable?
Why does the observer always have the impression of being in a definite state like $\left|m_{2n}\right\rangle $ rather than a superposition like $\left(\left|m_{2n}\right\rangle \pm\left|m_{2n+1}\right\rangle \right)/\sqrt{2}$?

This difficulty arises from the fact that the state after the measurement in Eq. (\ref{medidavN}) is still a coherent superposition of the possible measurement results, allowing the ``branches'' of the universe to interfere with each other.
The solution found to this problem lies in performing the same kind of operation that the measurement apparatus exerted on the system in Eq. (\ref{decoerencia}), but this time it is an external environment that makes the coherences vanish.
Once this environment $E$ becomes entangled with the measurement apparatus, the complete state becomes:
\[
	\left|\phi^{\left(S+M+E\right)}\right\rangle
	= \sum_{n} c_{n} \left|s_{n}\right\rangle \left|m_{n}\right\rangle \left|e_{n}\right\rangle,
\]
where we used the same notation as Eq. (\ref{chain}).
In this case, the joint state of the system $S$ and the measurement apparatus $M$ can no longer be represented by a state vector.
Instead, it becomes a diagonal density matrix, where all the coherences have vanished:
\[
	\hat{\rho}^{\left(S+M\right)}
	=\sum_{n}\left|c_{n}\right|^{2}\left|s_{n}\right\rangle \left|m_{n}\right\rangle \left\langle m_{n}\right|\left\langle s_{n}\right|.
\]

In the words of Wojciech Hubert Zurek (born in 1951), who pioneered this field, ``\emph{in a certain sense it is the environment of the apparatus which participates in deciding what the apparatus measures}'' \cite{Zurek4}.
Zurek popularized the term \emph{decoherence} \cite{Zurek1} for this phenomenon and inserted the fundamental role of the environment in the description, making the decoherence program a question of great importance \cite{Fortin,Omnes7,Schlosshauer,Schlosshauer1,Zurek,Zurek1,Zurek2}
that has already taken up an existence independent of Everett's interpretation of quantum mechanics.
Omnès \cite{Omnes1,Omnes2,Omnes3,Omnes4,Omnes5,Omnes6,Omnes}, for example, uses it together with the consistent histories formalism \cite{Griffiths,Dowker,Griffiths2} to arrive at an enhanced version of the orthodox interpretation.
And, despite having received also a series of criticisms \cite{Kastner}, decoherence plays an increasingly important role in the theory of quantum information, where defeating the undesired effects of the environment remains an important challenge to be overcome in the construction of quantum computers \cite{Shor,Zurek}.

We will discuss more consequences of the theory of open quantum systems to quantum measurements later in this article.
For the moment, we will illustrate our explanation of von Neumann's model by applying it to a specific modern question -- weak measurements.

\section{Weak measurements}

Despite the fact that von Neumann's model describes the measurement as a dynamical process, so far we have not studied its exact evolution in time before the measurement ends.
If we return to the interaction Hamiltonian from Eq. (\ref{intvN}) and remove the condition that the measurement is impulsive, we end up with an incomplete measurement, a situation where the coherences are smaller, but not yet zero.
This situation, where $\varepsilon \tau$ is not sufficiently large, is what is called a \emph{weak measurement}\cite{Aharonov1,Duck}.

We can write the final state of the system after von Neumann's measurement following the model from Eq. (\ref{finalm0}):
\begin{equation}
	\hat \Delta \left| \phi^{(S+M)} (0) \right\rangle =
	\int_{-\infty}^\infty \mathrm{d} q \;
	\exp \left\{
		- \varepsilon \tau \hat S^{(S)} \frac{\partial}{\partial q}
	\right\}
	\phi_0 (q) \left| \phi^{(S)} (0) \right\rangle \left| q \right\rangle,
	\label{preselected}
\end{equation}
where we employed the definition of $\left| \phi^{(S)} (0) \right\rangle$ given in Eq. (\ref{expvN}).

Next, we introduce a \emph{post-selection}.
This means that we will only consider the final state of the measurement apparatus that is the relative state corresponding to a certain state $\left| \Phi^{(S)}_\text{out} \right\rangle$ of the measured system.
Experimentally, this is done by performing a complete, strong measurement after the weak measurement and discarding the run unless the result is $\left| \Phi^{(S)}_\text{out} \right\rangle$.
In the calculations, the post-selected state of the measurement apparatus, $\left| \Phi^{(M)}_\text{post} \right\rangle$, is found by applying $\left\langle \Phi^{(S)}_\text{out} \right|$ from the left to Eq. (\ref{preselected}):
\begin{equation}
	\left| \Phi^{(M)}_\text{post} \right\rangle \equiv
	\left\langle \Phi^{(S)}_\text{out} \right|
	\hat \Delta \left| \phi^{(S+M)} (0) \right\rangle =
	\int_{-\infty}^\infty \mathrm{d} q \;
	\left\langle \Phi^{(S)}_\text{out} \right|
	\exp \left\{
		- \varepsilon \tau \hat S^{(S)} \frac{\partial}{\partial q}
	\right\}
	\left| \phi^{(S)} (0) \right\rangle
	\phi_0 (q) \left| q \right\rangle.
	\label{aux3-1}
\end{equation}

Now we use the fact that the measurement is weak, so that $\varepsilon$ is small and we can expand the exponential of the interaction Hamiltonian into a power series:
\begin{equation}
	\left\langle \Phi^{(S)}_\text{out} \right|
		\exp \left\{
			- \varepsilon \tau \hat S^{(S)} \frac{\partial}{\partial q}
		\right\}
	\left| \phi^{(S)} (0) \right\rangle
	=
	\left\langle \Phi^{(S)}_\text{out} \right|
	\left. \phi^{(S)} (0) \right\rangle
	\left[
		\sum_{n=0}^\infty \frac{(-\varepsilon\tau)^n}{n!}
		\left( S^n \right)_\text{w}
		\frac{\partial^n}{\partial q^n}
	\right],
	\label{aux4-1}
\end{equation}
where we defined:
\begin{eqnarray}
	\left(S^{n}\right)_\text{w}
	\equiv
	\frac{ \left\langle \Phi_\text{out}^{(S)} \right| \left[ \hat{S}^{(S)} \right]^n \left| \phi^{(S)} (0) \right\rangle }{ \left\langle \Phi_\text{out}^{(S)} \right| \left. \phi^{(S)} (0) \right\rangle  }.
	\label{defvf}
\end{eqnarray}

The term with $n=1$ is known as weak value, and will be represented by $S_\mathrm{w}$.
In terms of it, the sum in Eq. (\ref{aux4-1}) can be written as:
\begin{equation}
	\sum_{n=0}^\infty \frac{(-\varepsilon\tau)^n}{n!}
	\left( S^n \right)_\text{w}
	\frac{\partial^n}{\partial q^n}
	=
	\exp \left\{ - \varepsilon \tau S_\text{w} \frac{\partial}{\partial q} \right\}
	+ \sum_{n=0}^\infty \left( -\varepsilon\tau \right)^{n}
	\frac{ \left[ \left(S^{n}\right)_\text{w} -S_\text{w}^{n} \right]}{n!}
	\frac{\partial^n}{\partial q^n}.
	\label{aux5-1}
\end{equation}

\begin{figure}[htb]
	\includegraphics[width=0.495\textwidth]{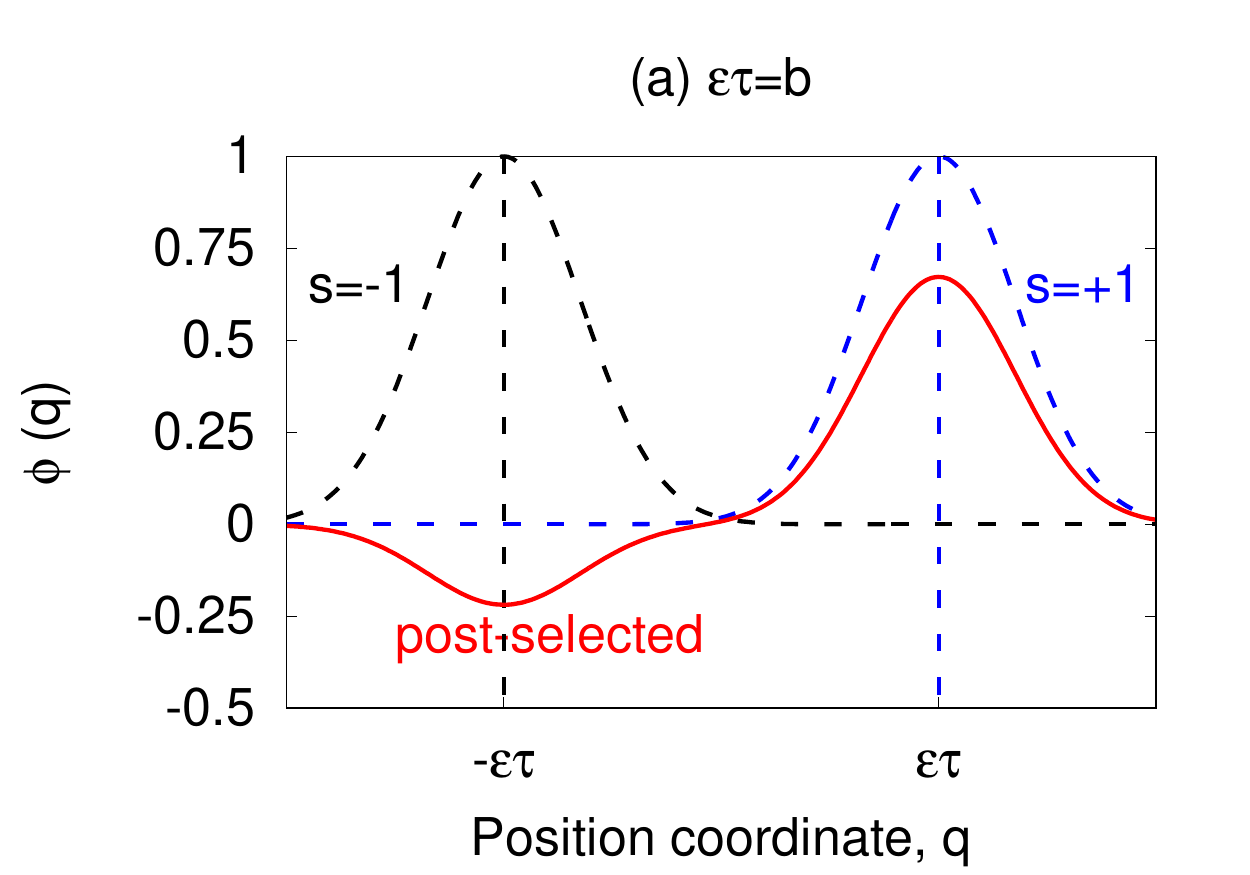}
	\includegraphics[width=0.495\textwidth]{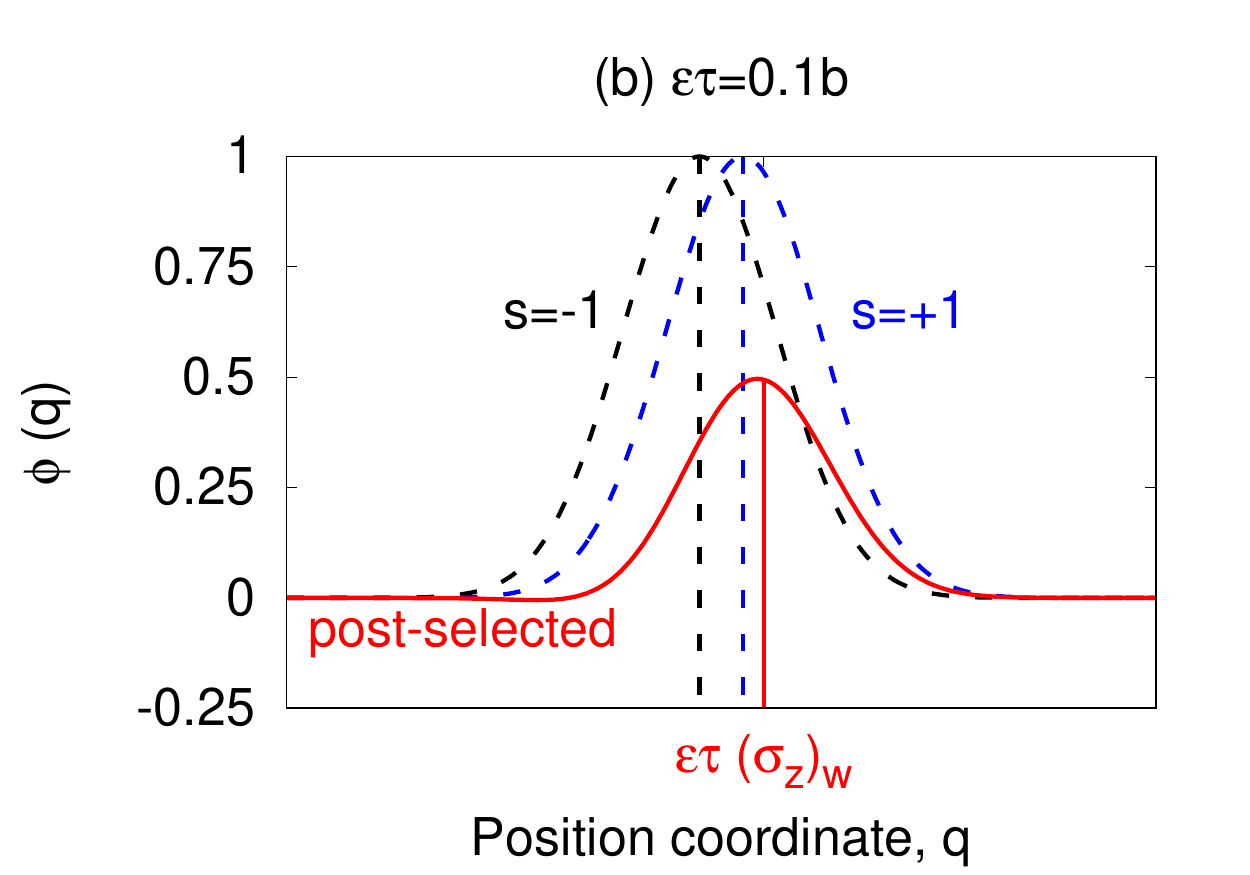}
	\caption{Wave function of an apparatus measuring $\hat \sigma_z$ for (a) strong and (b) weak measurements.
	Dashed lines represent the wave functions corresponding to the two possible results, and the continuous line represents the wave function after the post-selection.
	The initial state of the apparatus is a Gaussian centered at the origin with compact support $[-b,b]$, and the pre and post-selected states are given by Eq. (\ref{preAndPost}) with $\theta= 0.35 \pi$.
	After a weak measurement, the non-normalized post-selected state is approximately a Gaussian centered at $\varepsilon\tau(\sigma_z)_\mathrm{w}$, where $(\sigma_z)_\text{w}$ is greater than the maximum allowed value of the measurement.
	\label{weakIllustrated}}
\end{figure}

Using the same reasoning employed in Eq. (\ref{taylor}), we can conclude that the exponential of the derivative present in Eq. (\ref{aux5-1}) displaces the wave function:
\begin{equation}
	\exp \left\{ - \varepsilon \tau S_\text{w} \frac{\partial}{\partial q} \right\}
	\phi_0 (q) =
	\sum_{n=0}^\infty \frac{1}{n!} \left( - \varepsilon \tau S_\text{w} \right)^n
	\frac{\partial^n}{\partial q^n} \phi_0 (q)
	= \phi_0 \left( q - \varepsilon \tau S_\text{w} \right).
	\label{taylor2}
\end{equation}
Replacing Eqs. (\ref{taylor2}), (\ref{aux5-1}), and (\ref{aux4-1}) in Eq. (\ref{aux3-1}), we find:
\begin{equation}
	\left| \Phi^{(M)}_\text{post} \right\rangle =
	\int_{-\infty}^\infty \mathrm{d} q \;
	\left[
	\phi_0 (q-\varepsilon\tau S_\text{w})
	+
	\sum_{n=0}^\infty \left( -\varepsilon\tau \right)^{n}
		\frac{ \left[ \left(S^{n}\right)_\text{w} -S_\text{w}^{n} \right]}{n!}
		\frac{\partial^n}{\partial q^n}
	\phi_0 (q)
	\right]
	\left| q \right\rangle.
	\label{aux6-1}
\end{equation}
The terms in the sum on the right-hand side of Eq. (\ref{aux6-1}) vanish for $n<2$.
Noticing that, we can define a \emph{weak uncertainty} for $n=2$, the highest order that does not vanish:
\begin{equation}
	\left(\Delta S\right)_\text{w}
	\equiv
	\sqrt{\left|\left(S^{2}\right)_\text{w} - S_\text{w}^{2} \right|}.
\end{equation}
As long as $\varepsilon$ is sufficiently small so that the highest order term becomes negligible:
\[
	\left|
		\frac{1}{2}
		\varepsilon^2 \tau^2
		\left(  \Delta S \right)_\mathrm{w}^2
		\frac{\partial^2 \phi_0}{\partial q^2}
	\right|
	\ll 1,
\]
we can discard the sum in Eq. (\ref{aux6-1}), including the smaller higher-order terms, finding:
\[
	\left| \Phi^{(M)}_\text{post} \right\rangle \approx
	\int_{-\infty}^\infty \mathrm{d} q \;
	\phi_0 (q-\varepsilon\tau S_\text{w})
	\left| q \right\rangle.
\]
Fig. \ref{weakIllustrated} illustrates how this displacement occurs.

It is interesting to notice that, according to Eq. (\ref{defvf}), the weak value is a complex number.
We can make this fact explicit by defining:
\[
	S_\text{w} = S_\text{R}+iS_\text{I},\;
	\begin{cases}
		S_\text{R} & \equiv \text{Re} \left\{ S_\text{w}\right\} \\
		S_\text{I} & \equiv \text{Im} \left\{ S_\text{w}\right\}
	\end{cases}
\]
If the wave function $\phi_0(q)$ is real, the imaginary part of the weak value will not affect the probability distribution.
However, it will affect the probability distribution of the conjugate momentum $\left| p \right\rangle$:
\[
	\left| \Phi^{(M)}_\text{post} \right\rangle \approx
	\int_{-\infty}^\infty \mathrm{d} q \;
	\phi_0 (q-\varepsilon\tau S_\text{w})
	\int_{-\infty}^\infty \mathrm{d} p
	\left| p \right\rangle
	\left\langle p \right|
	\left. q \right\rangle
	=
	\int_{-\infty}^\infty \mathrm{d} p \;
	e^{-ip\varepsilon \tau S_\text{R}/\hbar}
	e^{p\varepsilon \tau S_\text{I}/\hbar}
	\tilde\phi_0 (p)
	\left| p \right\rangle,
\]
where $\tilde\phi_0 (p)$ is the Fourier transform of $\phi_0(q)$:
\[
	\tilde\phi_0 (p)
	\equiv
	\frac{1}{\sqrt{2\pi\hbar}}
	\int_{-\infty}^\infty \mathrm{d} q \;
	e^{-ipq/\hbar}
	\phi_0 (q).
\]
Hence, we see that the imaginary part of the weak value contributes as an exponential $e^{p\varepsilon \tau S_\text{I}/\hbar}$ multiplied by the wave function of the conjugate variable $\tilde\phi_0 (p)$.

Weak measurements have been used to derive some unexpected results in quantum mechanics.
One of them was presented in an article with the suggestive title ``\emph{How the result of a measurement of a component of the spin of a spin-1/2 particle can turn out to be 100?}'' \cite{Aharonov1}.
The answer involves nothing more than conventional quantum mechanics.
Suppose we write the initial (pre-selected) and final (post-selected) states for the quantum system using the notation from Eqs. (\ref{basez}) and (\ref{bases}):
\begin{equation}
	\begin{cases}
		\left|\phi_{0}^{\left(S\right)}\right\rangle  & =a_{+}\left|+\right\rangle +a_{-}\left|-\right\rangle \\
		\left|\Phi_\text{post}^{\left(S\right)}\right\rangle  & =b_{+}\left|+\right\rangle +b_{-}\left|-\right\rangle
	\end{cases},
	\;
	\left|a_{+}\right|^{2}+\left|a_{-}\right|^{2} = \left|b_{+}\right|^{2}+\left|b_{-}\right|^{2} = 1.
\end{equation}
In Table \ref{weakValues}, we see the expectation and weak values for this situation.
\begin{table}[htb]
\begin{tabular}{|c|c|c|c|}
\hline
Observable & Weak value & Expectation value of $\left| \phi_0^{\left(S\right)} \right\rangle$ & Expectation value of $\left| \Phi_\text{post}^{\left(S\right)} \right\rangle$ \\
\hline
$\hat{\sigma}_{x}$ & $\displaystyle \frac{a_{-}b_{+}^{*}+a_{+}b_{-}^{*}}{a_{+}b_{+}^{*}+a_{-}b_{-}^{*}}$ & $2\mathrm{Re}\left\{ a_{+}a_{-}^{*}\right\} $ & $2\mathrm{Re}\left\{ b_{+}b_{-}^{*}\right\} $\tabularnewline
\hline
$\hat{\sigma}_{y}$ & $\displaystyle i\frac{a_{+}b_{-}^{*}-a_{-}b_{+}^{*}}{a_{+}b_{+}^{*}+a_{-}b_{-}^{*}}$ & $-2\mathrm{Im}\left\{ a_{+}a_{-}^{*}\right\} $ & $-2\mathrm{Im}\left\{ b_{+}b_{-}^{*}\right\} $\tabularnewline
\hline
$\hat{\sigma}_{z}$ & $\displaystyle \frac{a_{+}b_{+}^{*}-a_{-}b_{-}^{*}}{a_{+}b_{+}^{*}+a_{-}b_{-}^{*}}$ & $\left|a_{+}\right|^{2}-\left|a_{-}\right|^{2}$ & $\left|b_{+}\right|^{2}-\left|b_{-}\right|^{2}$\tabularnewline
\hline
\end{tabular}
\caption{Comparison between weak values and expectation values. \label{weakValues}}
\end{table}

Suppose we choose the following pre-selected and post-selected states:
\begin{equation}
	\left|\phi_0^{\left(S\right)}\right\rangle =\frac{\left(\cos\theta+\sin\theta\right)\left|+\right\rangle +\left(\cos\theta-\sin\theta\right)\left|-\right\rangle }{\sqrt{2}},\;
	\left|\Phi_\text{post}^{\left(S\right)}\right\rangle =\frac{\left|+\right\rangle +\left|-\right\rangle }{\sqrt{2}}.
	\label{preAndPost}
\end{equation}
In this case, and according to Table \ref{weakValues}, the weak values will be the following:
\begin{equation}
	\left(\hat{\sigma}_{x}\right)_\text{w} = 1,\;
	\left(\hat{\sigma}_{y}\right)_\text{w} = i\tan\theta,\;
	\left(\hat{\sigma}_{z}\right)_\text{w} = \tan\theta.
\end{equation}

This situation allows us to analyze two aspects of the weak values.
First, we notice that the weak value of $\hat{\sigma}_{y}$ is imaginary.
Second, if $\theta\rightarrow \pi/2$, so that $\left|\phi_{0}^{(S)}\right\rangle \rightarrow (\left|+\right\rangle -\left|-\right\rangle )/\sqrt{2}$, the weak values will diverge:
\[
	\left(\hat{\sigma}_{z}\right)_{w}=-i\left(\hat{\sigma}_{y}\right)_{w}\rightarrow\infty.
\]
In this situation, where the pre and post-selected states tend towards orthogonality, the result of the measurement of the spin component of a spin-$1/2$ particle can turn out to be much larger than the usual limit \cite{Aharonov1}.
For it to be 100, suffices to choose $\theta\approx89,427^{o}$.

We see that weak measurements can be understood in the usual framework of quantum mechanics, as long as we consider von Neumann's measurement model.
They are important for foundational arguments~\cite{Danan} and also show parallels with the more practical atomic collision theory~\cite{Castro2}.
However, despite having the appearance of an intrisically quantum phenomenon, similar results can be obtained also in classical experiments, as long as we generalize the concept of weak value.
This was done by Christopher Ferrie and Joshua Combes~\cite{Ferrie}, and relies on realizing that the first weak value from Eq. (\ref{defvf}) can be written purely in terms of probabilities.
Indeed, consider:
\[
	S_\mathrm{w} = \frac{1}{Q} \frac{\left\langle \Phi_\text{out}^{(S)} \right| Q \hat S^{(S)} \left| \phi^{(S)} (0) \right\rangle}{ \left\langle \Phi_\text{out}^{(S)} \right| \left. \phi^{(S)} (0) \right\rangle },
\]
where $Q$ is a random variable dependent on the pre and post-selected states.
This variable will represent some form of estimate of the state of the system at the intermediary moments between preparation and measurement.
This is analogous to the position of the wave function of the measurement apparatus in the quantum prescription.
For this reason, we will further assume that $Q$ can only take the discrete values $\pm 1$, just like $\hat S^{(S)}$.
Using this fact (which implies $Q^{-1}=Q$), and further restricting $S_\mathrm{w}$ to real values, so that it is equal to the average of itself and its complex conjugate, we find:
\[
	S_\mathrm{w} = \frac{Q}{2} \frac{\left\langle \Phi_\text{out}^{(S)} \right| Q \hat S^{(S)}  \left| \phi^{(S)} (0) \right\rangle}{ \left\langle \Phi_\text{out}^{(S)} \right| \left. \phi^{(S)} (0) \right\rangle }
	+ \frac{Q}{2} \frac{\left\langle\phi^{(S)} (0) \right| Q \hat S^{(S)} \left| \Phi_\text{out}^{(S)} \right\rangle}{ \left\langle \phi^{(S)} (0) \right| \left. \Phi_\text{out}^{(S)} \right\rangle },
\]
or, using the fact that $Q^2$ is always one, we can sum over all $Q$ and divide by $2$:
\begin{align*}
	S_\mathrm{w} = & \frac{1}{2} \sum_{Q\in\{-1,1\}}\frac{Q}{2} \frac{\left\langle \Phi_\text{out}^{(S)} \right| Q \hat S^{(S)} \left| \phi^{(S)} (0) \right\rangle \left\langle \phi^{(S)} (0) \right| \left. \Phi_\text{out}^{(S)} \right\rangle}{ \left| \left\langle \Phi_\text{out}^{(S)} \right| \left. \phi^{(S)} (0) \right\rangle \right|^2 } \\
	& + \frac{1}{2} \sum_{Q\in\{-1,1\}} \frac{Q}{2} \frac{ \left\langle \Phi_\text{out}^{(S)} \right| \left. \phi^{(S)} (0) \right\rangle \left\langle\phi^{(S)} (0) \right| Q \hat S^{(S)}  \left| \Phi_\text{out}^{(S)} \right\rangle}{ \left| \left\langle \Phi_\text{out}^{(S)} \right| \left. \phi^{(S)} (0) \right\rangle \right|^2 }.
\end{align*}
Assuming again that $\varepsilon \tau \ll 1$, we can approximate $\varepsilon \tau Q \hat S^{(S)}/2$ for $e^{\varepsilon \tau Q \hat S^{(S)}/2}-1$, so that:
\[
	S_\mathrm{w} \approx \sum_{Q\in\{-1,1\}} \frac{Q}{\varepsilon \tau} \frac{\left| \left\langle \Phi_\text{out}^{(S)} \right| 2^{-1/2} e^{\varepsilon \tau Q \hat S^{(S)}/2} \left| \phi^{(S)} (0) \right\rangle \right|^2}{ \left| \left\langle \Phi_\text{out}^{(S)} \right| \left. \phi^{(S)} (0) \right\rangle \right|^2 },
\]
where we have eliminated sums over linear $Q$.
The operators $\hat K_Q^{(S)} \equiv 2^{-1/2} e^{\varepsilon \tau Q \hat S^{(S)}/2}$, with $Q=\pm 1$, represent two processes of the kind that we will describe in Sec. \ref{secPOVMs} -- it is possible to verify that they satisfy the condition from Eq. (\ref{KrausID}) up to first order in $\varepsilon \tau$.
These processes do not significantly change the state of the system, but have the effect of slightly projecting it in the eigenstate of $\hat S^{(S)}$ with eigenvalue $Q$.
In this sense, these processes labeled by $Q$ are not very different from a weak measurement.

Using this information, we can see that the term in the denominator is the probability of measuring $\left| \Phi_\text{out}^{(S)} \right\rangle$ when the state is in $\left| \phi^{(S)} (0) \right\rangle$, whereas the term in the numerator represents the probability of the same happening after the system undergoes a process labelled by a certain value of $Q$ (as opposed to $-Q$).

This allows us to represent the weak value entirely in terms of probabilites of pre-selecting $\phi$, post-selecting $\Phi$, and attributing a certain value $Q$ to an intermediary random variable:
\begin{equation}
	S_\mathrm{w} \approx \sum_{Q\in\{-1,1\}} \frac{Q}{\varepsilon \tau} \frac{p(Q,\Phi|\phi)}{ p(\Phi|\phi) } = \sum_{Q\in\{-1,1\}} \frac{Q}{\varepsilon \tau} p(Q|\Phi,\phi).
	\label{generalizedWeak}
\end{equation}
Notice that by this point we have eliminated all references to quantum notation, and these probabilities could be applied to variables associated with a classical process.
Ferrie and Combes~\cite{Ferrie} illustrate this with a simple process of coin toss, where heads and tails are represented by values $\pm 1$.

Suppose a regular classical coin is flipped and the results are observed by a reliable observer and by an unreliable one.
The reliable observer takes two notes of the result of a single coin flip, first representing the result as $\phi$, and later as $\Phi$.
Between the two measurements of the result of the coin flip, the unreliable observer makes a measurement of the state of the coin, which is registered as the variable $Q$.
This value, however, can be wrong with a probability $(1-\varepsilon\tau)/2$.
Moreover, the unreliable observer can also flip the coin after the observation, with a probability $1-\delta/(1+Q\varepsilon\tau)$.
This final flip allows the number in the denominator to be non-zero even when $\phi \ne \Phi$.
For example, if $\phi = 1$ and $\Phi = -1$, we have:
\begin{align*}
	p(Q=1,\Phi=-1|\phi=1) = & \frac{1+\varepsilon\tau}{2} \left( 1- \frac{\delta}{1+\varepsilon\tau} \right) = \frac{1+\varepsilon\tau-\delta}{2}, \\
	p(Q=-1,\Phi=-1|\phi=1) = & \frac{1-\varepsilon\tau}{2} \left( 1- \frac{\delta}{1-\varepsilon\tau} \right) = \frac{1-\varepsilon\tau-\delta}{2},
\end{align*}
so that the probability of finding $\Phi=-1$ and $\phi=1$, regardless of the value of $Q$, becomes:
\[
	p(\Phi=-1|\phi=1) = p(Q=1,\Phi=-1|\phi=1) + p(Q=-1,\Phi=-1|\phi=1)
	= (1-\delta).
\]
Replacing these results in Eq. (\ref{generalizedWeak}), we find:
\[
	S_\mathrm{w} \approx
	\frac{1}{\varepsilon \tau} \frac{p(Q=1,\Phi=-1|\phi=1)}{ p(\Phi=-1|\phi=1) }
	- \frac{1}{\varepsilon \tau} \frac{p(Q=-1,\Phi=-1|\phi=1)}{ p(\Phi=-1|\phi=1) }
	= \frac{1}{1-\delta}.
\]
By making $1-\delta$ small, we can make the weak value much larger than the limits $\pm 1$ of the coin toss results.

The fact to note here is that weak values can occur beyond the strict quantum domain, as long as there is some uncertainty in the acquisition of information in the intermediary measurement between pre-selection and post-selection, and this uncertainty can affect the final outcome.
We also remark that the generalized definition of weak value given in Eq. (\ref{generalizedWeak}) required a different kind of measurement process than what we had considered so far.
We will further discuss these generalized measurements in the next sections.

\section{Projective measurements}

Wolfgang Pauli (1900--1958) proposed a classification of measurements into those of the \emph{first kind} (or \emph{ideal}) and the \emph{second kind}.
The former are those measurements that can be repeated in sequence on the same system without changing the result \cite{Jammer2}.
All other measurements are of the \emph{second kind} \cite{Espagnat,Jammer2}.

A measurement that follows von Neumann's model is a measurement of the first kind---as we have showed in the previous sections and discussed in Appendix \ref{subsec:Non-demolition}, it does not cause the demolition of the system.
von Neumann's model also represents a \emph{projective measurement}.
To explain what this means, consider a state $\left|\phi^{\left(S\right)}\right\rangle$ of the system $S$ that can be decomposed into the orthonormal eigenbasis $\left| s_n \right\rangle$ of the observable $\hat S^{(S)}$:
\[
	\left| \phi^{(S)} \right\rangle =
	\sum_n c_n \left| s_n \right\rangle.
\]

According to Born's rule, the probability of measuring the eigenvalue $s_n$ is given by $\left| c_n \right|^2$.
Another way to express this probability is:
\begin{equation}
	P_{s_{n}} =
	\left| \left\langle \phi^{(S)} \right| \left. s_n \right\rangle \right|^2
	= \left\langle \phi^{\left(S\right)}\right| \hat{\Pi}_{n}^{(S)} \left|\phi^{\left(S\right)}\right\rangle,
	\label{proj1}
\end{equation}
where we defined the operator $\hat \Pi_n^{(S)}$ as:
\begin{equation}
	\hat{\Pi}_{n}^{(S)} \equiv
	\left|s_{n}\right\rangle \left\langle s_{n}\right|.
	\label{defproj}
\end{equation}

The operator $\hat \Pi_n^{(S)}$ is a projector, which means that it is Hermitian and idempotent:
\[
	\left[ \hat \Pi_n^{(S)} \right]^2 =
	\hat \Pi_n^{(S)}.
\]
Both facts that are easily verifiable from Eq. (\ref{defproj}).

If we obtained $s_{n}$ as the result, which will happen with a probability $P_n$, the state $\left|\phi^{\left(S\right)}\right\rangle$ will become $\left| s_n \right\rangle$.
In terms of the density matrix notation, this can be represented by a convex combination of pure states $\left| s_n \right\rangle\left\langle s_n \right|$, each of them with a weight $P_n$:
\begin{equation}
	\left| \phi^{(S)} \right \rangle \left\langle \phi^{(S)} \right|
	\to
	\sum_n P_n
	\left| s_n\right\rangle
	\left\langle s_n \right|.
	\label{aux2}
\end{equation}

Now, suppose the initial state is represented by a density matrix $\hat \rho^{(S)}$.
A density matrix is a positive-definite Hermitian operator with unit trace, and therefore can be decomposed in terms of probabilities $p_i$ and an orthonormal set of state vectors $\left\{ \left| \phi_i^{(S)} \right\rangle \right\}$:
\begin{equation}
	\hat \rho^{(S)} =
	\sum_i p_i \left| \phi_i^{(S)} \right\rangle \left\langle \phi_i^{(S)} \right|.
	\label{rho0}
\end{equation}
As this represents a mixture where there is a probability $p_i$ of finding a system in the state $\left| \phi_i^{(S)} \right\rangle$, the probability of measuring $s_n$ is:
\begin{equation}
	P_n
	= \sum_i p_i \left| \left\langle \phi_i^{(S)} \right| \left. s_n \right\rangle \right|^2
	= \left\langle s_n \right| \left( \sum_i p_i \left| \phi_i^{(S)} \right\rangle \left\langle \phi_i^{(S)} \right| \right) \left| s_n \right\rangle
	= \mathrm{Tr} \left\{
		\hat \Pi_n^{(S)} \hat \rho^{(S)} (0)
	\right\}.
	\label{aux3}
\end{equation}

Using Eqs. (\ref{aux2}) and (\ref{aux3}), we can conclude that the final state of the system after Process 1 will be:
\begin{equation}
	\sum_n \left( \sum_i p_i \left| \left\langle \phi_i^{(S)} \right| \left. s_n \right\rangle \right|^2 \right)
	\left| s_n \right\rangle \left\langle s_n \right| =
	\sum_n
	\left| s_n \right\rangle \left\langle s_n \right|
	\left( \sum_i p_i  \left| \phi_i^{(S)} \right\rangle \left\langle \phi_i^{(S)} \right| \right)
	\left| s_n \right\rangle \left\langle s_n \right|.
	\label{auxiliary}
\end{equation}
Recognizing Eq. (\ref{defproj}) and Eq. (\ref{rho0}) in Eq. (\ref{auxiliary}), we find the final expression for the state after Process 1:
\begin{equation}
	\hat \rho^{(S)} \to
	\sum_n \hat \Pi_n^{(S)} \hat \rho^{(S)} \hat \Pi_n^{(S)}.
	\label{proj3}
\end{equation}

Eqs. (\ref{aux3}) and (\ref{proj3}) characterize a \emph{projective measurement} \cite{CohenTannoudji}, or \emph{projection valued measure} \cite{Peres}.
It has this name because the operator multiplied by the density matrix to obtain the probabilities in Eq. (\ref{aux3}) is a projector.
Do notice that any sum of commuting projectors of the kind given in Eq. (\ref{defproj}) would also be a projector, and therefore characterize a projective measurement.

\section{Non-projective measurements}

\subsection{POVMs}
\label{secPOVMs}

Projection valued measures can be generalized into \emph{positive-operator valued measures} (POVMs).
This concept was described mathematically in 1966 by Sterling Berberian (born in 1926) \cite{Berberian} within the context of spectral theory, and applied to a concrete physical case one year later by Josef--Maria Jauch (1914--1974) and Constantin Piron (1932--2012)\cite{Jauch1967}.
It represents measurements that might be neither projective nor of the first kind.

This generalization consists in realizing that it is possible to extract probabilities from Eq. (\ref{aux3}) even if the operator inside the trace is not a projector.
It suffices to have a \emph{positive operator} $\hat{A}_{n}^{(S)}$ multiplied by the density matrix $\hat{\rho}^{\left(S\right)} (0)$ to obtain a result that is positive, like a probability should be:
\begin{equation}
	P_{n}=
	\mathrm{Tr}\left\{ \hat{A}_{n}^{(S)} \hat{\rho}^{\left(S\right)} \right\} \ge 0.
	\label{probPOVM}
\end{equation}
It is easy to verify the inequality above if we write the density matrix as in Eq. (\ref{rho0}).
The trace in Eq. (\ref{probPOVM}) then becomes:
\[
	\mathrm{Tr}\left\{ \hat{\rho}^{\left(S\right)}\hat{A}_{n}^{(S)} \right\} =
	\sum_i p_i \left\langle \phi_i^{(S)} \right| \hat A_n^{(S)}  \left| \phi_i^{(S)} \right\rangle.
\]
From the definition of a positive matrix, $\left\langle \phi_i^{(S)} \right| \hat A_n^{(S)} \left| \phi_i^{(S)} \right\rangle\ge 0$, and Eq. (\ref{probPOVM}) holds.

Moreover, in order for the $P_{n}$  to be probabilities, their sum must add up to one.
For this reason, we have to impose the condition that the sum of the $\hat A_n^{(S)}$ is the identity:
\begin{equation}
	\sum_{n} \hat{A}_n^{(S)} = \hat{1}^{(S)},
	\label{idPOVM}
\end{equation}
so that
\[
	\sum_n P_{n}
	=\mathrm{Tr}\left\{ \sum_{n}\hat{A}_{n}^{(S)} \hat{\rho}^{\left(S\right)} \right\} =\mathrm{Tr}\left\{ \hat 1^{(S)} \hat{\rho}^{\left(S\right)} \right\}
	=1.
\]
Eq. (\ref{idPOVM}) includes the tacit assumption that the eigenvalues of $\hat A_n^{(S)}$ are limited\cite{Jauch1967}.

We prove in Appendix \ref{decomposition} that every Hermitian positive operator $\hat{A}_{n}^{(S)}$ can be written as:
\begin{equation}
	\hat{A}_{n}^{(S)}=\left[\hat{K}_{n}^{(S)}\right]^{\dagger}\hat{K}_{n}^{(S)},
	\label{AKK}
\end{equation}
where the operator $\hat{K}_{n}^{(S)}$ might not necessarily be Hermitian.
In terms of it, Eq. (\ref{probPOVM}) becomes:
\begin{equation}
	P_{n}
	=\mathrm{Tr}\left\{
		\hat{K}_{n}^{(S)} \hat{\rho}^{\left(S\right)}\left[\hat{K}_{n}^{(S)}\right]^{\dagger}
	\right\}.
	\label{probPOVM2}
\end{equation}
The significance of the operators $\hat K_n^{(S)}$ will be explained later in this section.
For the moment, let us focus on illustrating the meaning of a POVM by providing an example.

\subsection{Interpreting the POVMs}

The meaning of probabilities like the ones from Eq. (\ref{probPOVM2}) is explained by  Asher Peres (1934--2005) as corresponding to the situation where ``\emph{the number of available preparations and that of available outcomes may be different from each other, and also different from the dimensionality of Hilbert space}''\cite{Peres}.
In other words, POVMs allow us to generalize projective measurements into the measurement of states that are not orthogonal, and therefore not mutually exclusive.
For example, their projectors might no longer commute.

To illustrate the situation\cite{Brandt},
consider a spin-$1/2$ particle whose state can be expressed in terms of the eigenbasis $\left| \pm \right\rangle$ of $\hat \sigma_z$ or in terms of the eigenbasis $\left| \pm_x \right\rangle$ of $\hat \sigma_x$---see Sec. \ref{EPRsection} for a definition of these bases.
The projectors
\[
	\hat \Pi_z^{(S)} = \left| + \right\rangle \left\langle + \right|, \;\;\;
	\hat \Pi_x^{(S)} = \left| +_x \right\rangle \left\langle +_x \right|
\]
represent apparatuses that detect with 100\% probability when a system is in the state $\left| + \right\rangle$ or $\left| +_x \right\rangle$, respectively:
\[
	\mathrm{Tr} \left\{ \hat \Pi_z^{(S)} \left| + \right\rangle \left\langle + \right| \right\} = 1, \;\;\;
	\mathrm{Tr} \left\{ \hat \Pi_x^{(S)} \left| +_x \right\rangle \left\langle +_x \right| \right\} = 1.
\]
Therefore, they represent measurements of the type that von Neumann described, projective and of the first kind, although they measure states that are not orthogonal.
For this reason, they cannot be performed at the same time, as their projectors do not commute:
\[
	\left[ \hat \Pi_z^{(S)}, \hat \Pi_x^{(S)} \right]
	= \frac{1}{\sqrt{2}} \left\{
		\left| + \right\rangle \left\langle +_x \right|
		- \left| +_x \right\rangle \left\langle + \right|
	\right\}
	\ne \hat 0.
\]

Moreover, it is not possible to form a POVM with $\hat \Pi_z^{(S)}$ and $\hat \Pi_x^{(S)}$.
Every POVM must satisfy Eq. (\ref{idPOVM}), and so constructing a POVM with $\hat \Pi_z^{(S)}$ and $\hat \Pi_x^{(S)}$ would imply finding a positive operator (or sum of positive operators) $\hat A^{(S)}$ that satisfies:
\begin{equation}
	\hat A^{(S)} = \hat 1^{(S)} - \hat \Pi_z^{(S)} - \hat \Pi_x^{(S)}.
	\label{A1}
\end{equation}
However, if we apply $\left\langle + \right|$ and $\left| + \right\rangle$ to Eq. (\ref{A1}), we find:
\[
	\left\langle + \right| \hat A^{(S)} \left| + \right\rangle
	= \left\langle + \right| \hat 1^{(S)} \left| + \right\rangle - \left\langle + \right| \hat \Pi_z^{(S)} \left| + \right\rangle - \left\langle + \right| \hat \Pi_x^{(S)} \left| + \right\rangle
	= 1 - 1 - \frac{1}{2} = - \frac{1}{2} < 0.
\]
Therefore, the operator $\hat A^{(S)}$ cannot be a positive operator.

However, we can still have a POVM that simultaneously detects $\left| + \right\rangle$ and $\left| +_x \right\rangle$ as long as we abandon ideal detectors and consider instead faulty apparatuses that detect $\left| + \right\rangle$ with a probability $p_z$ and $\left| +_x \right\rangle$ with a probability $p_x$:
\[
	\hat A_1^{(S)} = p_z \hat \Pi_z^{(S)}, \;\;\;
	\hat A_2^{(S)} = p_x \hat \Pi_x^{(S)}.
\]
Notice that $\hat A_1^{(S)}$ and $\hat A_2^{(S)}$ are no longer projectors, just positive operators.
Even if the system were initially in the state they were designed to detect, the probability of the detector firing is less than 100\%, a fact that shows that this is a measurement of the second kind:
\[
	\mathrm{Tr} \left\{ \hat A_1^{(S)} \left| + \right\rangle \left\langle + \right| \right\} = p_z < 1, \;\;\;
	\mathrm{Tr} \left\{ \hat A_2^{(S)} \left| +_x \right\rangle \left\langle +_x \right| \right\} = p_x <1.
\]
For these to add up to 100\%, we will need a third operator $\hat A_3^{(S)}$, which corresponds to the case where neither detector fires:
\begin{equation}
	\hat A_3^{(S)} = \hat 1^{(S)} - \hat A_1^{(S)} - \hat A_2^{(S)}.
	\label{defA3}
\end{equation}
The probabilities $p_x$ and $p_z$ have to be chosen appropriately so that $\hat A_3^{(S)}$ remains a positive operator.
We demonstrate in Appendix \ref{A3} that the choice that maximizes the probability of a correct detection in both devices is $p_x=p_z=2-\sqrt{2} \approx 58.6\%$.

Therefore, it is possible to create a POVM that corresponds to two imperfect detectors trying to measure incompatible states simultaneously, as long as we include the possibility of both of them failing.
While this illustrates what kind of experimental set-up can be described by a POVM, it does not correlate with a precise evolution of the system after the measurement.
In this example, the system under observation can even evade detection, in which case there is no final state to the system.

However, the dynamics of some open systems correspond to probabilities given by POVMs, so that the final state can be expressed in terms of Kraus operators.

\subsection{Kraus operators}

Karl Kraus (1938--1988) and collaborators \cite{Hellwig,Hellwig1,Kraus,Kraus1}, introduced a formalism to describe the dynamics of an open quantum system \cite{Maziero,Breuer} where operators just like the ones from Eq. (\ref{probPOVM2}) are responsible for its evolution.
For this reason, the operators $\hat K^{(S)}$ are known as Kraus operators.

To derive these operators from an open-systems point of view, let us consider a main quantum system $S$ in which we are interested, and an external system $E$ interacting with it.
Before they start to interact, the joint state of both systems is the tensor product of the density matrices of each individual system:
\[
	\hat{\rho}^{\left(S+E\right)}\left(0\right)
	=\hat{\rho}^{\left(S\right)}\left(0\right)\hat{\rho}^{\left(E\right)}\left(0\right).
\]
If these two systems are isolated from any other external interference, their evolution via Process 2 after an interval $t$ will be given in terms of some unitary time evolution operator $\hat{U}^{(S+E)} \left(t\right)$ that acts on both $S$ and $E$.
This is proven in Eq. (\ref{eq:TimeEvolution}) from Appendix \ref{LiouvilleVN}:
\begin{align*}
	\hat{\rho}^{\left(S+E\right)}\left(t\right)
	=&\hat{U}^{(S+E)} \left(t\right) \hat{\rho}^{\left(S+E\right)}\left(0\right) \left[ \hat{U}^{(S+E)} \left(t\right) \right]^\dagger \\
	=&\hat{U}^{(S+E)} \left(t\right) \left[\hat{\rho}^{\left(S\right)}\left(0\right)\hat{\rho}^{\left(E\right)}\left(0\right)\right] \left[ \hat{U}^{(S+E)} \left(t\right) \right]^\dagger.
\end{align*}
By taking the partial trace of the total density matrix over the Hilbert space of the system $E$, we find the time evolution of the reduced density matrix of the system of interest:
\begin{equation}
	\hat{\rho}^{\left(S\right)}\left(t\right)
	=\mathrm{Tr}_{E}\left\{ \hat{\rho}^{\left(S+E\right)}\left(t\right)\right\}
	 =\mathrm{Tr}_{E}\left\{ \hat{U}^{(S+E)} \left(t\right)
	 	\left[\hat{\rho}^{\left(S\right)}\left(0\right)\hat{\rho}^{\left(E\right)}\left(0\right)\right] \left[ \hat{U}^{(S+E)} \left(t\right) \right]^\dagger
	 \right\}.
	\label{trace1}
\end{equation}
Suppose $\left\{ \left|s_n\right\rangle \right\}$ is an orthonormal basis of the Hilbert space of $S$ while $\left\{ \left|e_n\right\rangle \right\}$ is an orthonormal basis of the Hilbert space of $E$.
We can write the time evolution operator $\hat{U}^{(S+E)} \left(t\right)$ as a sum of its elements:
\begin{equation}
	\hat{U}^{(S+E)} \left(t\right)
	= \sum_{j,k,r,s} c_{j,k,r,s}(t) \left| s_j \right\rangle  \left| e_k \right\rangle
	\left\langle s_r \right| \left\langle e_s \right|,
	\label{Ut}
\end{equation}
where the coefficients $c_{j,k,r,s}(t)$ are chosen appropriately so that this operator is unitary:
\begin{equation}
	\sum_{j,k}c_{j,k,m,n}^{*}(t)c_{j,k,r,s}(t)
	=\delta_{m,r}\delta_{n,s}.
	\label{Unitary}
\end{equation}

Using Eq. (\ref{Ut}) in Eq. (\ref{trace1}) and taking the trace in the basis $\left\{ \left|e_k\right\rangle \right\}$, we find:
\begin{multline}
	\hat{\rho}^{\left(S\right)}\left(t\right)
	= \sum_k \sum_{j,r,s}\sum_{j^{\prime},r^{\prime},s^{\prime}}
	c_{j,k,r,s}(t) c_{j^{\prime},k,r^{\prime},s^{\prime}}^{*} (t)
	\left(\left| s_j \right\rangle \left\langle s_r\right| \hat{\rho}^{\left(S\right)}\left(0\right) \left| s_{r^\prime} \right\rangle \left\langle s_{j^\prime} \right|\right) \\ \times
	\left\langle e_s\right| \hat{\rho}^{\left(E\right)} \left(0\right) \left| e_{s^\prime} \right\rangle.
	\label{rhoAopened}
\end{multline}
The reduced density matrix $\hat{\rho}^{\left(E\right)} (0)$ can be diagonalized in its eigenbasis $\left\{ \left|\xi_{i}^{\left(E\right)}\right\rangle \right\}$, just like we did for $\rho^{(S)} (0)$ in Eq. (\ref{rho0}):
\[
	\hat{\rho}^{\left(E\right)}\left(0\right)
	= \sum_{i} p_{i} \left|\xi_{i}^{\left(E\right)}\right\rangle \left\langle \xi_{i}^{\left(E\right)}\right|,
\]
where the eigenvalues $p_{i}$ are positive and add up to one.
Then, we can define a new matrix $\sqrt{\hat{\rho}^{\left(E\right)}\left(0\right)}$, which is diagonal in the same basis, but whose eigenvalues are the square root of the eigenvalues of $\hat{\rho}^{\left(E\right)}\left(0\right)$:
\[
	\sqrt{\hat{\rho}^{\left(E\right)}\left(0\right)} \equiv
	\sum_{i}\sqrt{p_{i}}\left|\xi_{i}^{\left(E\right)}\right\rangle \left\langle \xi_{i}^{\left(E\right)}\right|.
\]
Clearly, the original density matrix $\hat{\rho}^{\left(E\right)}\left(0\right)$ is the product of two of its square roots $\sqrt{\hat{\rho}^{\left(E\right)}\left(0\right)}$:
\begin{equation}
	\hat{\rho}^{\left(E\right)}\left(0\right)
	= \sqrt{\hat{\rho}^{\left(E\right)}\left(0\right)} \sqrt{\hat{\rho}^{\left(E\right)}\left(0\right)}
	= \sqrt{\hat{\rho}^{\left(E\right)}\left(0\right)}
	\sum_m \left| e_m \right\rangle \left\langle e_m \right|
	\sqrt{\hat{\rho}^{\left(E\right)}\left(0\right)}.
	\label{SqrtRhoB}
\end{equation}
Replacing Eq. (\ref{SqrtRhoB}) in Eq. (\ref{rhoAopened}), we find:
\begin{equation}
	\hat{\rho}^{\left(S\right)}\left(t\right)
	= \sum_{k,m} \hat{K}_{k,m}^{\left(S\right)} \left(t\right)
	\hat{\rho}^{\left(S\right)} \left(0\right)
	\left[ \hat{K}_{k,m}^{\left(S\right)}\left(t\right) \right]^\dagger,
	\label{rhoAt}
\end{equation}
where we defined the Kraus operators $\hat K_{k,m}^{(S)} (t)$ as:
\begin{equation}
	\hat{K}_{k,m}^{\left(S\right)}\left(t\right)
	\equiv \sum_{j,r,s} c_{j,k,r,s} \left(t\right)
	\left\langle e_s \right| \sqrt{\hat{\rho}^{\left(E\right)}\left(0\right)} \left| e_m \right\rangle
	\left| s_j \right\rangle \left\langle s_r \right|.
	\label{KrausOps}
\end{equation}

The probability of the evolution due to each Kraus operator in Eq. (\ref{rhoAt}) is given by Eq. (\ref{probPOVM2}), which we derived for a POVM.
Moreover, an operator formed by $\left[ \hat K^{(S)}_{k,m} \right]^\dagger \hat K_{k,m}^{(S)}$ will necessarily be positive.
We only have to show that these operators add up to the identity to prove that a set of Kraus operators will always be associated with a positive-operator valued measure:
\begin{align*}
	\sum_{k,m} \left[ \hat{K}_{k,m}^{\left(S\right)}\left(t\right) \right]^\dagger \hat{K}_{k,m}^{\left(S\right)}\left(t\right) =&
	\sum_{k,m}
	\left[ \sum_{j,r,s} c^*_{j,k,r,s} \left(t\right)
		\left\langle e_m \right| \sqrt{\hat{\rho}^{\left(E\right)}\left(0\right)} \left| e_s \right\rangle
		\left| s_r \right\rangle \left\langle s_j \right|
	\right] \\
	 & \times
	 \left[ \sum_{j^\prime,r^\prime,s^\prime} c_{j^\prime,k,r^\prime,s^\prime} \left(t\right)
	 	\left\langle e_{s^\prime} \right| \sqrt{\hat{\rho}^{\left(E\right)}\left(0\right)} \left| e_{m} \right\rangle
	 	\left| s_{j^\prime} \right\rangle \left\langle s_{r^\prime} \right|
	\right] \\
 	=&
 	\sum_{k} \sum_{j,r,s} \sum_{r^{\prime},s^{\prime}}
 	c_{j,k,r,s}^{*} \left(t\right) c_{j,k,r^{\prime},s^{\prime}} \left(t\right)
 	\left\langle e_{s^\prime }\right| \hat{\rho}^{\left(E\right)}\left(0\right) \left|e_s\right\rangle
 	\left| s_r \right\rangle \left\langle s_{r^\prime} \right|.
\end{align*}
Using the identity from Eq. (\ref{Unitary}), we conclude that:
\[
	\sum_{k,m} \left[ \hat{K}_{k,m}^{\left(S\right)}\left(t\right) \right]^\dagger \hat{K}_{k,m}^{\left(S\right)}\left(t\right) =
	\sum_{r,s} \left\langle e_s \right| \hat{\rho}^{\left(E\right)} \left(0\right)\left| e_s \right\rangle
	\left| s_r \right\rangle \left\langle s_r \right|.
\]
Noticing that $\sum_{s}\left\langle e_s \right| \hat{\rho}^{\left(E\right)}\left(0\right) \left| e_s \right\rangle$ is simply the trace of $\hat{\rho}^{\left(E\right)}$, which is $1$ for every density matrix, we arrive at the identity we wished to prove:
\begin{equation}
	\sum_{k,m} \left[ \hat{K}_{k,m}^{\left(S\right)}\left(t\right) \right]^\dagger \hat{K}_{k,m}^{\left(S\right)}\left(t\right) =
	\sum_{r} \left| s_r\right\rangle \left\langle s_r \right|=
	\hat{1}^{\left(S\right)}.
	\label{KrausID}
\end{equation}
Hence, the probabilities add up to one, as expected of a POVM.

When there is only one Kraus operator, it becomes equivalent to a unitary time evolution operator $\hat U^{(S)} (t)$ and the system evolves according to Eq. (\ref{eq:TimeEvolution}), a sign that this is an isolated quantum system.
The presence of more than one Kraus operator indicates that there is more into play -- such as the interaction with an external system or measurement apparatus -- a sign that we are dealing with an open quantum system.

\section{Measurement master equations}

In the previous section, we analyzed how POVMs generalized von Neumann's projective measurements and how their dynamics could be described by Kraus operators.
In this section, we will see how these dynamics can be calculated at every instant using differential equations.
We will employ a statistical tool widely used in the treatment of irreversible processes that involve interacting systems, the \emph{master equation} \cite{Breuer,Caldeira,Reif}.
Classically, an equation of this kind provides the time evolution of the probabilities associated with a given property  \cite{Reif}, but in the quantum case it will offer a time evolution of the reduced density operator \cite{Breuer,Caldeira}.

\subsection{The Lindblad equation}

Once again, we will consider the situation where our main system $S$ interacts with an external environment $E$.
Their evolution according to Process 2 will be governed by a total Hamiltonian:
\begin{equation}
	\hat{H}
	=\hat{H}^{\left(S\right)}+\hat{H}^{\left(E\right)}+\hat{H}^{\left(S+E\right)},
	\label{hamil}
\end{equation}
where the term $\hat{H}^{\left(S\right)}$ acts only on the system $S$, $\hat{H}^{\left(E\right)}$ acts only on $E$, and $\hat{H}^{\left(S+E\right)}$ contains the interaction terms that act on both.

As derived in Appendix \ref{LiouvilleVN}, the time evolution of the density matrix of the joint system is given by Eq. (\ref{eqLvN}), the Liouville-von Neumann equation:
\begin{equation}
	\frac{\mathrm{d}}{\mathrm{d}t}\hat{\rho}^{\left(S+E\right)}\left(t\right)
	=-\frac{i}{\hbar}
	\left[\hat{H}^{\left(S\right)}+\hat{H}^{\left(E\right)}+\hat{H}^{\left(S+E\right)},\hat{\rho}^{\left(S+E\right)}\left(t\right)\right].
	\label{liouvon1}
\end{equation}
However, the evolution of the reduced density matrix in which we are interested, $\hat{\rho}^{\left(S\right)} \left(t\right)$, will be given by the partial trace of Eq. (\ref{liouvon1}), which is not always trivial.

To simplify Eq. (\ref{liouvon1}), we introduce the interaction-picture Hamiltonian \cite{CohenTannoudji}:
\begin{equation}
	\hat{H}^{(S+E)}_I \left(t\right)\equiv
	e^{i\left(\hat{H}^{\left(S\right)}+\hat{H}^{\left(E\right)}\right)t/\hbar}
	\hat{H}^{\left(S+E\right)}
	e^{-i\left(\hat{H}^{\left(S\right)}+\hat{H}^{\left(E\right)}\right)t/\hbar},
	\label{int}
\end{equation}
and the interaction-picture density matrix:
\begin{equation}
	\hat{\rho}_I^{(S+E)}\left(t\right)
	\equiv e^{i\left(\hat{H}^{\left(S\right)}+\hat{H}^{\left(E\right)}\right)t/\hbar}
	\hat{\rho}^{\left(S+E\right)}\left(t\right)
	e^{-i\left(\hat{H}^{\left(S\right)}+\hat{H}^{\left(E\right)}\right)t/\hbar}.
	\label{rotrans}
\end{equation}
Using Eqs. (\ref{int}) and (\ref{rotrans}), we can re-write Eq. (\ref{liouvon1}) in a simpler form:
\begin{equation}
	\frac{\mathrm{d}}{\mathrm{d}t} \hat{\rho}_I^{(S+E)}\left(t\right)
	=-\frac{i}{\hbar}
	\left[ \hat{H}_I^{(S+E)}\left(t\right), \hat{\rho}_I^{(S+E)} \left(t\right) \right].
	\label{liouvon2}
\end{equation}
To find the master equation for the system $S$ alone, we take an iterative approach.
We integrate Eq. (\ref{liouvon2}) in time, and replace the result in Eq. (\ref{liouvon2}) itself.
After this, we take the partial trace over the degrees of freedom of the environment, obtaining an equation for the reduced density operator $\hat{\rho}_I^{\left(S\right)}\left(t\right)$:
\[
	\frac{\mathrm{d}}{\mathrm{d}t}\hat{\rho}_I^{\left(S\right)}\left(t\right)=
	-\frac{1}{\hbar^{2}} \int_{0}^{t}\mathrm{d}t^{\prime} \;
	\mathrm{Tr}_{E} \left\{
		\left[
			\hat{H}_I^{(S+E)} \left(t\right),\left[\hat{H}_I^{(S+E)}\left(t^{\prime}\right), \hat{\rho}^{\left(S+E\right)}_I \left(t^\prime\right) \right]
		\right]
	\right\}.
\]
If $t-t^{\prime}$ is sufficiently small, the difference between $\hat{\rho}^{\left(S+E\right)}_I \left(t\right)$ and $\hat{\rho}^{\left(S+E\right)}_I \left(t^{\prime}\right)$ can be ignored.
Moreover, we can assume that the system and the environment are almost not entangled, and that the environment has almost not changed since the beginning of the interaction---an assumption known as the Born approximation\cite{CohenTannoudji}:
\[
	\hat{\rho}^{\left(S+E\right)}_I \left(t^\prime\right)
	\approx
	\hat{\rho}^{\left(S+E\right)}_I \left(t\right)
	\approx
	\hat{\rho}_I^{\left(S\right)}\left(t\right)\hat{\rho}_I^{\left(E\right)} (0).
\]
If we extend the integral to infinity, which is known as the Markov approximation \cite{Breuer}, we find the Born--Markov master equation:
\begin{equation}
	\frac{\mathrm{d}}{\mathrm{d}t}\hat{\rho}_I^{\left(S\right)}\left(t\right)=
	-\frac{1}{\hbar^{2}} \int_{0}^{\infty}\mathrm{d}t^{\prime} \;
	\mathrm{Tr}_{E} \left\{
		\left[
			\hat{H}_I^{(S+E)} \left(t\right),\left[\hat{H}_I^{(S+E)}\left(t^{\prime}\right), \hat{\rho}_I^{\left(S\right)}\left(t\right)\hat{\rho}_I^{\left(E\right)} (0) \right]
		\right]
	\right\}.
	\label{bornmarkovfinal}
\end{equation}
It is possible\cite{Brasil4} to develop Eq. (\ref{bornmarkovfinal}) into the most general kind of Markovian master equation for a density matrix, the Lindblad equation\cite{Lindblad,Adler,Adler1,Peres1,Weinberg,Distler}:
\begin{equation}
	\frac{\mathrm{d}}{\mathrm{d}t}\hat{\rho}^{\left(S\right)}\left(t\right)
	=-\frac{i}{\hbar}\left[\hat{H}^{\left(S\right)},\hat{\rho}^{\left(S\right)}\left(t\right)\right]
	+\gamma \sum_j \left[
		\hat{L}_{j}^{\left(S\right)}\hat{\rho}^{\left(S\right)}\left(t\right)\hat{L}_{j}^{\left(S\right)\dagger}
		-\frac{1}{2}\left\{
			\hat{L}_{j}^{\left(S\right)\dagger}\hat{L}_{j}^{\left(S\right)},\hat{\rho}^{\left(S\right)}\left(t\right)
		\right\}
	\right],
	\label{lindgeral}
\end{equation}
where $\gamma$ is a constant scalar and the Lindblad operators $\hat L_j^{(S)}$ give the non-unitary dynamics for the system $S$.
These operators form the Lindbladian, the last term on the right-hand side of Eq. (\ref{lindgeral}).
Its presence is the only difference between the Lindblad equation from Eq. (\ref{lindgeral})
and the Liouville--von Neumann equation from Eq. (\ref{eqLvN}):
the Lindbladian is responsible for the effects that the external environment causes on the system.

It is of particular interest to us the situation where the Lindblad operators represent the action of an external measurement apparatus on the system~\cite{Brasil5,Brasil2,Brasil4}.
With this, we can obtain the dynamics of a generalized POVM measurement in the same way that the von Neumann model was capable of providing the dynamics of a projective measurement.

\subsection{The Markovian master equation for measurements}

We will follow the derivation of the Markovian master equation for continuous measurements that is due to James Cresser and collaborators \cite{Cresser}.
Suppose that our quantum system of interest $S$ usually evolves according to Process 2, as prescribed by Eq. (\ref{eq:TimeEvolution}).
However, the system has a probability $\lambda$ per unit of time of undergoing a POVM, in which case it evolves according to the Kraus operators from Eq. (\ref{rhoAt}).
After a small interval of time $\Delta t$, the system will have a probability $1-\lambda \Delta t$ of having evolved according to Process 2, and a probability $\lambda \Delta t$ of having undergone a POVM:
\[
	\hat \rho^{(S)} (t+\Delta t) =
	\left( 1 - \lambda \Delta t \right)
	\left[
		e^{-i \hat H^{(S)} \Delta t/\hbar} \hat\rho^{(S)} (t) e^{i \hat H^{(S)} \Delta t/\hbar}
	\right]
	+ \lambda \Delta t
	\sum_i \hat K_i^{(S)} \hat \rho^{(S)} (t) \left[ \hat K_i^{(S)} \right]^\dagger.
\]
Keeping only terms up to the first order in $\Delta t$ and expanding the exponentials into their Taylor series, we find:
\begin{multline*}
	\hat \rho^{(S)} (t+\Delta t) - \hat \rho^{(S)} (t) =
	- \frac{i}{\hbar} \Delta t \left[ \hat H^{(S)}, \hat\rho^{(S)} (t) \right]
	- \lambda \Delta t \hat\rho^{(S)} (t)
	+ \lambda \Delta t
		\sum_i \hat K_i^{(S)} \hat \rho^{(S)} (t) \left[ \hat K_i^{(S)} \right]^\dagger \\
	+ O \left( \Delta t^2 \right).
\end{multline*}
If we divide both sides by $\Delta t$ and take the limit of $\Delta t \to 0$, we find a Markovian master equation that governs this kind of system:
\begin{equation}
	\frac{\mathrm{d}}{\mathrm{d} t} \hat \rho^{(S)} (t) =
	- \frac{i}{\hbar} \left[ \hat H^{(S)} , \hat \rho^{(S)} (t) \right]
	+ \lambda \left\{
		\sum_i \hat K_i^{(S)} \hat \rho^{(S)} (t) \left[ \hat K_i^{(S)} \right]^\dagger
		- \hat \rho^{(S)} (t)
	\right\}.
	\label{CresserEq}
\end{equation}
If we choose the Lindblad operators in Eq. (\ref{lindgeral}) as being equal to the Kraus operators of some POVM ($\hat L_i^{(S)} = \hat K_i^{(S)}$) and apply the identity from Eq. (\ref{KrausID}), we obtain Eq. (\ref{CresserEq}).
Therefore, Eq. (\ref{CresserEq}) is a specific case of the Lindblad equation that describes a Markovian continuous measurement\cite{Jacobs1}.

However, it is important to notice that the dynamics of these finite-time measurements do not include the ``\emph{reduction and backaction effects}'' \cite{Klyshko} on the system.
After all, this is something that is not present in all interpretations of quantum mechanics.
For example, Everett's many-worlds notably lack it.

If one wants to introduce this additional backaction term, it is necessary to employ techniques from It\=o calculus \cite{Jacobs1,Wiseman}.
But many interesting phenomena do not require this addition, including some with practical applications.
We will illustrate this with the protection of a quantum system in a way that is analogous to the Quantum Zeno Effect \cite{Brasil,Brasil3}.

\section{Application: Quantum Zeno effect}

The Quantum Zeno Effect is named after the Ancient Greek philosopher Zeno of Elea, who proposed a series of paradoxes on the impossibility of movement by dividing the path into an infinite series of small fractions.
The analogous phenomenon in quantum mechanics occurs when a quantum system under continuous observation does not evolve in time \cite{Misra,Omnes1}.

This phenomenon was theoretically proposed in 1977 by Baidyanath Misra (born in 1937) and Ennackal Chandy George Sudarshan (1931--2018)\cite{Misra}.
Their original demonstration using projective measurements is given below, followed by another demonstration that employs the master equation for continuous measurements.

\subsection{Quantum Zeno Effect via projective measurements}

Let us consider that the system $S$ is initially in some pure state represented by a density matrix $\hat\rho^{(S)} \left(0\right)$, which is also a projector $\hat \Pi^{(S)}$:
\begin{equation}
	\hat\rho^{(S)} \left(0\right)
	= \left|\psi\right\rangle \left\langle \psi\right|
	\equiv \hat \Pi^{(S)}.
	\label{eq:RhoIsPi}
\end{equation}
After a given period of time, the probability $P\left(t\right)$ of measuring the system and finding it still in the initial state is given by Eq. (\ref{aux3}):
\begin{equation}
	P\left(t\right)
	=\mathrm{Tr}\left\{ \hat{\Pi}^{(S)} \hat \rho^{(S)} \left(t\right) \right\}.
	\label{eq:ProbabilityPi}
\end{equation}
If the system is found to be in the state $\hat{\Pi}^{(S)}$, its subsequent evolution will once again follow Process 2, and the probability of measuring it still in the state $\hat{\Pi}^{(S)}$ after another period $t$ will be given by Eq. (\ref{eq:ProbabilityPi}).
The probability $P_{n}\left(nt\right)$ of the system being found in the state $\hat{\Pi}^{(S)}$ in $n$ measurements equally spaced by periods of time $t$ will be simply the $n$th power of Eq. (\ref{eq:ProbabilityPi}):
\begin{equation}
	P_{n}\left(nt\right)
	=\left[P\left(t\right)\right]^{n}.
	\label{eq:pN}
\end{equation}
Fixing a time $T=nt$, we can take the limit where $n\to\infty$ to find the probability of the system still being in the initial state after a period $T$ of constant projective measurements:
\begin{equation}
	P_\infty \left(T\right)
	=\lim_{n\to\infty}P_{n}\left(T\right)
	=\lim_{n\to\infty}\left[ P \left(T/n\right) \right]^n.
	\label{eq:PT}
\end{equation}
Using the solution given in Eq. (\ref{eq:TimeEvolution}) to describe the state of $\hat \rho^{(S)}$ after a period $t$ of evolution due to Process 2, we can re-write Eq. (\ref{eq:pN}) in terms of the time evolution operator $\hat U^{S} (t)$:
\begin{equation}
	P_\infty \left(T\right)=
	\lim_{n\to\infty}
	\left[
		\mathrm{Tr}
		\left\{
			\hat{\Pi}^{(S)} \hat{U}^{(S)} \left(T/n\right)
			\hat{\rho}^{(S)} \left(0\right)
			\hat{U}^{(S)}\left(-T/n\right)
		\right\}
	\right]^{n},
	\label{eq:step1}
\end{equation}
where we used the fact that:
\[
	\left[ \hat U^{(S)} (T/n) \right]^\dagger =
	\hat U^{(S)} (-T/n).
\]
Using Eq. (\ref{eq:RhoIsPi}), it is possible to re-write the trace in Eq. (\ref{eq:step1}) as an inner product:
\[
	P_\infty \left(T\right)=
	\lim_{n\to\infty}
	\left[
		\left\langle \psi\right|
			\hat{U}^{(S)} \left(T/n\right)
			\hat{\rho}^{(S)} \left(0\right)
			\hat{U}^{(S)}\left(-T/n\right)
		\left|\psi\right\rangle
	\right]^{n}.
\]
Eq. (\ref{eq:RhoIsPi}) also allows us to replace the density matrix with a projector:
\begin{equation}
	P_\infty \left(T\right)=
	\lim_{n\to\infty}
	\left[
		\left\langle \psi\right|
			\hat{U}^{(S)} \left(T/n\right)
		\left|\psi\right\rangle
		\left\langle \psi\right|
			\hat{U}^{(S)}\left(-T/n\right)
		\left|\psi\right\rangle
	\right]^{n}.
	\label{eq:pnnt}
\end{equation}

We can write Eq. (\ref{eq:pnnt}) in the compact form:
\begin{equation}
	P_\infty \left(T\right)
	= F\left(-T\right) F\left(T\right),
	 	\label{eq:finalProbabilityFinal}
\end{equation}
so that we only have to study the form of the function $F(T)$ to determine the probabilities:
\begin{equation}
	F\left(T\right)
	\equiv
	\lim_{n\to\infty}
	\left[
		\left\langle \psi\right|
		\hat{U}^{(S)} \left(T/n\right)
		\left|\psi\right\rangle
	\right]^{n}.
	\label{eq:FT3}
\end{equation}
Now, notice that, given positive real numbers $n_{1},n_{2},t$, the following identity holds:
\begin{equation}
	\left[
		\left\langle \psi\right| \hat{U}^{(S)}\left(t\right) \left|\psi\right\rangle
	\right]^{n_{1}}
	\left[
		\left\langle \psi\right| \hat{U}^{(S)} \left(t\right) \left|\psi\right\rangle
	\right]^{n_{2}} =
	\left[
		\left\langle \psi\right| \hat{U}^{(S)} \left(t\right) \left|\psi\right\rangle
	\right]^{n_{1}+n_{2}}.
	\label{T1T2}
\end{equation}
Defining $T_{1}\equiv n_{1}t >0$ and $T_{2}\equiv n_{2}t >0$, so that
\[
	t = \frac{T_1}{n_1} = \frac{T_2}{n_2} = \frac{T_1 + T_2}{n_1 + n_2},
\]
Eq. (\ref{T1T2}) becomes:
\begin{multline*}
	\left[\left\langle \psi\right|
		\hat{U}^{(S)} \left(T_{1}/n_{1}\right)
	\left|\psi\right\rangle \right]^{n_{1}}
	\left[\left\langle \psi\right|
		\hat{U}^{(S)} \left(T_{2}/n_{2}\right)
	\left|\psi\right\rangle \right]^{n_{2}} \\
	=\left[\left\langle \psi\right|
		\hat{U}^{(S)}\left[\left(T_{1}+T_{2}\right)/\left(n_{1}+n_{2}\right)\right]
	\left|\psi\right\rangle \right]^{n_{1}+n_{2}}.
\end{multline*}
Taking the limits $n_{1}\to\infty$ and $n_{2}\to\infty$ and using
Eq. (\ref{eq:FT3}), we find:
\begin{equation}
	F\left(T_{1}\right)F\left(T_{2}\right)
	=F\left(T_{1}+T_{2}\right),
	\label{eq:SumIdentity}
\end{equation}
for any $T_1,T_2>0$.
Similarly, notice that, given some positive number $a>0$:
\begin{equation}
	F\left(aT\right)
	=\lim_{n\to\infty}
	\left[
		\left\langle \psi\right| \hat{U}^{(S)} \left(aT/n\right) \left|\psi\right\rangle
	\right]^{n}.
	\label{aT}
\end{equation}
Replacing $n$ with $m=n/a$, we find:
\begin{equation}
	F\left(aT\right)
	=\lim_{m\to\infty}
	\left[
		\left\langle \psi\right| \hat{U}^{(S)}\left(T/m\right) \left|\psi\right\rangle
	\right]^{am}
	=\left[F\left(T\right)\right]^{a}.
	\label{eq:PowerIdentity}
\end{equation}
Combining Eqs. (\ref{eq:SumIdentity}) and (\ref{eq:PowerIdentity}), we find that, for every $T_{1},T_{2},a,b\ge0$:
\begin{equation}
	\left[F\left(T_{1}\right)\right]^{a}\left[F\left(T_{2}\right)\right]^{b}
	=F\left(aT_{1}+bT_{2}\right).
	\label{eq:LinearIdentity}
\end{equation}
Taking the logarithm on both sides of Eq. (\ref{eq:LinearIdentity}), we have:
\[
	\log\left[F\left(aT_{1}+bT_{2}\right)\right]
	=a\log\left[F\left(T_{1}\right)\right]
	+b\log\left[F\left(T_{2}\right)\right],
\]
which shows that the logarithm of $F\left(T\right)$ is a linear function that can be expressed in terms of two scalar constants $A$ and $B$:
\[
	\log F\left(T\right)=A+BT.
\]
According to Eq. (\ref{eq:FT3}), $F\left(0\right)=1$, which means we must have $A=0$.
Hence, $F\left(T\right)$ becomes:
\begin{equation}
	F\left(T\right)
	=e^{\mathrm{Re}\left\{ B\right\} T}
	e^{i\mathrm{Im}\left\{ B\right\} T},\;\;\;
	T\ge0.
	\label{eq:FTpos}
\end{equation}

It is possible to repeat this argument exchanging $t$ by $-t$ in Eq. (\ref{T1T2}) and $T$ by $-T$ in Eq. (\ref{aT}), so that Eq. (\ref{eq:LinearIdentity}) is derived for $T_{1},T_{2}\le0$.
In this way, we arrive at an expression that is analogous to Eq. (\ref{eq:FTpos}), but in terms of a different scalar $C$:
\begin{equation}
	F\left(T\right)=
	e^{\mathrm{Re}\left\{ C\right\} T}e^{i\mathrm{Im}\left\{ C\right\} T},\;\;\;
	T\le0.
	\label{eq:FTneg}
\end{equation}
However, from the definition given in Eq. (\ref{eq:FT3}), we can see that
\begin{align*}
	F\left(-T\right)
	=& \lim_{n\to\infty}
		\left[
			\left\langle \psi\right|
			\hat{U}^{(S)} \left(-T/n\right)
			\left|\psi\right\rangle
		\right]^{n} \\
	=& \lim_{n\to\infty}
		\left\{
			\left\langle \psi\right|
			\left[ \hat{U}^{(S)} \left(T/n\right) \right]^\dagger
			\left|\psi\right\rangle
		\right\}^{n} \\
	=& F^{*}\left(T\right),
\end{align*}
which is only possible for every $T>0$ if
\begin{align*}
	\mathrm{Re}\left\{ B\right\}  & =-\mathrm{Re}\left\{ C\right\} ,\\
	\mathrm{Im}\left\{ B\right\}  & =\mathrm{Im}\left\{ C\right\} .
\end{align*}

Misra and Sudarshan employed arguments from group theory to prove that $B$ is a purely imaginary number. \cite{Misra}
Here we will do something simpler \cite{Facchi}
and write the time evolution operator explicitly in terms of the Hamiltonian $\hat{H}^{(S)}$:
\begin{equation}
	\hat{U}^{(S)} \left(T/n\right)
	=\exp\left\{ -\frac{i}{\hbar}\frac{T}{n}\hat{H}^{(S)}\right\}
	=\sum_{m=0}^{\infty}\frac{1}{m!}
	\left(-\frac{i}{\hbar}\frac{T}{n}\right)^{m}\left[ \hat{H}^{(S)} \right]^{m}.
	\label{expH}
\end{equation}
Replacing Eq. (\ref{expH}) in Eq. (\ref{eq:FT3}), we find:
\begin{equation*}
	F\left(T\right)
	=\lim_{n\to\infty}
	\left[
		\sum_{m=0}^{\infty}\frac{1}{m!}\left(-\frac{i}{\hbar}\frac{T}{n}\right)^{m}\left\langle \psi\right|\left[ \hat{H}^{(S)} \right]^{m}\left|\psi\right\rangle
	\right]^{n}.
\end{equation*}
Terms with $m>2$ will vanish in the limit when $n\to\infty$, because they are multiplied by $n^{-2}$ or lower, which vanish even when multiplied by $n$ in the binomial.
The only terms left are $m=0$ and $m=1$:
\begin{equation}
	F\left(T\right)
	=\lim_{n\to\infty}
	\left[
		1+\left(-\frac{i}{\hbar}\frac{T}{n}\right)
		\left\langle \psi\right|\hat{H}^{(S)}\left|\psi\right\rangle
	\right]^{n}=\exp\left\{ -\frac{i}{\hbar}T
	\left\langle \psi\right|\hat{H}^{(S)}\left|\psi\right\rangle \right\} .
\end{equation}
Therefore,
\begin{equation}
	B=C=
	-\frac{i}{\hbar}\left\langle \psi\right|\hat{H}^{(S)}\left|\psi\right\rangle ,
	\label{eq:BImag}
\end{equation}
which is a purely imaginary number, because $\left\langle \psi\right|\hat{H}^{(S)}\left|\psi\right\rangle $ is the expectation value of a Hermitian operator, and therefore real.

Then, we can replace Eqs. (\ref{eq:FTpos}) and (\ref{eq:FTneg}) in Eq. (\ref{eq:finalProbabilityFinal}) and use the result from Eq. (\ref{eq:BImag}) to conclude that:
\begin{equation}
	P_\infty \left(T\right)
	= e^{\mathrm{Re}\left\{ B\right\} T}e^{\mathrm{Re}\left\{ C\right\} T}
	= e^{2\mathrm{Re}\left\{ -i\left\langle \psi\right|\hat{H}^{(S)}\left|\psi\right\rangle/\hbar \right\} T}
	= e^{0}
	=1.
\end{equation}
Hence, the probability of finding the system in the initial state becomes $1$ as the projective observations become more closely spaced.
This is the Quantum Zeno Effect: a system under constant observation of its initial state will never evolve to a different state.

\subsection{Quantum Zeno Effect via continuous measurements}

Misra and Sudarshan discussed the implications of the existence of such continuous measurements in their original
article:
\begin{quote}
	``Is it a curious but innocent mathematical result or does it have something to say about the foundation of quantum theory?
	Does it, for example, urge us to have a principle in the formulation of quantum theory that forbids the continuous observation of an observable that is not a constant of motion?''\cite{Misra}
\end{quote}
They conclude that quantum mechanics may lack the proper mathematical methods to describe this type of measurement and to calculate the probability that a system will change its state within a period of time:
\begin{quote}
	``The quantum Zeno's paradox shows that the seemingly natural approach to this problem discussed in the preceding sections leads to bizarre and physically unacceptable answers.
	We thus lack a trustworthy quantum-theoretic algorithm for computing such probabilities.
	Until such a trustworthy algorithm is developed the completeness of quantum theory must remain in doubt.''\cite{Misra}
\end{quote}
Since then, however, experimental evidence seems to be demonstrating that the Quantum Zeno Effect occurs in nature and that the quantum theory is correct~\cite{Itano} -- although not without some controversy~\cite{AntiZeno}.

In terms of formalism, the development of master equations such as Eq. (\ref{CresserEq}) have allowed the dynamical description of a system under constant observation, including means of obtaining the Quantum Zeno Effect \cite{Brasil3}.
As a simple example of this, suppose we have a spin-$1/2$ system initially in the state $\left| + \right\rangle$, but subjected to a Hamiltonian that flips this state:
\begin{equation}
	\hat H^{(S)}
	= \hbar \omega_0 \left(
		\left| + \right\rangle \left\langle - \right|
		+ \left| - \right\rangle \left\langle + \right|
	\right)
	= \hbar \omega_0 \hat \sigma_x,
	\label{lastH}
\end{equation}
where $\omega_0$ is a frequency.
Concomitantly, we observe it with an apparatus that continuously detects whether the system has changed its initial state, which corresponds to the following Kraus operators:
\begin{equation}
	\hat K_1^{(S)} = \left | + \right\rangle \left\langle + \right|,
	\;\;\;
	\hat K_2^{(S)} = \left | - \right\rangle \left\langle - \right|.
	\label{lastKraus}
\end{equation}
It is easy to see that they add up to the identity, as demanded by Eq. (\ref{KrausID}).

Replacing Eqs. (\ref{lastH}) and (\ref{lastKraus}) in Eq. (\ref{CresserEq}), we find:
\begin{equation}
	\frac{\mathrm{d}}{\mathrm{d} t} \hat \rho^{(S)} (t) =
	- i\omega_0 \left[ \hat \sigma_x , \hat \rho^{(S)} (t) \right]
	+ \frac{1}{2} \lambda \left\{
		\hat \sigma_z \hat \rho^{(S)} (t) \hat \sigma_z
		- \hat \rho^{(S)} (t)
	\right\},
	\label{finalCresser}
\end{equation}
where we used the following identity to simplify the Lindbladian term in Eq. (\ref{finalCresser}):
\begin{align*}
	\hat \sigma_z \hat \rho^{(S)} (t) \hat \sigma_z - \hat \rho^{(S)} (t) =&
	\left(  \left| + \right\rangle \left\langle + \right| - \left| - \right\rangle \left\langle - \right| \right)  \hat \rho^{(S)} (t) \left(  \left| + \right\rangle \left\langle + \right| - \left| - \right\rangle \left\langle - \right| \right)
	-\hat \rho^{(S)} (t) \\
	=&
	2 \left| + \right\rangle \left\langle + \right| \hat \rho^{(S)} (t) \left| + \right\rangle \left\langle + \right|
	+2 \left| - \right\rangle \left\langle - \right| \hat \rho^{(S)} (t) \left| - \right\rangle \left\langle - \right| \\
	&-2 \left| + \right\rangle \left\langle + \right| \hat \rho^{(S)} (t) \left| + \right\rangle \left\langle + \right|
	-2 \left| - \right\rangle \left\langle - \right| \hat \rho^{(S)} (t) \left| - \right\rangle \left\langle - \right| \\
	&-2 \left| - \right\rangle \left\langle - \right| \hat \rho^{(S)} (t) \left| + \right\rangle \left\langle + \right|
	-2 \left| + \right\rangle \left\langle + \right| \hat \rho^{(S)} (t) \left| - \right\rangle \left\langle - \right| \\
	=&
	2 \left| + \right\rangle \left\langle + \right| \hat \rho^{(S)} (t) \left| + \right\rangle \left\langle + \right|
	+2 \left| - \right\rangle \left\langle - \right| \hat \rho^{(S)} (t) \left| - \right\rangle \left\langle - \right| \\
	&-2 \left( \left| + \right\rangle \left\langle + \right| + \left| - \right\rangle \left\langle - \right| \right) \hat \rho^{(S)} (t) \left( \left| + \right\rangle \left\langle + \right| + \left| - \right\rangle \left\langle - \right| \right) \\
	=&
	2 \sum_i \hat K_i^{(S)} \hat \rho^{(S)} (t) \left[ \hat K_i^{(S)} \right]^\dagger
	- 2 \hat \rho^{(S)} (t).
\end{align*}

The probability of the system being detected in the same state as the initial state by a projective measurement at the instant $t$ is found by multiplying the density matrix by $\left| + \right\rangle \left\langle + \right|$ and taking the trace:
\[
	P_\infty (t)
	= \mathrm{Tr} \left\{ \hat \rho^{(S)} (t) \left| + \right\rangle \left\langle + \right| \right\}
	= \left\langle + \right| \hat \rho^{(S)} (t) \left| + \right\rangle.
\]
Multiplying Eq. (\ref{finalCresser}) by $\left\langle \pm \right|$ and $\left| \pm \right\rangle$, we find the differential equations for the components of the density matrix:
\begin{align*}
	\frac{\mathrm{d}}{\mathrm{d} t} \left\langle + \right| \hat \rho^{(S)} (t) \left| + \right\rangle =&
	- i \omega_0 \left\langle - \right| \hat \rho^{(S)} (t) \left| + \right\rangle
	+ i \omega_0 \left\langle + \right| \hat \rho^{(S)} (t) \left| - \right\rangle, \\
	\frac{\mathrm{d}}{\mathrm{d} t} \left\langle + \right| \hat \rho^{(S)} (t) \left| - \right\rangle =&
	- i \omega_0 \left\langle - \right| \hat \rho^{(S)} (t) \left| - \right\rangle
	+ i \omega_0 \left\langle + \right| \hat \rho^{(S)} (t) \left| + \right\rangle
	- \lambda \left\langle + \right| \hat \rho^{(S)} (t) \left| - \right\rangle, \\
	\frac{\mathrm{d}}{\mathrm{d} t} \left\langle - \right| \hat \rho^{(S)} (t) \left| + \right\rangle =&
	- i \omega_0 \left\langle + \right| \hat \rho^{(S)} (t) \left| + \right\rangle
	+ i \omega_0 \left\langle - \right| \hat \rho^{(S)} (t) \left| - \right\rangle
	- \lambda \left\langle - \right| \hat \rho^{(S)} (t) \left| + \right\rangle, \\
	\frac{\mathrm{d}}{\mathrm{d} t} \left\langle - \right| \hat \rho^{(S)} (t) \left| - \right\rangle =&
	- i \omega_0 \left\langle + \right| \hat \rho^{(S)} (t) \left| - \right\rangle
	+ i \omega_0 \left\langle - \right| \hat \rho^{(S)} (t) \left| + \right\rangle.
\end{align*}
In matrix form, this system of four linear equations can be written as:
\begin{equation*}
	\frac{\mathrm{d}}{\mathrm{d} t}
	\left( \begin{array}{c}
		\left\langle + \right| \hat \rho^{(S)} (t) \left| + \right\rangle \\
		\left\langle + \right| \hat \rho^{(S)} (t) \left| - \right\rangle \\
		\left\langle - \right| \hat \rho^{(S)} (t) \left| + \right\rangle \\
		\left\langle - \right| \hat \rho^{(S)} (t) \left| - \right\rangle
	\end{array} \right)
	=
	\left( \begin{array}{cccc}
		0 & i \omega_0 & -i\omega_0 & 0 \\
		i\omega_0 & -\lambda & 0 & -i\omega_0 \\
		-i\omega_0 & 0 & -\lambda & i\omega_0 \\
		0 & -i\omega_0 & i\omega_0 & 0
	\end{array} \right)
	\left( \begin{array}{c}
		\left\langle + \right| \hat \rho^{(S)} (t) \left| + \right\rangle \\
		\left\langle + \right| \hat \rho^{(S)} (t) \left| - \right\rangle \\
		\left\langle - \right| \hat \rho^{(S)} (t) \left| + \right\rangle \\
		\left\langle - \right| \hat \rho^{(S)} (t) \left| - \right\rangle
	\end{array} \right),
\end{equation*}
whose solution for the initial state $\left| + \right\rangle\left\langle + \right|$ is the matrix exponential:
\begin{equation}
	\left( \begin{array}{c}
		\left\langle + \right| \hat \rho^{(S)} (t) \left| + \right\rangle \\
		\left\langle + \right| \hat \rho^{(S)} (t) \left| - \right\rangle \\
		\left\langle - \right| \hat \rho^{(S)} (t) \left| + \right\rangle \\
		\left\langle - \right| \hat \rho^{(S)} (t) \left| - \right\rangle
	\end{array} \right)
	=
	\exp \left\{
	\left( \begin{array}{cccc}
		0 & i \omega_0 & -i\omega_0 & 0 \\
		i\omega_0 & -\lambda & 0 & -i\omega_0 \\
		-i\omega_0 & 0 & -\lambda & i\omega_0 \\
		0 & -i\omega_0 & i\omega_0 & 0
	\end{array} \right) t
	\right\}
	\left( \begin{array}{c}
		1 \\ 0 \\ 0 \\ 0
	\end{array} \right).
	\label{finalSolution}
\end{equation}

\begin{figure}[htb]
	\includegraphics[width=.8\textwidth]{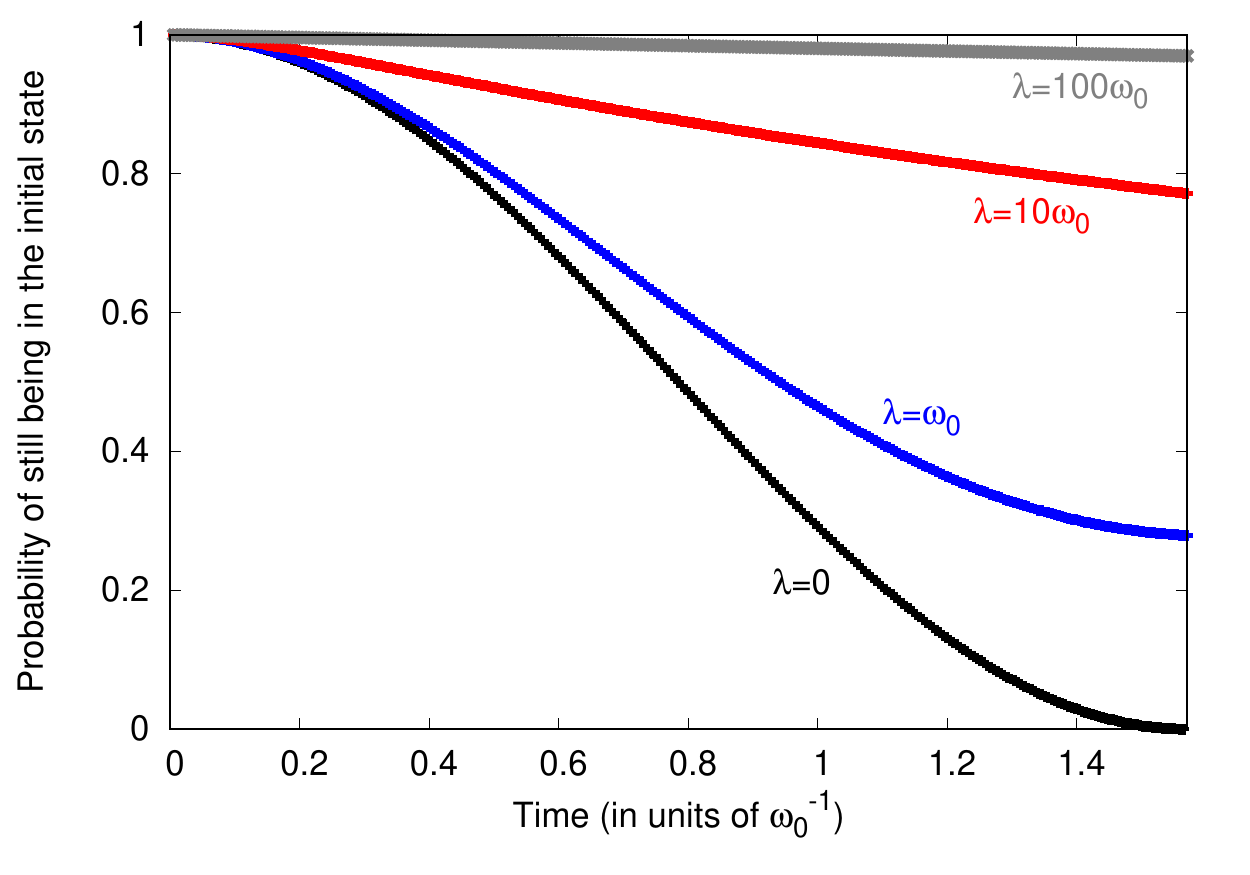}
	\caption{Probability $P_\infty (t)$ of keeping a system under continuous measurement in its initial state, calculated by solving numerically Eq. (\ref{finalSolution}).
	The unperturbed system ($\lambda=0$) transitions to a different state, but as the intensity $\lambda$ of the interaction with the measurement apparatus increases, the system almost does not leave the initial state, analogously to the Quantum Zeno Effect for discrete projective measurements.
	\label{figZeno} }
\end{figure}

The exponential of the matrix in Eq. (\ref{finalSolution}) is not easy to be calculated analytically, but its numerical simulation showing the Quantum Zeno Effect can be seen in Fig. \ref{figZeno}.
With this, we see that the formalism of continuous measurements allows us to quantify the Quantum Zeno Effect at each stage of the system's evolution with less hassle than the very formal result obtained before for projective measurements.

\section{Summary}

The emergence of quantum mechanics in the end of the 19th century and beginning of the 20th century resulted in the quantization of magnitudes that used to be continuously measured.
After a few years, this discontinuity was recognized as typical of atomic processes in general.
Meanwhile, new experiments and theoretical advances by Einstein and de Broglie revealed that atomic systems also displayed both wave and particle aspects.
This inspired Schr\"odinger's introduction of the wave function, which would be interpreted according to the rule of Born-Dirac, which turned measurements into probabilistic processes.
If in classical physics we could talk about an isolated system, in the orthodox interpretation of quantum mechanics it is necessary to consider the measurement apparatus and its interaction with the observed system.
According to the uncertainty principle, this imposes intrinsic limits to the precision of what can be measured.

These developments were concerned with what one \textit{cannot} do with a quantum measurement, and constitute the bulk of the problem of measurement that persists to this day.
The main difficulty is that the measurement in quantum mechanics necessarily requires the interaction between a quantum system and a classical apparatus, in a process that neither theory can fully describe.
But if the apparatus is considered quantum as well, it is possible to understand its dynamics and what \textit{can} be done with quantum measurements.

This is the step taken by von Neumann with his influential model of measurement, which has had lasting influence.
von Neumann boldly treated the measurement apparatus as a quantum system, and described the system-measurer interaction as unitary, thus constructing a theory for the transmission of information from the atomic world to our classical macroscopic instruments.

Treating the apparatus as a quantum object sparked a fruitful discussion about where the quantum world ends.
While Wigner believed that the answer lied in the consciousness of the observer, which would make it a privileged system, Bohm attributed classical features to the intensity of the interaction between the apparatus and the observer, which would make the interference terms vanish.
Everett went further, treating the whole universe as a wave function, where worlds containing all the possible outcomes would co-exist without interacting with each other.
These different views are summarized in Table \ref{interpretations}.

\begin{table}[hbt]
\begin{tabular}{|c|c|c|c|}
	\hline
	\bf Interpretation &
		\begin{tabular}{c} \bf Observed \\ \bf system $(S)$ \end{tabular} & \begin{tabular}{c} \bf Measurement \\ \bf apparatus $(M)$ \end{tabular} & \begin{tabular}{c} \bf Human \\ \bf observer $(E)$ \end{tabular} \\
	\hline
	\bf Orthodox &
		\begin{tabular}{c} Superposition \\  \end{tabular} &
		\begin{tabular}{c} Incoherent \\ mixture \end{tabular} &
		\begin{tabular}{c} Incoherent \\ mixture \end{tabular} \\
	\hline
	\begin{tabular}{c}
		\bf Bohm's formulation \\
		\bf of von Neumann's \\
		\bf model
	\end{tabular} &
		\begin{tabular}{c} Superposition \\  \end{tabular} &
		\begin{tabular}{c}
			Superposition  when \\
			 interacting with $S$, \\
			 incoherent mixture \\
			 when interacting \\ with $E$
		\end{tabular} &
		\begin{tabular}{c} Incoherent \\ mixture \end{tabular} \\
	\hline
	\bf Wigner &
		\begin{tabular}{c} Superposition \\  \end{tabular} &
		\begin{tabular}{c} Superposition \\  \end{tabular} &
		\begin{tabular}{c} Consciousness\\ destroys \\ superposition \end{tabular} \\
	\hline
	\bf Everett &
		\begin{tabular}{c} Superposition \\  \end{tabular} &
		\begin{tabular}{c} Superposition \\   \end{tabular} &
		\begin{tabular}{c} Superposition \\   \end{tabular} \\
	\hline
\end{tabular}
\caption{Which parts of von Neumann's chain can be in superposition states and which parts cannot be more than incoherent mixtures, according to the interpretations mentioned in this article. \label{interpretations}}
\end{table}

A discussion of the finer points of Everett's interpretation led to the decoherence program, now used independently of the many-worlds interpretations to describe errors in quantum computers, for examples.
This is not the only practical contribution of von Neumann's model: the Quantum Phase Estimation algorithm, for example, proposes using an analogous interaction between two discrete systems to determine the energy of a quantum state encoded as qubits \cite{chemistry}.
Moreover, reducing the intensity of the interaction during von Neumann's measurement led to weak measurements, which are under intense research due to the surprising effects that they cause, such as obtaining results that lie much beyond the expected limits.

This is a field still open to investigations, as exemplified by the several books \cite{Braginsky1,Busch,Jacobs,Mensky1,Mensky}, reports \cite{Bassi,Namiki} and collected works \cite{Tombesi,WheelerZurek} on the subject.
We have not intended to write a complete treatise on quantum measurements, but rather to provide a broad introduction to the subject centered on the main paradigm introduced by von Neumann, discussing both the orthodoxy and some attempts to break with it.
Other approaches that, for concision, were not included here include quantum state diffusion \cite{Percival}; dynamical reduction models \cite{Bassi} like the original proposal by Giancarlo Ghirardi, Alberto Rimini and Tulio Weber (GRW) \cite{Ghirardi} and related developments \cite{Ghirardi1,Pearle}; superselection rules \cite{Wick,Wick1,Bub,Landsman,Giulini}; and consistent histories \cite{Dowker,Griffiths,Griffiths2}.
We hope that, with the tools we have provided, the readers will feel prepared to make an informed choice about where to continue pursuing their studies.

\appendix

\section{Quantum non-demolition measurements}
\label{subsec:Non-demolition}

A quantum non-demolition (QND) measurement is constructed so that, if its result is an eigenvalue $s_n$ of the observable $\hat{S}^{(S)}$, then any subsequent measurement of this observable will yield the same result.
This means that the experimental apparatus will not insert more uncertainty into the system beyond the uncertainty inherent to quantum mechanics.
Examples of QND measurements can be found in quantum optics \cite{Milburn,Walls1,Grangier} and in electromechanical systems \cite{Olimpio}.

An example of a measurement that does not satisfy this is the position of a free particle \cite{Unnikrishnan}.
There is no restriction in principle about how precise the position can be determined, but this will create some uncertainty $\Delta p$ in the associated momentum.
A new measurement of the position performed at a time $t$ after the first one will have an uncertainty corresponding to how much the particle may have moved during this interval, $\Delta x=t\left(\Delta p\right)/m$, where
$m$ is the mass of the free particle.
Hence, a new measurement of the position will not necessarily yield the same value as the original measurement.
On the other hand, a measurement of the momentum will not be affected by the uncertainty in the position, and can therefore be turned into a QND measurement.

More generally, suppose that we want to measure a system originally in a certain eigenstate $\left| s_n \right\rangle $ of $\hat{S}^{\left(S\right)}$, while the measuring apparatus is in the original state $\left| m_0 \right\rangle $.
If the measurement is a QND, the state after the interaction $\hat \Delta$ should still be an eigenstate with the same eigenvalue $s_n$:
\begin{equation}
	\hat{S}^{\left(S\right)}
	\left[
		\hat\Delta
		\left(
			\left| s_n \right\rangle \left| m_0 \right\rangle
		\right)
	\right]
	= s_n
	\left[
		\hat\Delta \left(\left|s_n\right\rangle \left| m_0 \right\rangle \right)
	\right].
	\label{eq:eigenvalue1}
\end{equation}
But as $\hat{S}^{\left(S\right)} \left|s_n\right\rangle = s_n \left|s_n\right\rangle$, Eq. (\ref{eq:eigenvalue1}) can be re-written as:
\begin{equation}
	\hat{S}^{\left(S\right)}
	\hat\Delta
	\left(
		\left|s_n\right\rangle \left| m_0 \right\rangle
	\right)
	=
	\hat\Delta
	\left(
	\hat{S}^{\left(S\right)} \left|s_n\right\rangle
	\left| m_0 \right\rangle
	\right).
	\label{eq:eigenvalue2}
\end{equation}
In order for Eq. (\ref{eq:eigenvalue2}) to be true for any $s_n$, the following identity must be satisfied:
\begin{equation}
	\left[ \hat{S}^{\left(S\right)}, \hat\Delta \right]
	\left| m_0 \right\rangle
	=0.
	\label{eq:Condition1}
\end{equation}
Eq. (\ref{eq:Condition1}) is the necessary and sufficient condition for a measurement to be considered a QND measurement \cite{Braginsky}.
A sufficient condition that satisfies Eq. (\ref{eq:Condition1}) is if $\hat{S}^{\left(S\right)}$ and $\hat\Delta$ commute:
\begin{equation}
	\left[ \hat{S}^{\left(S\right)}, \hat\Delta \right] = \hat{0}.
	\label{eq:Condition2}
\end{equation}
Furthermore, as $\hat \Delta = e^{-i\hat H \tau/\hbar}$, Eq. (\ref{eq:Condition2}) will be satisfied if the observable commutes with the Hamiltonian, which is another sufficient, but not necessary condition:
\begin{equation}
	\left[\hat{S}^{\left(S\right)},\hat{H}\right]
	=\hat{0}.
	\label{eq:Condition3}
\end{equation}
Eq. (\ref{eq:Condition3}) is equivalent to saying that $\left\langle \hat{S}\right\rangle $ is a constant of motion, which means that the eigenvalue $s_n$ will not change in time.

Returning to the example given in the beginning of this section, we can see that, for a free particle, the Hamiltonian of the system is:
\[
\hat{H}^{\left(S\right)}=\frac{\hat{P}^{2}}{2m}.
\]
Unless the interaction Hamiltonian $\hat{H}^{\left(S+M\right)}$ is chosen very carefully, the total Hamiltonian $\hat{H}=\hat{H}^{\left(S\right)}+\hat{H}^{\left(M\right)}+\hat{H}^{\left(S+M\right)}$ will not commute with the position operator $\hat{Q}$, because the term $\hat{H}^{\left(S\right)}$ does not commute with it.
This means that the condition from Eq. (\ref{eq:Condition3}) will not be satisfied.
On the other hand, operator $\hat{P}$ commutes with this $\hat{H}^{\left(S\right)},$ and will also commute with the interaction term as long as $\hat{H}^{\left(S+M\right)}$ takes a form similar to Eq. (\ref{intvN}).
Therefore, a measurement of the position causes the demolition of the system, while a measurement of the momentum can be a QND measurement.

\section{von Neumann's derivation of the density matrix formalism}
\label{vnMatrix}

Here we reproduce the procedure employed in von Neumann's textbook\cite{Neumann} using modern notation.
Write the observable $\hat{Q}^{\left(S\right)}$ of the system $S$ in matrix form, using the eigenbasis $\left\{ \left| s_n \right\rangle \right\}$:
\[
	\hat{Q}^{\left(S\right)}
	=
	\sum_{m,n}
	Q_{m,n} \left|s_{m}\right\rangle \left\langle s_{n}\right|.
\]
Now, divide the elements of the matrix into diagonal and off-diagonal:
\begin{equation}
	\hat{Q}^{(S)}  =
	\sum_m
	Q_{m,m}
		\left|s_{m}\right\rangle \left\langle s_{m}\right|
	+ \sum_{m>n} \left\{
		Q_{m,n} \left|s_{m}\right\rangle \left\langle s_{n}\right|
		+ Q_{n,m} \left|s_{n}\right\rangle \left\langle s_{m}\right|
	\right\}.
\label{Qmn}
\end{equation}

Next, define three new sets of observables:
\begin{align*}
	\hat{A}_{m}^{\left(S\right)} &
		\equiv\left|s_{m}\right\rangle \left\langle s_{m}\right|,\\
	\hat{B}_{m,n}^{\left(S\right)} &
		\equiv\left|s_{m}\right\rangle \left\langle s_{n}\right|+\left|s_{n}\right\rangle \left\langle s_{m}\right|,\\
	\hat{C}_{m,n}^{\left(S\right)} &
		\equiv i\left|s_{m}\right\rangle \left\langle s_{n}\right|-i\left|s_{n}\right\rangle \left\langle s_{m}\right|,
\end{align*}
and replace them in Eq. (\ref{Qmn}), obtaining the following decomposition of $\hat Q^{(S)}$:
\begin{equation}
	\hat{Q}^{\left(S\right)} =
	\sum_m
	Q_{m,m} \hat{A}_{m}^{\left(S\right)}
	+ \frac{1}{2} \sum_{m>n}
	\left\{
		\left( Q_{m,n} + Q_{n,m} \right) \hat{B}_{m,n}^{\left(S\right)}
		-i \left( Q_{m,n} - Q_{n,m} \right) \hat{C}_{m,n}^{\left(S\right)}
	\right\} .
	\label{decomposition}
\end{equation}

Using Eq. (\ref{decomposition}), the expectation value of any observable $\hat{Q}^{\left(S\right)}$ can be calculated from the expectation values of the auxiliary observables $\hat{A}_{m}^{\left(S\right)}$, $\hat{B}_{m,n}^{\left(S\right)}$, and $\hat{C}_{m,n}^{\left(S\right)}$.
Thus, von Neumann defines the density operator $\hat{\rho}^{\left(S\right)}$
as the Hermitian matrix formed by these expectation values:
\[
	\hat{\rho}^{\left(S\right)} \equiv
	\sum_m
	\left\langle \hat{A}_{m}^{\left(S\right)}\right\rangle \hat A_m^{(S)}
	+ \frac{1}{2}\sum_{m>n}
	\left\{
		\left\langle \hat{B}_{m,n}^{\left(S\right)} \right\rangle
		\hat B_{m,n}^{(S)}
		-
		\left\langle \hat{C}_{m,n}^{\left(S\right)} \right\rangle
		\hat C_{m,n}^{(S)}
	\right\} ,
\]
so that the expectation value of any observable $\hat{Q}^{\left(S\right)}$ will be given by the trace of it multiplied by the density matrix, just as in Eq. (\ref{expectation}):
\begin{align*}
	\left\langle \hat{Q}^{\left(S\right)}\right\rangle
	 = &
		\sum_m
		Q_{m,m} \left\langle \hat{A}_{m}^{\left(S\right)} \right\rangle
		+ \frac{1}{2} \sum_{m>n}
		\left\{
			\left( Q_{m,n} + Q_{n,m} \right) \left\langle \hat{B}_{m,n}^{\left(S\right)} \right\rangle \right. \\
		& \left.
			-i \left( Q_{m,n} - Q_{n,m} \right) \left\langle \hat{C}_{m,n}^{\left(S\right)} \right\rangle
		\right\} \\
	 = &
		\sum_m
		\mathrm{Tr}_{S} \left\{
		\hat Q^{(S)} \hat{A}_{m}^{\left(S\right)}
		\right\}
		 \left\langle \hat{A}_{m}^{\left(S\right)} \right\rangle
		+ \frac{1}{2} \sum_{m>n}
		\left\{
		\mathrm{Tr}_{S} \left\{
			\hat Q^{(S)}
			\hat{B}_{m,n}^{\left(S\right)} \right\}
			\left\langle \hat{B}_{m,n}^{\left(S\right)} \right\rangle \right.
			\\ & \left.
			- \mathrm{Tr}_{S} \left\{ \hat Q^{(S)} \hat{C}_{m,n}^{\left(S\right)} \right\}
			\left\langle \hat{C}_{m,n}^{\left(S\right)} \right\rangle
		\right\} \\
	= & \mathrm{Tr}_{S} \left\{
		\hat{\rho}^{\left(S\right)} \hat{Q}^{\left(S\right)}
	\right\}.
\end{align*}

\section{Splitting a positive operator}
\label{decomposition}

Suppose we want to split a positive Hermitian operator $\hat A^{(S)}$ into the product of two operators $\hat B^{(S)}$ and $\hat C^{(S)}$:
\begin{equation}
	\hat A^{(S)} = \hat B^{(S)} \hat C^{(S)}.
	\label{id1}
\end{equation}
This is clearly feasible if we choose $\hat B^{(S)} = \hat A^{(S)}$ and $\hat C^{(S)} = \hat 1^{(S)}$ (or vice-versa), but we are seeking non-trivial solutions.

The fact that $\hat A^{(S)}$ is Hermitian and positive means that it can be diagonalized in its eigenbasis $\left| a_i \right\rangle$, leaving its positive eigenvalues $a_i$ in the diagonal:
\begin{equation}
	\hat A^{(S)} = \sum_i a_i \left| a_i \right\rangle \left\langle a_i \right|.
	\label{decompA}
\end{equation}
We can write $\hat B^{(S)}$ and $\hat C^{(S)}$ in this same basis, although they are not necessarily diagonal:
\begin{equation}
	\hat B^{(S)} = \sum_{i,j} b_{i,j} \left| a_i \right\rangle \left\langle a_j \right|, \;\;\;
	\hat C^{(S)} = \sum_{i,j} c_{i,j} \left| a_i \right\rangle \left\langle a_j \right|.
	\label{decompBC}
\end{equation}

Replacing Eqs. (\ref{decompA}) and (\ref{decompBC}) in Eq. (\ref{id1}), we find:
\[
	\sum_i a_i \left| a_i \right\rangle \left\langle a_i \right|
	= \sum_{i,j} \sum_k b_{i,k} c_{k,j}^* \left| a_i \right\rangle \left\langle a_j \right|.
\]
This is true only if the coefficients that multiply $\left| a_i \right\rangle \left\langle a_j \right|$ on both sides are equal:
\begin{equation}
	\delta_{i,j} a_i = \sum_k b_{i,k} c_{k,j}^*.
	\label{condition}
\end{equation}
The trivial solution to this is $b_{i,k} = \delta_{i,k} a_i$ and $c_{k,j} = \delta_{k,j}$, which corresponds to $\hat B^{(S)} = \hat A^{(S)}$ and $\hat C^{(S)} = \hat 1^{(S)}$.
However, as $a_i>0$, it is far more interesting to choose
\begin{equation}
	b_{i,k} = \delta_{i,k} e^{i \phi_i} \sqrt{a_i}, \;\;\;
	c_{k,j} = \delta_{k,j} e^{-i \phi_i} \sqrt{a_i},
	\label{solution}
\end{equation}
where $\phi_i$ is a phase.
Eq. (\ref{solution}) clearly satisfies Eq. (\ref{condition}), but it also means that $\hat B^{(S)} = \left[ \hat C^{(S)} \right]^\dagger$.
Therefore, every positive Hermitian operator $\hat A^{(S)}$ can be written as the product:
\[
	\hat A^{(S)} = \left[ \hat C^{(S)} \right]^\dagger \hat C^{(S)}.
\]

\section{Deriving the POVM}
\label{A3}

We want to make sure that the $\hat A_3^{(S)}$ defined in Eq. (\ref{defA3}) is a positive operator.
This means that, for every state vector
\[
	\left| \psi \right\rangle = \cos \theta \left| + \right\rangle + \sin \theta \left| - \right\rangle,
\]
the expectation value of $\hat A_3^{(S)}$ must be positive:
\[
	\left\langle \psi \right| \hat A_3^{(S)} \left| \psi \right\rangle =
	\left\langle \psi \right| \left. \psi \right\rangle
	- \left\langle \psi \right| \hat A_1^{(S)} \left| \psi \right\rangle
	- \left\langle \psi \right| \hat A_2^{(S)} \left| \psi \right\rangle
	= 1 - p_z \left| \cos \theta \right|^2 - p_x \frac{1}{2} \left| \cos \theta + \sin \theta \right|^2 \ge 0.
\]
Assuming that both detectors are equally  important to our experiment, we choose $p_x = p_z$.
Then, a little algebra turns
\[
	1 - p_z \left| \cos \theta \right|^2 - p_z \frac{1}{2} \left| \cos \theta + \sin \theta \right|^2 \ge 0
\]
into the inequality
\begin{equation}
	p_z f(\theta)
	\le 1,
	\label{appInequality}
\end{equation}
where we defined the real function
\[
	f (\theta) \equiv \cos^2 \theta + \frac{1}{2} \left( \cos \theta + \sin \theta \right)^2
	= 1 + \frac{1}{2} \cos(2\theta) + \frac{1}{2} \sin \left( 2 \theta \right).
\]
The value of $p_z$ must be chosen so that Eq. (\ref{appInequality}) is satisfied for every state vector, and hence for every value of $\theta$.
This is true as long as it is satisfied for the maximum value of $f(\theta)$.

The extremes of the function $f(\theta)$ occur when its derivative is zero:
\[
	\frac{\mathrm{d}}{\mathrm{d} \theta} f(\theta) =
	- \sin \left( 2 \theta \right)
	+ \cos \left( 2 \theta \right) = 0,
\]
which is to say that
\[
	\sin 2\theta = \cos 2\theta,
\]
meaning that $2\theta = \pi/4$.
A second derivative shows that this is indeed a maximum of the function:
\[
	\left. \frac{\mathrm{d}^2}{\mathrm{d} \theta^2} f(\theta) \right|_{2\theta=\pi/4} =
	- 2 \left. \cos \left( 2 \theta \right) \right|_{2\theta=\pi/4}
	- 2 \left. \sin \left( 2 \theta \right) \right|_{2\theta=\pi/4}
	= - 2 \frac{1}{\sqrt{2}} - 2 \frac{1}{\sqrt{2}} <0.
\]
Hence, the largest value of $p_z$ that we can choose, the one that will maximize the precision of our devices, is:
\[
	p_z = \frac{1}{f(2\theta=\pi/4)} = \frac{1}{1+1/\sqrt{2}} = \frac{\sqrt{2}}{\sqrt{2}+1} =
	2 - \sqrt{2}.
\]

\section{The Liouville--von Neumann equation}
\label{LiouvilleVN}

Here we will derive the evolution of a density matrix via Process 2.
Any density matrix can be expressed as a convex combination of pure states:
\[
	\hat \rho (t) = \sum_i p_i \left| \Psi_i (t) \right\rangle \left\langle \Psi_i (t) \right|,
\]
where the $p_i$ are probabilities and the $\left| \Psi_i \right\rangle$ are state vectors.

To find the equation of motion for Process 2, it suffices to take the time derivative of a density matrix:
\[
	\frac{\mathrm{d} \hat{\rho} (t)}{\mathrm{d}t}
	= \sum_{i}p_{i}\frac{\mathrm{d}\left|\Psi_{i} (t) \right\rangle }{\mathrm{d}t}\left\langle \Psi_{i} (t) \right|
	+ \sum_{i}p_{i}\left|\Psi_{i} (t) \right\rangle \frac{\mathrm{d}\left\langle \Psi_{i} (t) \right|}{\mathrm{d}t}.
\]
Replacing the Schr\"{o}dinger equation:
\[
	i \hbar \frac{\mathrm{d}\left| \Psi_{i} (t) \right\rangle}{\mathrm{d}t}
	= \hat H \left| \Psi_{i} (t) \right\rangle
\]
and its Hermitian conjugate:
\[
	-i \hbar \frac{\mathrm{d}\left\langle \Psi_{i} (t) \right|}{\mathrm{d}t}
	=  \left\langle \Psi_{i} (t) \right| \hat H,
\]
we find the Liouville--von Neumann equation \cite{CohenTannoudji,Neumann} in terms of the Hamiltonian $\hat H$:
\begin{equation}
	\frac{\mathrm{d} \hat{\rho} (t)}{\mathrm{d}t}
	=-\frac{i}{\hbar}\left[\hat{H},\hat{\rho}\left(t\right)\right].
	\label{eqLvN}
\end{equation}
The solution for Eq. (\ref{eqLvN}) is given by the time evolution operator $\hat{U} \left(t\right)$,
\begin{equation}
	\hat{\rho}\left(t\right)
	= \hat{U}\left(t\right) \hat{\rho} \left(0\right) \left[\hat{U}\left(t\right) \right]^{\dagger},
	\label{eq:TimeEvolution}
\end{equation}
which is defined as the operator that satisfies
\begin{equation}
	\frac{\mathrm{d} \hat{U} (t)}{\mathrm{d}t}
	=-\frac{i}{\hbar} \hat{H} \hat{U} (t),
	\label{defU}
\end{equation}
and which becomes the following when the Hamiltonian $\hat H$ is time-independent:
\begin{equation*}
	\hat U (t) =
	\exp \left\{
		- \frac{i}{\hbar} \hat H t
	\right\}.
\end{equation*}
If we take the time derivative of Eq. (\ref{eq:TimeEvolution}) and replace Eq. (\ref{defU}) in it, we obtain the Lioville-von Neumann equation as expressed in Eq. (\ref{eqLvN}), thus proving that Eq. (\ref{eq:TimeEvolution}) is indeed the solution.

\section*{Acknoledgements}
	The authors thank Prof. Reginaldo de Jesus Napolitano and Prof.  Miled Hassan Youssef Moussa (IFSC/USP) for useful discussions, and Dr. Oscar Salomon Duarte Mu\~{n}oz for comments on the manuscript.
	O. P. de S\'a  Neto acknowledges support from ``EDITAL FAPEPI/MCT/CNPq No. 007/2018: Programa de Infraestrutura para Jovens Pesquisadores/ Programa Primeiros Projetos (PPP)''.
	C. A. Brasil acknowledges support from Coordena\c{c}\~ao de Aperfei\c{c}oamento de Pessoal de N\'ivel Superior (CAPES) under the Programa Nacional de P\'os-Doutorado (PNPD), for the post-doctoral project developed at IFSC/ USP under the supervision of Prof. Reginaldo de Jesus Napolitano.

\bibliographystyle{unsrt}



\begin{thebibliography}{99}

\bibitem{Laplace}
P.-S. de{ }Laplace.
\newblock {\em A philosophical essay on probabilities}.
\newblock John Wiley \& Sons, New York, 1902.

\bibitem{Pais}
A.~Pais.
\newblock {\em Subtle is the {L}ord... -- The science and the life of {A}lbert
  {E}instein}.
\newblock Oxford University Press, Oxford, 1982.

\bibitem{Shor}
P.~W. Shor.
\newblock Scheme for reducing decoherence in quantum computer memory.
\newblock {\em Physical Review A}, 52, 1995.

\bibitem{Briegel}
H.~J. Briegel and R.~Raussendorf.
\newblock Persistent entanglement in arrays of interacting particles.
\newblock {\em Physical Review Letters}, 86:910--913, 2001.

\bibitem{Briegel1}
H.~J. Briegel, D.~E. Browner, W.~D\"{u}r, R.~Raussendorf, and M.~V.~D. Nest.
\newblock Measurement-based quantum computation.
\newblock {\em Nature Physics}, 5:19--26, 2009.

\bibitem{Eisberg1}
R.~Eisberg and R.~Resnick.
\newblock {\em Quantum Physics of Atoms, Molecules, Solids, Nuclei, and
  Particles}.
\newblock John Wiley \& Sons, New York, 1985.

\bibitem{Messiah}
A.~Messiah.
\newblock {\em Quantum mechanics, vol. 1}.
\newblock North-Holland Publishing Company, Amsterdam, 1961.

\bibitem{Schiff}
L.~I. Schiff.
\newblock {\em Quantum mechanics}.
\newblock McGraw-Hill Book Company, New York, 1968.

\bibitem{Eisberg2}
R.~Eisberg and R.~Resnick.
\newblock {\em Fundamentals of Modern Physics}.
\newblock John Wiley \& Sons, New York, 1961.

\bibitem{Merzbacher}
E.~Merzbacher.
\newblock {\em Quantum mechanics}.
\newblock John Wiley \& Sons Inc, New York, 1998.

\bibitem{Ballentine3}
L.~E. Ballentine.
\newblock {\em Quantum mechanics: A modern development}.
\newblock World Scientific Publishing, Singapore, 1988.

\bibitem{Peres}
A.~Peres.
\newblock {\em Quantum theory: {C}oncepts and methods}.
\newblock Kluwer Academic Publishers, New York, 2002.

\bibitem{Griffiths2}
R.~B. Griffiths.
\newblock {\em Consistent Quantum Theory}.
\newblock Cambridge University Press, Cambridge, 2002.

\bibitem{Purcell}
E.~M. Purcell.
\newblock {\em Electricity and magnetism -- {B}erkeley Physics Course vol. 2}.
\newblock McGraw--Hill Book Company, New York, 1985.

\bibitem{Sears}
F.~W. Sears and G.~L. Salinger.
\newblock {\em Thermodynamics, kinetic theory, and statistical thermodynamics}.
\newblock Addison-Wesley Publishing Company, Reading, 1982.

\bibitem{French1}
A.~P. French and E.~F. Taylor.
\newblock {\em Special relativity}.
\newblock W. W. Norton and Company Inc, New York, 1968.

\bibitem{Bucher}
M.~Bucher.
\newblock Rise and premature fall of the old quantum theory, 2008.
\newblock arXiv:0802.1366.

\bibitem{Hermann}
A.~Hermann.
\newblock {\em The genesis of quantum theory 1899--1913}.
\newblock MIT Press, Cambridge, 1971.

\bibitem{Jammer}
M.~Jammer.
\newblock {\em The conceptual development of quantum mechanics}.
\newblock McGraw--Hill Book Company, New York, 1966.

\bibitem{Feldens}
B.~Feldens, P.~M.~C. Dias, and W.~M.~S. Santos.
\newblock E assim se fez o quantum... / let there be quantum...
\newblock {\em Revista Brasileira de Ensino de F\'{i}sica}, 22:2602, 2010.

\bibitem{Tomonaga}
S.-I. Tomonaga.
\newblock {\em Quantum mechanics vol. 1}.
\newblock North-Holland Publishing Company, Amsterdam, 1968.

\bibitem{Castro}
L.~A. Castro, C.~A. Brasil, and R.~d.~J. Napolitano.
\newblock Elliptical orbits in the phase-space quantization.
\newblock {\em Revista Brasileira de Ensino de F\'{i}sica}, 38:e3318, 2016.

\bibitem{Abro}
A.~d'Abro.
\newblock {\em The rise of the new physics}.
\newblock Dover Publications Inc, New York, 1951.

\bibitem{Duncan1}
A.~Duncan and M.~Janssen.
\newblock The trouble with the orbits: {T}he {S}tark effect in the old and the
  new quantum theory.
\newblock {\em Studies in History and Philosophy of Science}, 48:68--83, 2014.

\bibitem{Duncan2}
A.~Duncan and M.~Janssen.
\newblock The {S}tark effect in the {B}ohr--{S}ommerfeld theory and in
  {S}chr\"{o}dinger's wave mechanics, 2014.
\newblock arXiv:1404.5341.

\bibitem{Parente}
F.~A.~G. Parente, A.~C.~F. dos{ }Santos, and A.~C. Tort.
\newblock Os 100 anos do \'{a}tomo de {B}ohr / one hundred years of {B}ohr's atom.
\newblock {\em Revista Brasileira de Ensino de F\'{i}sica}, 35:4301, 2013.

\bibitem{Studart}
N.~Studart.
\newblock A invenção do conceito de quantum de energia segundo {P}lanck.
\newblock {\em Revista Brasileira de Ensino de F\'{i}sica}, 22:523--535, 2000.

\bibitem{Planck1orig}
M.~Planck.
\newblock \"{U}ber eine verbesserung der wienschen spektralgleichung.
\newblock {\em Verhandlungen der Deutschen Physicalishen Gesellschaft},
  2:202--204, 1900.

\bibitem{Planck2orig}
M.~Planck.
\newblock Ueber das gesetz der energieverteilung im normalspectrum.
\newblock {\em Annalen der Physik}, 309:553--563, 1901.

\bibitem{Haar}
D.~ter{ }Haar.
\newblock {\em The old quantum theory}.
\newblock Pergamon Press Ltd, Oxford, 1967.

\bibitem{Kuhn}
T.~S. Kuhn.
\newblock {\em Black-body theory and the quantum discontinuity, 1894--1912}.
\newblock University of Chicago Press, Chicago, 1987.

\bibitem{Beck}
A.~Beck and P.~Havas.
\newblock {\em The collected papers of Albert Einstein vol. 2 - The Swiss
  Years: Writings, 1900-1909}.
\newblock Princeton University Press, Princeton, 1989.

\bibitem{Arons}
A.~Arons and M.~B. Peppard.
\newblock Einstein's proposal of the photon concept -- a translation of the
  {A}nnalen der {P}hysik paper of 1905.
\newblock {\em American Journal of Physics}, 33:367--374, 1965.

\bibitem{Thomson1}
J.~J. Thomson.
\newblock The magnetic properties of systems of corpuscles describing circular
  orbits.
\newblock {\em Philosophical Magazine Series 6}, 6:673--693, 1903.

\bibitem{Thomson2}
J.~J. Thomson.
\newblock On the structure of the atom: An investigation of the stability and
  periods of oscillation of a number of corpuscles arranged at equal intervals
  around the circumference of a circle; with application of the results to the
  theory of atomic structure.
\newblock {\em Philosophical Magazine Series 6}, 7:237--265, 1904.

\bibitem{Rutherford}
E.~Rutherford.
\newblock The structure of the atom.
\newblock {\em Philosophical Magazine and Journal of Science}, 27:488--498,
  1914.

\bibitem{Nagaoka}
H.~Nagaoka.
\newblock Motion of particles in an ideal atom illustrating the line and band
  spectra and the phenomena of radioactivity.
\newblock {\em Bulletin of the Mathematical and Physical Society of Tokyo},
  2:140--141, 1904.

\bibitem{Nagaoka1}
H.~Nagaoka.
\newblock On a dynamical system illustrating the spectrum lines and the
  phenomena of radio-activity.
\newblock {\em Nature}, 69:392--393, 1904.

\bibitem{Nagaoka2}
H.~Nagaoka.
\newblock Kinetics of a system of particles illustrating the line and band
  spectrum and the phenomena of radio-activity.
\newblock {\em Philosophical Magazine}, 7:445--455, 1904.

\bibitem{Heisenberg2}
W.~Heisenberg.
\newblock {\em The physical principles of the quantum theory}.
\newblock Dover, Mineola, 1949.

\bibitem{Laloe}
F.~Lalo\"{e}.
\newblock Do we really understand quantum mechanics? {S}trange correlations,
  paradoxes, and theorems.
\newblock {\em American Journal of Physics}, 69:655--701, 2001.

\bibitem{Espagnat}
B.~d'Espagnat.
\newblock {\em Conceptual foundations of quantum mechanics}.
\newblock Perseus Books, Reading, 1999.

\bibitem{Liboff}
R.~L. Liboff.
\newblock The correspondence principle revisited.
\newblock {\em Physics Today}, 37:50--55, 1984.

\bibitem{Makowski}
A.~J. Makowski.
\newblock A brief survey of various formulations of the correspondence
  principle.
\newblock {\em European Journal of Physics}, 27:1133--1139, 2006.

\bibitem{Hassoun}
G.~Q. Hassoun and D.~H. Kobe.
\newblock Synthesis of the {P}lanck and {B}ohr formulations of the
  correspondence principle.
\newblock {\em American Journal of Physics}, 57:658--662, 1969.

\bibitem{Bhattacharyya}
B.~Bhattacharyya.
\newblock Looking back into {B}ohr's atom.
\newblock {\em European Journal of Physics}, 27:497--500, 2006.

\bibitem{Bohr1}
N.~Bohr.
\newblock On the constitution of atoms and molecules.
\newblock {\em Philosophical Magazine}, 26:1--25, 1913.

\bibitem{Ishiwara3}
K.~Pelogia and C.~A. Brasil.
\newblock Analysis of the {J}un {I}shiwara's `{T}he universal meaning of the
  quantum of action'.
\newblock {\em European Physical Journal H}, 43:507--521, 2017.

\bibitem{Eckert}
M.~Eckert.
\newblock How {S}ommerfeld extended {B}ohr's model of the atom (1913--1916).
\newblock {\em European Physical Journal H}, 39:141--156, 2014.

\bibitem{Sommerfeld1}
A.~Sommerfeld.
\newblock On the theory of the {B}almer series.
\newblock {\em European Physical Journal H}, 39:157--177, 2014.

\bibitem{Sommerfeld2}
A.~Sommerfeld.
\newblock The fine structure of hydrogen and hydrogen-like lines.
\newblock {\em European Physical Journal H}, 39:179--204, 2014.

\bibitem{Sommerfeld3}
A.~Sommerfeld.
\newblock Zur quantentheorie der spektrallinien.
\newblock {\em Annalen der Physik}, 51:1--94, 1916.

\bibitem{Sommerfeld4}
A.~Sommerfeld.
\newblock Zur quantentheorie der spektrallinien.
\newblock {\em Annalen der Physik}, 51:125--167, 1916.

\bibitem{Wilson}
W.~Wilson.
\newblock The quantum-theory of radiation and line spectra.
\newblock {\em Philosophical Magazine}, 29:795--802, 1915.

\bibitem{Wilson2}
W.~Wilson.
\newblock The quantum of action.
\newblock {\em Philosophical Magazine}, 31:156--162, 1916.

\bibitem{Ishiwara}
J.~Ishiwara.
\newblock Die universelle bedeutung des wirkungsquantums.
\newblock {\em Proceedings of Tokyo Mathematico--Physical Society}, 8:106--116,
  1915.

\bibitem{Ishiwara2}
J.~Ishiwara.
\newblock The universal meaning of the quantum of action.
\newblock {\em European Physical Journal H}, 43:523--536, 2017.

\bibitem{Waerden}
B.~L. van~der Waerden.
\newblock {\em Sources of quantum mechanics}.
\newblock Dover Publications Inc, Mineola, 2007.

\bibitem{Planck5}
M.~Planck.
\newblock The nobel prize in physics 1918.
\newblock http://www.nobelprize.org/nobel\_prizes/physics
  /laureates/1918/planck-lecture.
\newblock Accessed on: September 17, 2015.

\bibitem{Martins}
R.~A. Martins and P.~S. Rosa.
\newblock {\em Hist\'{o}ria da teoria qu\^{a}ntica -- a dualidade onda part\'{i}cula, de
  {E}instein a de {B}roglie}.
\newblock Editora Livraria da F\'{i}sica, São Paulo, 2014.

\bibitem{deBroglie2}
L.~de{ }Broglie.
\newblock {\em An introduction to the study of wave mechanics}.
\newblock Methuen \& Co. Ltd, Essex, 1930.

\bibitem{deBroglieOrig}
L.~de{ }Broglie.
\newblock Recherches sur la th\'{e}orie des quanta.
\newblock {\em Annales de Physique}, 10:22--128, 1925.

\bibitem{deBroglieTranslated}
L.~de{ }Broglie.
\newblock {\em On the theory of quanta}.
\newblock Fondation Louis de Broglie, Paris, 2004.

\bibitem{MacKinnon}
E.~MacKinnon.
\newblock De {B}roglie's thesis: a critical retrospective.
\newblock {\em American Journal of Physics}, 44:1047--1055, 1976.

\bibitem{Jammer2}
M.~Jammer.
\newblock {\em The philosophy of quantum mechanics: the interpretations of
  quantum mechanics in historical perspective}.
\newblock Wiley, New York, 1974.

\bibitem{Einstein1orig}
A.~Einstein.
\newblock Quantentheorie des einatomigen idealen {G}ases. {Z}weite
  {A}bhandlung.
\newblock {\em Sitzungsberichte der Preu{\ss}ischen Akademie der Wissenschaften
  (Berlin). Physikalisch-mathematische Klasse}, pages 3--14, 1925.

\bibitem{BKS}
N.~Bohr, H.~K. Kramers, and J.~C. Slater.
\newblock The quantum theory of radiation.
\newblock {\em Philosophical Magazine}, 47:785--802, 1924.

\bibitem{Heisenberg}
W.~Heisenberg.
\newblock {\em Physics and philosophy -- {T}he revolution in modern science}.
\newblock Harper Perennial, New York, 2007.

\bibitem{Joas}
C.~Joas and C.~Lehner.
\newblock The classical roots of wave mechanics: {S}chr\"{o}dinger's
  transformations of the optical-mechanical analogy.
\newblock {\em Studies in History and Philosophy of Modern Physics},
  40:338--351, 2009.

\bibitem{Koberle}
R.~K\"{o}berle.
\newblock Sobre a gênese da mec\^{a}nica ondulat\'{o}ria.
\newblock {\em Revista Brasileira de F\'{i}sica}, 9:243--273, 1979.

\bibitem{Wessels}
L.~Wessels.
\newblock Schr\"{o}dinger's route to wave mechanics.
\newblock {\em Studies in History and Philosophy of Science}, 10:311--340,
  1977.

\bibitem{Fetter}
A.~L. Fetter and J.~D. Walecka.
\newblock {\em Theoretical mechanics of particles and continua}.
\newblock Dover, Mineola, 2003.

\bibitem{Schroedinger2}
E.~Schr\"odinger.
\newblock Quantisierung als eigenwertproblem.
\newblock {\em Annalen der Physik}, 384:489--527, 1926.

\bibitem{Schroedinger1}
E.~Schr\"odinger.
\newblock Quantisierung als eigenwertproblem.
\newblock {\em Annalen der Physik}, 385:361--376, 1926.

\bibitem{Schrodinger}
E.~Schr\"{o}dinger.
\newblock {\em Collected papers on wave mechanics}.
\newblock Chelsea Pub. Co., New York, 1978.

\bibitem{CohenTannoudji}
C.~C. Cohen-Tannoudji, B.~Diu, and F.~Lalo\"{e}.
\newblock {\em Quantum mechanics}.
\newblock Wiley, New York, 1977.

\bibitem{Im}
G.~S. Im.
\newblock Experimental constraints on formal quantum mechanics: The emergence
  of {B}orn's quantum theory of collision processes in {G}\"{o}ttingen,
  1924--1927.
\newblock {\em Archive for History of Exact Sciences}, 50:73--101, 1995.

\bibitem{BornOrig}
M.~Born.
\newblock Zur quantenmechanik der sto{\ss}vorg\"{a}nge.
\newblock {\em Zeitschrift f\"{u}r Physik}, 37:863--867, 1926.

\bibitem{WheelerZurek}
J.~A. Wheeler and W.~H. Zurek.
\newblock {\em Quantum theory and measurement}.
\newblock Princeton University Press, Princeton, 1983.

\bibitem{BornLong}
M.~Born.
\newblock Quantenmechanik der sto{\ss}vorg\"{a}nge.
\newblock {\em Zeitschrift f\"{u}r Physik}, 38:803--827, 1926.

\bibitem{Ludwig}
G.~Ludwig.
\newblock {\em Wave mechanics}.
\newblock Pergamon Press, Oxford, 1968.

\bibitem{Butkov}
E.~Butkov.
\newblock {\em Mathematical physics}.
\newblock Addison--Wesley, Reading, 1968.

\bibitem{Morse}
P.~Morse and H.~Feshbach.
\newblock {\em Methods of theoretical physics}.
\newblock McGraw--Hill, New York, 1953.

\bibitem{Gottfried}
K.~Gottfried.
\newblock P. a. m. dirac and the discovery of quantum mechanics.
\newblock {\em American Journal of Physics}, 79:261--266, 2011.

\bibitem{Dirac3}
P.~A.~M. Dirac.
\newblock On the theory of quantum mechanics.
\newblock {\em Proceedings of the Royal Society of London A}, 112:661--677,
  1926.

\bibitem{Bunge}
M.~Bunge.
\newblock Survey of the interpretations of quantum mechanics.
\newblock {\em American Journal of Physics}, 24:272--286, 1956.

\bibitem{Hartle}
J.~B. Hartle.
\newblock Quantum mechanics of individual systems.
\newblock {\em American Journal of Physics}, 36:704--712, 1968.

\bibitem{Ballentine}
L.~E. Ballentine.
\newblock The statistical interpretation of quantum mechanics.
\newblock {\em Reviews of Modern Physics}, 42:358--380, 1970.

\bibitem{Hanson}
N.~R. Hanson.
\newblock {C}openhagen interpretation of quantum theory.
\newblock {\em American Journal of Physics}, 27:1--15, 1959.

\bibitem{Stapp}
H.~P. Stapp.
\newblock The {C}openhagen interpretation.
\newblock {\em American Journal of Physics}, 40:1098--1116, 1972.

\bibitem{Neumann}
J.~von{ }Neumann.
\newblock {\em Mathematical foundations of quantum mechanics}.
\newblock Princeton University Press, Princeton, 1953.

\bibitem{Howard}
D.~Howard.
\newblock Who invented the `{C}openhagen interpretation'? a study in
  mythology.
\newblock {\em Philosophy of Science}, 71:669--682, 2004.

\bibitem{Pauli}
W.~Pauli, L.~Rosenfeld, and V.~Weisskopf.
\newblock {\em Niels {B}ohr and the development of physics -- {E}ssays
  dedicated to {N}iels {B}ohr on the occasion of his seventieth birthday}.
\newblock Pergamon Press, London, 1955.

\bibitem{Bohr4}
N.~Bohr.
\newblock The quantum postulate and the recent development of atomic theory.
\newblock {\em Nature}, 121:580--590, 1928.

\bibitem{Charria}
J.~Rold\'{a}n-Charria.
\newblock Indivisibility, complementarity and ontology: A {B}ohrian
  interpretation of quantum mechanics.
\newblock {\em Foundations of Physics}, 44:1336--1356, 2014.

\bibitem{Omnes1}
R.~Omn\`{e}s.
\newblock {\em The interpretation of quantum mechanics}.
\newblock Princeton University Press, Princeton, 1994.

\bibitem{Bohr7}
N.~Bohr.
\newblock {\em Atomic physics and human knowledge}.
\newblock John Wiley \& Sons, New York, 1958.

\bibitem{Omnes2}
R.~Omn\`{e}s.
\newblock {\em Quantum philosophy}.
\newblock Princeton University Press, Princeton, 1994.

\bibitem{Henderson}
J.~R. Henderson.
\newblock Classes of {C}openhagen interpretations: {M}echanisms of collapse as
  typologically determinative.
\newblock {\em Studies in History and Philosophy of Modern Physics}, 41:1--8,
  2010.

\bibitem{Heisenberg1}
W.~Heisenberg.
\newblock {\em Physics and beyond -- encounters and conversations}.
\newblock Harper \& Row, New York, 1971.

\bibitem{Seth1}
S.~Seth.
\newblock Crafting the quantum: {A}rnold {S}ommerfeld and the older quantum
  theory.
\newblock {\em Studies in History and Philosophy of Science}, 39:335--349,
  2008.

\bibitem{Seth2}
S.~Seth.
\newblock Zeideutigkeit about {Z}weideutigkeit: {S}ommerfeld, {P}auli and the
  methodological origins of quantum mechanics.
\newblock {\em Studies in History and Philosophy of Science}, 40:303--315,
  2009.

\bibitem{Reichenbach}
H.~Reichenbach.
\newblock {\em Philosophic foundations of quantum mechanics}.
\newblock University of California Press, Berkeley, 1948.

\bibitem{HeisenbergOrig}
W.~Heisenberg.
\newblock \"{U}ber den anschaulichen inhalt der quantentheoretischen kinematik und
  mechanik.
\newblock {\em Zeitschrift f\"{u}r Physik}, 43:172--198, 1927.

\bibitem{Ditchburn}
R.~W. Ditchburn.
\newblock The uncertainty principle in quantum mechanics.
\newblock {\em Proceedings of the Royal Irish Academy}, 39:73--81, 1930.

\bibitem{Robertson}
H.~P. Robertson.
\newblock The uncertainty principle.
\newblock {\em Physical Review}, 34:163--164, 1929.

\bibitem{Robertson1}
H.~P. Robertson.
\newblock Proceedings of the american physical society, minutes of the new york
  meeting, feb. 21 and 22 -- {A} general formulation of the uncertainty
  principle and its classical interpretation.
\newblock {\em Physical Review}, 35:667A, 1930.

\bibitem{Robertson2}
H.~P. Robertson.
\newblock An indeterminacy relation for several observables and its classical
  interpretation.
\newblock {\em Physical Review}, 46:794--801, 1934.

\bibitem{photonOrig}
A.~Einstein.
\newblock \"{U}ber einen die erzeugung und verwandlung des lichtes betreffenden
  heuristischen gesichtspunkt.
\newblock {\em Annalen der Physik}, 17:132, 1905.

\bibitem{Bohr2}
N.~Bohr.
\newblock On the constitution of atoms and molecules.
\newblock {\em Philosophical Magazine}, 26:476--502, 1913.

\bibitem{Bohr3}
N.~Bohr.
\newblock On the constitution of atoms and molecules.
\newblock {\em Philosophical Magazine}, 26:857--875, 1913.

\bibitem{Planck3}
M.~Planck.
\newblock {\em Scientific autobiography and other papers}.
\newblock Williams \& Norgate Ltd, London, 1950.

\bibitem{Bohm}
D.~Bohm.
\newblock {\em Quantum theory}.
\newblock Dover Publications Inc, New York, 1989.

\bibitem{Peat}
F.~D. Peat.
\newblock {\em Infinity potential -- {T}he life and times of {D}avid {B}ohm}.
\newblock Basic Books, 1996.

\bibitem{Landau}
L.~D. Landau.
\newblock {\em Collected Papers of L. D. Landau}.
\newblock Pergamon Press Ltd, Oxford, 1965.

\bibitem{Fano}
U.~Fano.
\newblock Description of states in quantum mechanics by density matrix and
  operator techniques.
\newblock {\em Reviews of Modern Physics}, 29:74--93, 1957.

\bibitem{DeWitt2}
B.~S. De{W}itt and R.~N. Graham.
\newblock {\em The many-worlds interpretation of quantum mechanics: a
  fundamental exposition by Hugh Everett III}.
\newblock Princeton University Press, Princeton, 1973.

\bibitem{Jauch}
J.~M. Jauch.
\newblock {\em Foundations of quantum mechanics}.
\newblock Addison--Wesley Publishing Company, Reading, 1968.

\bibitem{Nielsen}
M.~A. Nielsen and I.~L. Chuang.
\newblock {\em Quantum Computation and Quantum Information}.
\newblock Cambridge University Press, Cambridge, 2000.

\bibitem{EPR}
A.~Einstein, B.~Podolsky, and N.~Rosen.
\newblock Can quantum-mechanical description of physical reality be considered
  complete?
\newblock {\em Physical Review}, 47:777--780, 1935.

\bibitem{Bohr6}
N.~Bohr.
\newblock Can quantum-mechanical description of physical reality be considered
  complete?
\newblock {\em Physical Review}, 48:696--702, 1935.

\bibitem{Mehra1}
J.~Mehra.
\newblock {\em The quantum principle: its interpretation and epistemology}.
\newblock D. Reidel Publishing Company, Dordrecht, 1974.

\bibitem{BornEinstein}
M.~Born.
\newblock {\em The {B}orn--{E}instein Letters}.
\newblock Macmillan, London, 1971.

\bibitem{Schrodinger1}
E.~Schr\"{o}dinger.
\newblock Discussion of probability relations between separated systems.
\newblock {\em Proceedings of the Cambridge Philosophical Society},
  31:555--563, 1935.

\bibitem{Schrodinger2}
E.~Schr\"{o}dinger.
\newblock Probability relations between separated systems.
\newblock {\em Proceedings of the Cambridge Philosophical Society},
  32:446--452, 1935.

\bibitem{catpaper}
J.~D. Trimmer.
\newblock The present situation in quantum mechanics: A translation of
  schr\"{o}dinger's `cat paradox' paper.
\newblock {\em Proceedings of the American Philosophical Society},
  124:323--338, 1980.

\bibitem{Everett}
H.~Everett{ }III.
\newblock `relative state' formulation of quantum mechanics.
\newblock {\em Reviews of Modern Physics}, 29:454--492, 1957.

\bibitem{Wheeler}
J.~A. Wheeler.
\newblock Assessment of {E}verett's 'relative state' formulation of quantum
  theory.
\newblock {\em Reviews of Modern Physics}, 29:463--465, 1957.

\bibitem{Freitas}
F.~Freitas and O.~Freire~Jr.
\newblock A formulação dos `estados relativos' da teoria qu\^{a}ntica.
\newblock {\em Revista Brasileira de Ensino de F\'{i}sica}, 30:1--15, 2008.

\bibitem{DeWitt3}
B.~S. De{W}itt.
\newblock {B}ryce {S}. {D}e{W}itt letter to {J}ohn {W}heeler, {W}heeler's copy.
\newblock http://ucispace.lib.uci.edu/handle/10575/1145.

\bibitem{Brasil1}
C.~A. Brasil and L.~A. Castro.
\newblock Understanding the pointer states.
\newblock {\em European Journal of Physics}, 36:065024, 2014.

\bibitem{Lombardi}
O.~Lombardi and L.~Vanni.
\newblock Medici\'{o}n cu\'{a}ntica y decoherencia: ¿qu\'{e} medimos cuando medimos?
\newblock {\em Scientiae Studia}, 8:273--291, 2010.

\bibitem{Zurek4}
W.~H. Zurek.
\newblock Pointer basis of quantum apparatus: Into what mixture does the wave
  packet collapse?
\newblock {\em Physical Review D}, 24:1516--1525, 1981.

\bibitem{Zurek1}
W.~H. Zurek.
\newblock Decoherence and the transition from quantum to classical.
\newblock {\em Physics Today}, 44:36, 1991.

\bibitem{Fortin}
S.~Fortin and L.~Vanni.
\newblock Quantum decoherence: a logical perspective.
\newblock {\em Foundations of Physics}, 44:1258--1268, 2014.

\bibitem{Omnes7}
R.~Omn\`{e}s.
\newblock Results and problems in decoherence theory.
\newblock {\em Brazilian Journal of Physics}, 35:207--210, 2005.

\bibitem{Schlosshauer}
M.~Schlosshauer.
\newblock Decoherence, the measurement problem, and interpretations of quantum
  mechanics.
\newblock {\em Reviews of Modern Physics}, 76:1267--1305, 2004.

\bibitem{Schlosshauer1}
M.~Schlosshauer.
\newblock {\em Decoherence and the quantum-to-classical transition}.
\newblock Springer, Berlin, 2007.

\bibitem{Zurek}
W.~H. Zurek.
\newblock Decoherence, einselection, and the quantum origins of the classical.
\newblock {\em Reviews of Modern Physics}, 75:715--775, 2003.

\bibitem{Zurek2}
W.~H. Zurek.
\newblock Decoherence and the transition from quantum to classical --
  revisited.
\newblock {\em Los Alamos Science}, 27:2--25, 2002.

\bibitem{Omnes3}
R.~Omn\`{e}s.
\newblock Logical reformulation of quantum mechanics. i. {F}oundations.
\newblock {\em Journal of Statistical Physics}, 53:893--932, 1988.

\bibitem{Omnes4}
R.~Omn\`{e}s.
\newblock Logical reformulation of quantum mechanics. ii. {I}nterferences and
  the {E}instein--{P}odolsky--{R}osen experiment.
\newblock {\em Journal of Statistical Physics}, 53:933--955, 1988.

\bibitem{Omnes5}
R.~Omn\`{e}s.
\newblock Logical reformulation of quantum mechanics. iii. {C}lassical limit
  and irreversibility.
\newblock {\em Journal of Statistical Physics}, 53:957--975, 1988.

\bibitem{Omnes6}
R.~Omn\`{e}s.
\newblock Logical reformulation of quantum mechanics. iv. {P}rojectors in
  semiclassical physics.
\newblock {\em Journal of Statistical Physics}, 57:357--382, 1989.

\bibitem{Omnes}
R.~Omn\`{e}s.
\newblock Consistent interpretations of quantum mechanics.
\newblock {\em Reviews of Modern Physics}, 64:329--382, 1992.

\bibitem{Griffiths}
R.~B. Griffiths.
\newblock Consistent histories and the interpretation of quantum mechanics.
\newblock {\em Journal of Statistical Physics}, 36:212--272, 1984.

\bibitem{Dowker}
F.~Dowker and A.~Kent.
\newblock On the consistent histories approach to quantum mechanics.
\newblock {\em Journal of Statistical Physics}, 82:1575--1646, 1996.

\bibitem{Kastner}
R.~Kastner.
\newblock `{E}inselection' of point observables: {T}he new {H}-theorem?,
  2014.
\newblock arXiv:1406.4126.

\bibitem{Aharonov1}
Y.~Aharonov, D.~Z. Albert, and L.~Vaidman.
\newblock How the result of a measurement of a component of the spin of a
  spin-$\frac{1}{2}$ particle can turn out to be 100.
\newblock {\em Physical Review Letters}, 60:1351--1354, 1988.

\bibitem{Duck}
I.~M. Duck, P.~M. Stevenson, and E.~C.~G. Sudarshan.
\newblock The sense in which a `weak measurement' of a spin-1/2 particle's
  spin component yields a value 100.
\newblock {\em Physical Review D}, 40:2112--2117, 1989.

\bibitem{Danan}
A.~Danan, D.~Farfurnik, S.~Bar-Ad, and L.~Vaidman.
\newblock Asking photons where they have been.
\newblock {\em Physical Review Letters}, 111:240402, 2013.

\bibitem{Castro2}
L.~A. Castro, C.~A. Brasil, and R.~d.~J. Napolitano.
\newblock Weak values in collision theory.
\newblock {\em Annals of Physics}, 392:272--286, 2018.

\bibitem{Ferrie}
C.~Ferrie and J.~Combes.
\newblock How the result of a single coin toss can turn out to be 100 heads.
\newblock {\em Physical Review Letters}, 113:120404, 2014.

\bibitem{Berberian}
S.~K. Berberian.
\newblock {\em Notes on Spectral Theory}.
\newblock D. Van Nostrand Company, Inc., Princeton, 1966.

\bibitem{Jauch1967}
J.~M. Jauch and C.~Piron.
\newblock Generalized localizability.
\newblock {\em Helvetica Physica Acta}, 40:559, 1967.

\bibitem{Brandt}
H.~E. Brandt.
\newblock Positive operator valued measure in quantum information processing.
\newblock {\em American Journal of Physics}, 67:434--439, 1999.

\bibitem{Hellwig}
K.-E. Hellwig and K.~Kraus.
\newblock Pure operations and measurements.
\newblock {\em Communications in Mathematical Physics}, 11:214--220, 1969.

\bibitem{Hellwig1}
K.-E. Hellwig and K.~Kraus.
\newblock Operations and measurements ii.
\newblock {\em Communications in Mathematical Physics}, 16:142--147, 1969.

\bibitem{Kraus}
K.~Kraus.
\newblock {\em States, effects and operations -- {F}undamental notions of
  quantum theory}.
\newblock Springer--Verlag, Berlin, 1983.

\bibitem{Kraus1}
K.~Kraus.
\newblock General state changes in quantum theory.
\newblock {\em Annals of Physics}, 64:311--335, 1971.

\bibitem{Maziero}
J.~Maziero.
\newblock A representação de kraus para a din\^{a}mica de sistemas qu\^{a}nticos
  abertos -- the kraus representation for the dynamics of open quantum systems.
\newblock {\em Revista Brasileira de Ensino de F\'{i}sica}, 38:e2307, 2016.

\bibitem{Breuer}
H.-P. Breuer and F.~Petruccione.
\newblock {\em The theory of open quantum systems}.
\newblock Oxford University Press, Oxford, 2002.

\bibitem{Caldeira}
A.~Caldeira.
\newblock {\em An introduction to macroscopic quantum phenomena and quantum
  dissipation}.
\newblock Cambridge University Press, Cambridge, 2014.

\bibitem{Reif}
F.~Reif.
\newblock {\em Fundamentals of statistical and thermal physics}.
\newblock Waveland Press Inc, Long Grove, 1965.

\bibitem{Brasil4}
C.~A. Brasil, F.~F. Fanchini, and R.~d.~J. Napolitano.
\newblock A simple derivation of the {L}indblad equation.
\newblock {\em Revista Brasileira de Ensino de F\'{i}sica}, 35:1303, 2013.

\bibitem{Lindblad}
G.~{L}indblad.
\newblock On the generators of quantum dynamical semigroups.
\newblock {\em Communications in Mathematical Physics}, 48:119--130, 1976.

\bibitem{Adler}
S.~L. Adler.
\newblock Derivation of the {L}indblad generator structure by use of the it\={o}
  stochastic calculus.
\newblock {\em Physics Letters A}, 265:58--61, 2000.

\bibitem{Adler1}
S.~L. Adler.
\newblock Corrigendum to: Derivation of the {L}indblad generator structure by
  use of the it\={o} stochastic calculus.
\newblock {\em Physics Letters A}, 267:212, 2000.

\bibitem{Peres1}
A.~Peres.
\newblock Classical interventions in quantum systems. i. {T}he measuring
  process.
\newblock {\em Physical Review A}, 61:022116, 2000.

\bibitem{Weinberg}
S.~Weinberg.
\newblock What happens in a measurement?
\newblock {\em Physical Review A}, 93:032124, 2016.

\bibitem{Distler}
J.~Distler and S.~Paban.
\newblock {v}on {N}eumann's formula, measurements and the {L}indblad equation,
  2017.
\newblock arXiv:1702.01724v1.

\bibitem{Brasil5}
C.~A. Brasil and R.~d.~J. Napolitano.
\newblock The master equation for the reduced open-system dynamics, including a
  {L}indbladian description of the finite-duration measurement.
\newblock {\em European Physical Journal Plus}, 126:91, 2011.

\bibitem{Brasil2}
C.~A. Brasil, L.~A. Castro, and R.~d.~J. Napolitano.
\newblock Protecting a quantum state from environmental noise by an
  incompatible finite-time measurement.
\newblock {\em Physical Review A}, 84:022112, 2011.

\bibitem{Cresser}
J.~D. Cresser, S.~M. Barnett, J.~Jeffers, and D.~T. Pegg.
\newblock Measurement master equation.
\newblock {\em Optics Communications}, 264:352--361, 2006.

\bibitem{Jacobs1}
K.~Jacobs and D.~A. Steck.
\newblock A straightforward introduction to continuous quantum measurement.
\newblock {\em Contemporary Physics}, 47:279--303, 2006.

\bibitem{Klyshko}
D.~N. Klyshko.
\newblock Reduction of the wave function: an operational approach.
\newblock {\em Physics Letters A}, 243:179--186, 1998.

\bibitem{Wiseman}
H.~W. Wiseman and G.~Milburn.
\newblock {\em Quantum measurement and control}.
\newblock Cambridge University Press, Cambridge, 2009.

\bibitem{Brasil}
C.~A. Brasil, L.~A. Castro, and R.~d.~J. Napolitano.
\newblock How much time does a measurement take?
\newblock {\em Foundations of Physics}, 43:642--655, 2015.

\bibitem{Brasil3}
C.~A. Brasil, L.~A. Castro, and R.~d.~J. Napolitano.
\newblock Efficient finite-time measurements under thermal regimes.
\newblock {\em European Physical Journal Plus}, 129:206, 2014.

\bibitem{Misra}
B.~Misra and E.~C.~G. Sudarshan.
\newblock The {Z}eno's paradox in quantum theory.
\newblock {\em Journal of Mathematical Physics}, 18:756, 1977.

\bibitem{Facchi}
P.~Facchi and S.~Pascazio.
\newblock Quantum {Z}eno phenomena: Pulsed versus continuous measurement.
\newblock {\em Fortschritte der Physik}, 49:941, 2001.

\bibitem{Itano}
W.~M. Itano, D.~J. Heinzen, J.~J. Bollinger, and D.~J. Wineland.
\newblock Quantum {Z}eno effect.
\newblock {\em Physical Review A}, 41:2295, 1990.

\bibitem{AntiZeno}
A.~G. Kofman and G.~Kurizki.
\newblock Acceleration of quantum decay processes by frequent observations.
\newblock {\em Nature}, 405:546--550, 2000.

\bibitem{chemistry}
S.~McArdle, S.~Endo, A.~Aspuru-Guzik, S.~Benjamin, and X.~Yuan.
\newblock Quantum computational chemistry, 2019.
\newblock arXiv:quant-ph/1808.10402v2.

\bibitem{Braginsky1}
V.~B. Braginsky and F.~Y. Khalili.
\newblock {\em Quantum measurement}.
\newblock Cambridge University Press, Cambridge, 1992.

\bibitem{Busch}
P.~Busch, P.~J. Lahti, and P.~Mittelstaedt.
\newblock {\em The quantum theory of measurement}.
\newblock Springer--Verlag, Berlin, 1996.

\bibitem{Jacobs}
K.~Jacobs.
\newblock {\em Quantum measurement theory and its applications}.
\newblock Cambridge University Press, Cambridge, 2014.

\bibitem{Mensky1}
M.~B. Mensky.
\newblock {\em Continuous quantum measurements and path integrals}.
\newblock IOP Publishing, London, 1993.

\bibitem{Mensky}
M.~B. Mensky.
\newblock {\em Quantum measurements and decoherence -- models and
  phenomenology}.
\newblock Springer-Science+Business, Dordrecht, 2000.

\bibitem{Bassi}
A.~Bassi and G.~C. Ghirardi.
\newblock Dynamical reduction models.
\newblock {\em Physics Reports}, 379:257--426, 2003.

\bibitem{Namiki}
M.~Namiki and S.~Pascazio.
\newblock Quantum theory of measurement based on the many-hilbert-space
  approach.
\newblock {\em Physics Reports}, 232:201--411, 1993.

\bibitem{Tombesi}
P.~Tombesi and D.~F. Walls.
\newblock {\em Quantum measurements in optics -- Proceedings of a {NATO}
  Advanced Research Workshop on Quantum Measurements in Optics}.
\newblock Springer Science+Business, New York, 1992.

\bibitem{Percival}
I.~Percival.
\newblock {\em Quantum state diffusion}.
\newblock Cambridge University Press, Cambridge, 1998.

\bibitem{Ghirardi}
G.~Ghirardi, R.~Rimini, and T.~Weber.
\newblock Unified dynamics for microscopic and macroscopic systems.
\newblock 34:470--491, 1986.

\bibitem{Ghirardi1}
G.~Ghirardi, P.~Pearle, and A.~Rimini.
\newblock Markov processes in {H}ilbert space and continuous spontaneous
  localization of systems of identical particles.
\newblock {\em Physical Review A}, 42:78--89, 1990.

\bibitem{Pearle}
P.~Pearle.
\newblock Combining stochastic dynamical state-vector reduction with
  spontaneous localization.
\newblock {\em Physical Review A}, 39:2277--2289, 1989.

\bibitem{Wick}
G.~C. Wick, A.~S. Wightman, and E.~P. Wigner.
\newblock The intrinsic parity of elementary particles.
\newblock {\em Physical Review}, 88:101--105, 1952.

\bibitem{Wick1}
G.~C. Wick, A.~S. Wightman, and E.~P. Wigner.
\newblock Superselection rule for charge.
\newblock {\em Physical Review D}, 1:3267--3267, 1970.

\bibitem{Bub}
J.~Bub.
\newblock How to solve the measurement problem of quantum mechanics.
\newblock {\em Foundations of Physics}, 18:701--722, 1988.

\bibitem{Landsman}
N.~P. Landsman.
\newblock Observation and superselection in quantum mechanics, 1994.
\newblock arXiv:hep-th/9411173.

\bibitem{Giulini}
D.~Giulini.
\newblock Superselection rules, 2007.
\newblock arXiv:0710.1516.

\bibitem{Milburn}
G.~J. Milburn and D.~F. Walls.
\newblock Quantum nondemolition measurements via quadratic coupling.
\newblock {\em Physical Review A}, 28:2065--2070, 1983.

\bibitem{Walls1}
D.~F. Walls, M.~J. Collet, and G.~J. Milburn.
\newblock Analysis of a quantum measurement.
\newblock {\em Physical Review D}, 32:3208, 1985.

\bibitem{Grangier}
P.~Grangier, J.~A. Levenson, and J.-P. Poizat.
\newblock Quantum non-demolition measurements in optics.
\newblock {\em Nature}, 396:537--542, 1998.

\bibitem{Olimpio}
O.~P. S\'{a}~Neto, M.~C. de{ }Oliveira, and G.~J. Milburn.
\newblock Temperature measurement and phonon number statistics of a
  nanoelectromechanical resonator.
\newblock {\em New Journal of Physics}, 17:093010, 2015.

\bibitem{Unnikrishnan}
C.~S. Unnikrishnan.
\newblock Quantum non-demolition measurements: concepts, theory and practice.
\newblock {\em Current Science}, 109:2052, 2015.

\bibitem{Braginsky}
V.~B. Braginsky and F.~Y. Khalili.
\newblock Quantum nondemolition measurements.
\newblock {\em Reviews of Modern Physics}, 68, 1996.

\end{thebibliography}


\end{document}